\documentclass[10pt,citeautoscript,twocolumn,floatfix,aps,prx,superscriptaddress]{revtex4-2}

\usepackage{amsmath}
\usepackage{graphicx}
\usepackage{epstopdf}
\usepackage{placeins}
\usepackage{natbib}
\usepackage{array}
\usepackage{braket}
\usepackage[usenames,dvipsnames]{color}
\usepackage{dcolumn}
\usepackage{bm}
\usepackage[hidelinks]{hyperref}
\usepackage{makecell}
\usepackage{tabularx}
\usepackage{siunitx,booktabs}
\usepackage{multirow}

\usepackage{array}

\newcommand{\PreserveBackslash}[1]{\let\temp=\\#1\let\\=\temp}
\newcolumntype{C}[1]{>{\PreserveBackslash\centering}p{#1}}
\newcolumntype{R}[1]{>{\PreserveBackslash\raggedleft}p{#1}}
\newcolumntype{L}[1]{>{\PreserveBackslash\raggedright}p{#1}}

\newcommand{\NbX}[1]{Nb$_3$#1$_8$}

\begin{document}

\title{From strong to weak correlations in breathing-mode kagome van der Waals materials: Nb$_3$(F,Cl,Br,I)$_8$ as a robust and versatile platform for many-body engineering}

\author{Joost~Aretz}
\author{Sergii Grytsiuk}
\affiliation{Institute for Molecules and Materials, Radboud University, Heijendaalseweg 135, 6525AJ Nijmegen, The Netherlands}

\author{Xiaojing Liu}
\affiliation{Zernike Institute for Advanced Materials, University of Groningen, 9747 AG Groningen, The Netherlands}

\author{Giovanna Feraco}
\affiliation{Zernike Institute for Advanced Materials, University of Groningen, 9747 AG Groningen, The Netherlands}
\author{Chrystalla Knekna}
\affiliation{Zernike Institute for Advanced Materials, University of Groningen, 9747 AG Groningen, The Netherlands}
\affiliation{Van der Waals-Zeeman Institute, Institute of Physics, University of Amsterdam, Science Park 904, 1098 XH, Amsterdam, The Netherlands}
\author{Muhammad Waseem}
\author{Zhiying Dan}
\affiliation{Zernike Institute for Advanced Materials, University of Groningen, 9747 AG Groningen, The Netherlands}

\author{Marco Bianchi}
\author{Philip Hofmann}
\affiliation{Department of Physics and Astronomy, Interdisciplinary Nanoscience Center (iNANO), Aarhus University, 8000 Aarhus C, Denmark}

\author{Mazhar N. Ali}
\affiliation{Kavli Institute of Nanoscience, Delft University of Technology,
Lorentzweg 1, 2628 CJ Delft, the Netherlands}

\author{Mikhail I. Katsnelson}
\affiliation{Institute for Molecules and Materials, Radboud University, Heijendaalseweg 135, 6525AJ Nijmegen, The Netherlands}
\affiliation{Constructor Knowledge Institute, Constructor University, Campus Ring 1, 28759 Bremen, Germany}

\author{Antonija Grubi\v{s}i\'{c}-\v{C}abo}
\affiliation{Zernike Institute for Advanced Materials, University of Groningen, 9747 AG Groningen, The Netherlands}

\author{Hugo U. R. Strand}
\affiliation{School of Science and Technology, \"Orebro University, SE-701 82 \"Orebro, Sweden}

\author{Erik G. C. P. van Loon}
\affiliation{NanoLund and Division of Mathematical Physics, Department of Physics, Lund University, Lund, Sweden}

\author{Malte~Rösner}
\email{m.roesner@science.ru.nl}
\affiliation{Institute for Molecules and Materials, Radboud University, Heijendaalseweg 135, 6525AJ Nijmegen, The Netherlands}

\begin{abstract}

By combining ab initio downfolding with cluster dynamical mean-field theory, we study the degree of correlations in monolayer, bilayer and bulk breathing-mode kagome van der Waals materials Nb$_3$(F,Cl,Br,I)$_8$. Our new material-specific many-body model library shows that in low-temperature bulk structures the Coulomb correlation strength steadily increases from I to F, allowing us to identify Nb$_3$I$_8$ as a weakly correlated insulator, Nb$_3$Br$_8$ and Nb$_3$Cl$_8$ as strongly correlated insulators, and Nb$_3$F$_8$ as a prototypical bulk Mott-insulator. Angle-resolved photoemission spectroscopy measurements comparing Nb$_3$Br$_8$ and Nb$_3$I$_8$ allow us to experimentally confirm these findings by revealing spectroscopic footprints of the degree of correlation. Our calculations uncover how the thickness and the stacking affect the degree of correlations and predict that the entire material family can be tuned into correlated charge-transfer or Mott-insulating phases upon doping. Our magnetic property analysis based on our model parameter library additionally confirms that inter-layer magnetic interactions drive the lattice phase transition to the low-temperature structures. The accompanying bilayer hybridization through inter-layer dimerization yields magnetic singlet-like ground states in the Cl, Br, and I compounds. We further prove that all low-temperature compounds are dynamically stable and that electron-phonon coupling to the low-energy subspace is suppressed. Our findings establish Nb$_3$X$_8$ as a robust, versatile, and tunable class for van der Waals-based Coulomb and Mott engineering with a rich phase diagram and allow us to speculate on the symmetry-breaking effects necessary for the recently observed Josephson diode effect in NbSe$_2$/Nb$_3$Br$_8$/NbSe$_2$ heterostructures.

\end{abstract}

\maketitle

\section{Introduction}

Correlation effects resulting from electron-electron (Coulomb) interactions are omnipresent in solids. Depending on the relative Coulomb interaction strength, they manifest in various ways, starting from renormalizations of quasiparticle dispersions in the case of weak correlations~\cite{onida02}, via induced shake-off/replica bands in the case of intermediate correlations~\cite{Bostwick_2010,caruso15}, or in form of completely different ground states in the case of strong correlations~\cite{Imada_1998,kotliar2004strongly}. Being able to tune the degree of correlations would therefore allow to engineer many-body properties of quantum materials~\cite{basov2017}. 

The layered structure of van der Waals (vdW) materials makes them uniquely tunable~\cite{geim2013van}, with opportunities such as field-effect-doping~\cite{novoselov2005two,ye2012}, substrate engineering~\cite{raja2017coulomb}, heterostructuring~\cite{geim2013van}, and moir\'e engineering~\cite{cao2018correlated}, but also via chemical modifications~\cite{basov2017}. Furthermore, within the library of layered vdW materials, we now have access to a variety of correlated ground states including magnetism~\cite{burch2018magnetism}, charge-density wave (CDW) order~\cite{wilson1975charge}, superconductivity~\cite{xi_ising_2016,cao2018unconventional}, as well as Mott-insulators~\cite{Perfetti2003,cao2018correlated,butler2020mottness}. 
However, the complex electronic structure of the systems in question and the inherent complexity and partial fragility of correlated electron physics make it challenging to fully understand the microscopic origin of the observed phenomena, let alone making predictive simulations. 
Thus, simple and robust correlated layered material systems with varying and tunable degrees of correlation strengths are highly desirable. 

Here, we show that the transition metal halide family \NbX{X} holds up to this promise. 
These so-called breathing-mode kagome vdW materials~\cite{wang2023quantum} host only one band per monolayer close to the Fermi level, which is cleanly separated from the rest of the electronic structure as a result of in-plane trimerization of the transition metal atoms. In multilayer structures this further competes with an out-of-plane dimerization, especially at low-temperatures.

\begin{figure}
    \centering
    \includegraphics[width=1\linewidth]{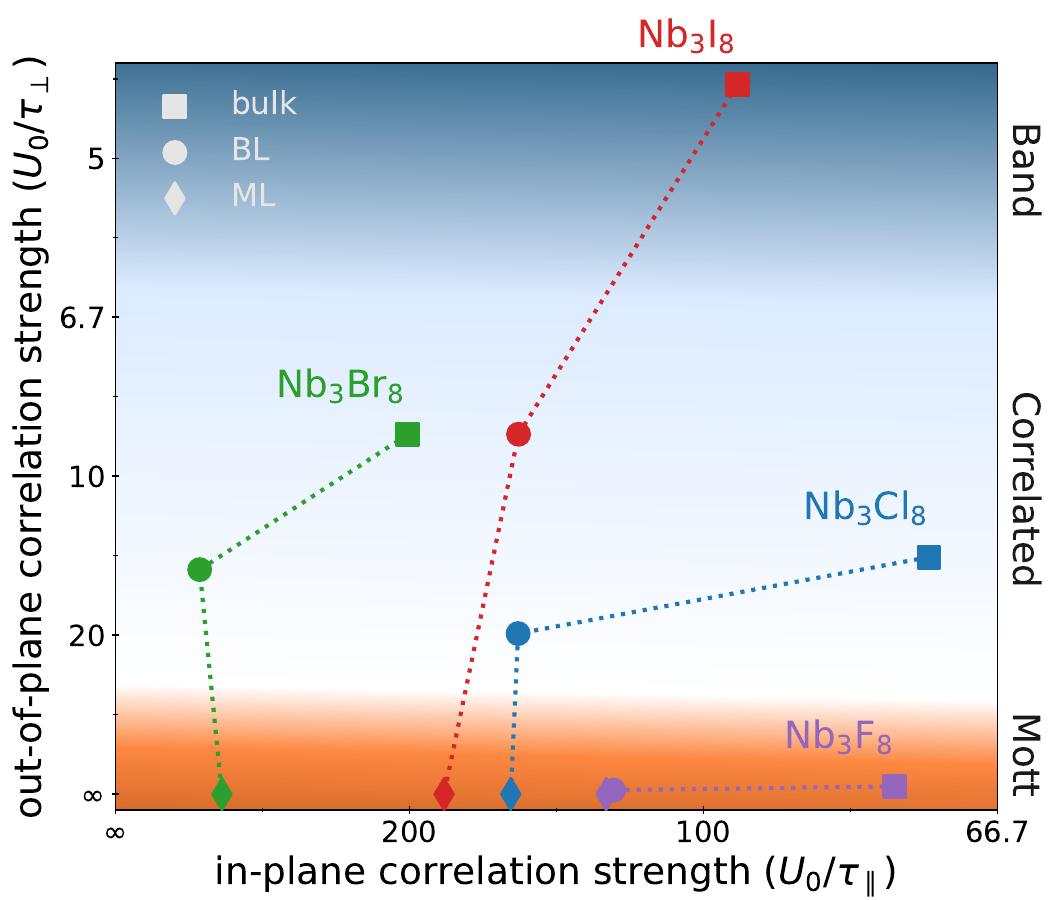}
    \caption{\textbf{Phase diagram of \NbX{X} for monolayer, bilayer, and bulk structures in the low-temperature phase.} The horizontal axis measures the in-plane correlation strength and the vertical axis the out-of-plane correlation strength with $\tau_\parallel$, $\tau_\perp$ and $U_0$ being material and structure specific effective in- and out-of-plane hopping and local Coulomb interaction matrix elements. The phase-space is divided into Mott, correlated, and band insulators.}
    \label{fig-new:phasediagram}
\end{figure}

Using ab initio downfolding to molecular orbitals, we derive material-specific minimal models, including single-particle (hopping) and many-body screened Coulomb interaction matrix elements for the entire family \NbX{X}, with $\text{X} \in \{\text{F},\text{Cl},\text{Br},\text{I}\}$ ranging from monolayer, bilayer, and to bulk structures. Studying them using cluster dynamical mean-field theory~\cite{Maier_2005} shows how the balance between kinetic and screened Coulomb interaction energies changes from \NbX{F} to \NbX{I} as well as from monolayer to bulk structures. In this way, we prove that \NbX{X} realizes several correlated phases ranging from weakly correlated band insulators to correlated and strongly correlated Mott insulators, which we summarize in the generalized phase diagram spanned by the in-plane versus out-of-plane correlation strengths, as shown in Fig.~\ref{fig-new:phasediagram}. We further clarify the distinction between the band insulator and correlated insulator through experimental angle-resolved photoemission spectroscopy and by theoretically investigating their response to charge doping. Finally, we present a fully ab initio investigation of the magnetic ground state properties of all low-temperature bulk compounds and specifically on the high- to low-temperature phase transition in \NbX{Cl} and \NbX{Br}. This finally allows us to suggest a new mechanism for time-reversal symmetry-breaking in finite stacks of \NbX{X}, which is needed to understand the experimentally observed Josephson diode effect in NbSe$_2$/\NbX{Br}/NbSe$_2$ heterostructures.

\subsection*{\NbX{X} Overview}

\NbX{X} belongs to the class of layered vdW kagome materials, which are intensively studied due to the interplay of their non-trivial electronic band topology and intriguing electron correlation phenomena~\cite{wang2023quantum,Haraguchi_2024_review}. \NbX{X} specifically belongs to the sub-class of breathing-mode kagome materials, with alternating in-plane bond lengths, as illustrated in Fig.~\ref{fig:structure}. Among them, \NbX{Cl} and \NbX{Br} have recently gained much attention due to their promises to represent prototypical single-band or few-band Mott insulators~\cite{hu_correlated_2023,Zhang_2023_PRB,Gao_2023,grytsiuk2024nb3cl8,date2024mott, yang2025gatetunableroomtemperaturemott}, possibly accompanied by strongly correlated magnetic ground states~\cite{hu_correlated_2023,Liu_2024,mangeri_2024,grytsiuk2024nb3cl8}. \NbX{Br} has received further attention due to its role in inducing field-free Josephson diode effects in NbSe$_2$/\NbX{Br}/NbSe$_2$ heterostructures~\cite{wu2022field}. Although some scenarios have been suggested~\cite{zhang_JDiode_2022}, it is not yet clear what microscopic behavior yields the required symmetry breakings and whether it is intrinsic to \NbX{Br} or related to the interfaces.

Experimentally, X-ray diffraction shows lattice phase transitions in \NbX{Cl} and \NbX{Br} from high- to low-temperature structures, which can be understood as a shift in the vdW layer stacking~\cite{NbCl_structC2,NbCl_expmagsusc}. Raman spectroscopy \cite{Jeff_2023} corroborates this interpretation but also indicates that phonon symmetries are not affected by the transition. Angle-resolved photoemission spectroscopy (ARPES) data shows gapped electronic structures with rather flat, albeit broadened, bands below the Fermi level for bulk \NbX{I}~\cite{regmi_spectroscopic_2022}, \NbX{Cl}~\cite{Sun_2022} and \NbX{Br}~\cite{Regmi_2023_prb,date2024mott} as well as their admixtures~\cite{Gao_2023}. 
Spectral gaps in \NbX{Cl} have also been reported from transport measurements \cite{Yoon_2020,yang2025gatetunableroomtemperaturemott}.
For Nb$_3$Cl$_{8-x}$Br$_x$, magnetic susceptibilities further show Curie-Weiss behavior at elevated temperatures in the high-temperature phases, accompanied by sudden drops upon entering the low-temperature phases~\cite{NbCl_structC2,NbCl_expmagsusc,Pasco_2019,Liu_2024,Sun_2022,Gao_2023}. This has been interpreted as a transition from a paramagnetic to a non-magnetic phase, and the finite susceptibility at very low temperatures is ascribed to magnetic impurities~\cite{NbCl_structC2,NbCl_expmagsusc}.

Theoretically, the modeling of these materials takes place in several steps. A first impression is given by Density Functional Theory (DFT), which has been used to study the monolayer, bilayer, and bulk structures, in both low- and high-temperature crystal structures~\cite{Conte_2020,Peng_2020,regmi_spectroscopic_2022,Zhang_2023_PRB,grytsiuk2024nb3cl8}. Starting with the monolayer, DFT shows a single, rather flat band crossing the Fermi level, which is half-filled and thus hosts exactly one electron per unit cell. 
At low temperatures and going to bilayer and bulk structures, there is an alternating strong and weak hybridization between adjacent layers, which leads to a dimerization in the out-of-plane direction~\cite{date2024mott,grytsiuk2024nb3cl8,Gao_2023,Zhang_2023_PRB}. 
In this case, DFT predicts two rather flat bands around the Fermi level, which show a small, but finite hybridization gap.
Since flat bands are indicators for low kinetic energies, the role of competing Coulomb interactions needs to be carefully taken into account. Mean-field DFT calculations are, however, known to underestimate the impact of the latter, especially in the case of $d$ orbitals which dominate the low-energy space in the case of \NbX{X}. Therefore, beyond mean-field theories, which take possibly strong Coulomb interactions accurately into account, are necessary. So far this has been done only inconsistently for monolayer \NbX{Cl}~\cite{hu_correlated_2023,Gao_2023,stepanov2024signatureschargeicestate}, generic bilayer models approximating \NbX{Cl} and \NbX{Br}~\cite{Zhang_2023_PRB}, as well as for simplified models of bulk \NbX{Cl}~\cite{grytsiuk2024nb3cl8} and \NbX{Br}~\cite{date2024mott} using different variants of higher-level many-body theory or alternative supercell approaches~\cite{xiong2024supercell}. All of these calculations hint towards an intriguing interplay between enhanced Coulomb interactions and kinetic energies, which seem to drive variants of Mott insulating states. For bulk structures, this especially requires a careful analysis of the impact of the out-of-plane dimerization, which has so far only been discussed qualitatively for bilayers of \NbX{Br} and \NbX{Cl}~\cite{Zhang_2023_PRB} as well as for bulk \NbX{Br}~\cite{date2024mott}. Most importantly, the role of screening and its impact to the actual material-specific Coulomb interactions along the \NbX{X} family members has not been studied yet.

\section{Results}

To systematically, material-specifically, and quantitatively study these effects, we calculate and investigate all kinetic and Coulomb interaction terms of \NbX{X} with X $\in$ \{F,Cl,Br,I\} in their monolayer, bilayer, and bulk forms and solve the resulting models using cluster dynamical mean-field theory. Since the high-temperature structure of \NbX{I} has not been observed yet and since correlation effects are most intriguing in the low-temperature phases, we mostly focus on the latter in the following and will only comment on differences in the high-temperature structures of \NbX{Cl} and \NbX{Br} where appropriate. We further put a special focus on the Coulomb interactions and, most importantly, their screening to uncover the exact microscopic origin of various resulting phases. We apply the same formalism to all studied compounds allowing for direct comparisons.

\subsection{\NbX{X} Crystal Structure\label{sec:structure}}

\begin{figure}[!t]
    \centering
    \includegraphics[width=0.4\textwidth]{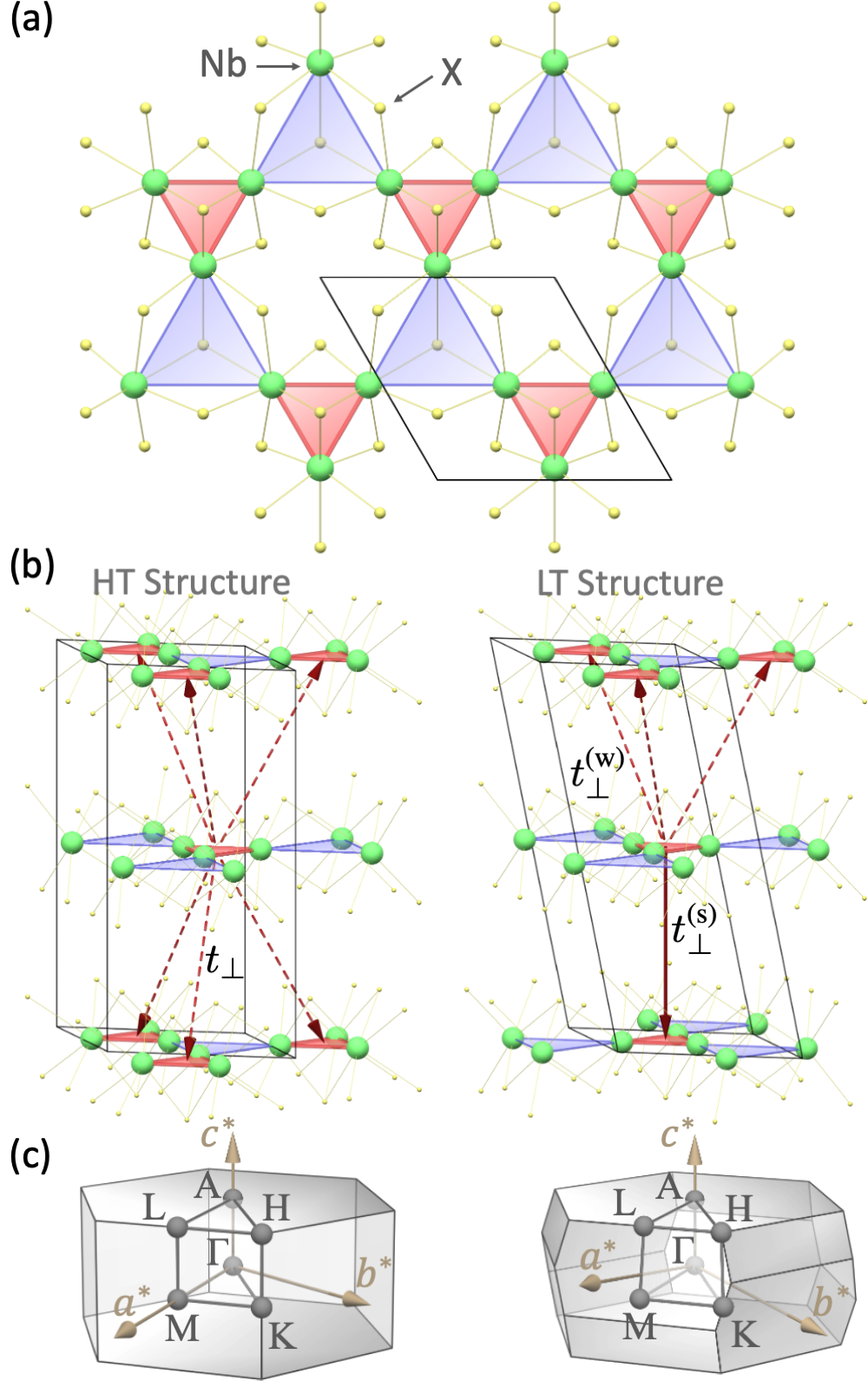}
    \caption{
    \textbf{Nb$_3$X$_8$ Crystal Structures.} (a) single layer (top view) 
    and (b) bulk structures (side view) at high and low temperatures, respectively. Darker red solid (dashed) arrows indicate strong  (weak) interlayer couplings $t_\perp^\text{s}$  ($t_\perp^\text{w}$) between the nearest trimers. The gray lines in (a) and (b) represent the primitive unit cell. (c) illustrates the Brillouin zones and high-symmetry points for both structures.}
    \label{fig:structure}
\end{figure}

The breathing-mode kagome structure depicted in Fig.~\ref{fig:structure}~(a) yields an in-plane trimerization of Nb atoms (see small triangles in red). In the out-of-plane direction, there exist two stackings, referred to as the high- (HT) and low-temperature (LT) structures, which are depicted in Fig.~\ref{fig:structure}~(b) together with their Brillouin zones in Fig.~\ref{fig:structure}~(c). The LT stacking leads to a dimerization in the out-of-plane direction with alternating strong and weak hybridization between adjacent layers, as indicated by dashed and solid line arrows in Fig.~\ref{fig:structure}~(b) and discussed in more detail below.
\NbX{Cl} and \NbX{Br} undergo a structural phase transition that changes the stacking from the HT to the LT structures at around $90\,$K and $380\,$K, respectively \cite{NbCl_expmagsusc, NbCl_structC2, Kim_2023, Pasco_2019}. While it is hypothesized that \NbX{I} exhibits a similar transition at even higher temperatures, this has yet to be confirmed experimentally. 

At high temperatures, crystals of both \NbX{Cl} and \NbX{Br} belong to the P$\bar{3}$m1 space group. However, at low temperatures, \NbX{Br} and \NbX{I} have been reported to adopt the R$\bar{3}$m space group \cite{Pasco_2019, Kim_2023}. In contrast, the LT crystal structure of \NbX{Cl} has been described inconsistently in different studies, with reported space groups of R$3$~\cite{NbCl_expmagsusc}, R$\bar{3}$m~\cite{Kim_2023, Jeff_2023, Pasco_2019}, and C2/m~\cite{NbCl_structC2}. For consistency we adopt in the following the R$\bar{3}$m point group for \NbX{Cl} and 
note that a symmetry reduction has minimal impact on the correlation effects and electronic structure~\cite{grytsiuk2024nb3cl8}. To the best of our knowledge, only Cl, Br, and I have been synthesized, but the addition of \NbX{F}* gives a unified view of the full series. To emphasize that \NbX{F}* is putative, we will use the starred notation. 

To get a homogeneous crystal structure set, we perform for all compounds in their LT bulk phases (and for \NbX{Cl} and \NbX{Br} also in their HT phases) a lattice and structural optimization keeping only the space group fixed to R$\bar{3}$m (P$\bar{3}$m1). We find all of them, including the putative \NbX{F}*, to be dynamically stable as further discussed in section \ref{sec:phonons}. The mono- and bilayer structures are taken from these relaxed bulk phases by introducing an appropriate vacuum spacing.

\subsection{Electronic Structure}

\subsubsection{In-Plane Trimerization as Origin of Robust Flat Bands}

\begin{figure}[t]
    \centering
    \includegraphics[width=0.45\textwidth]{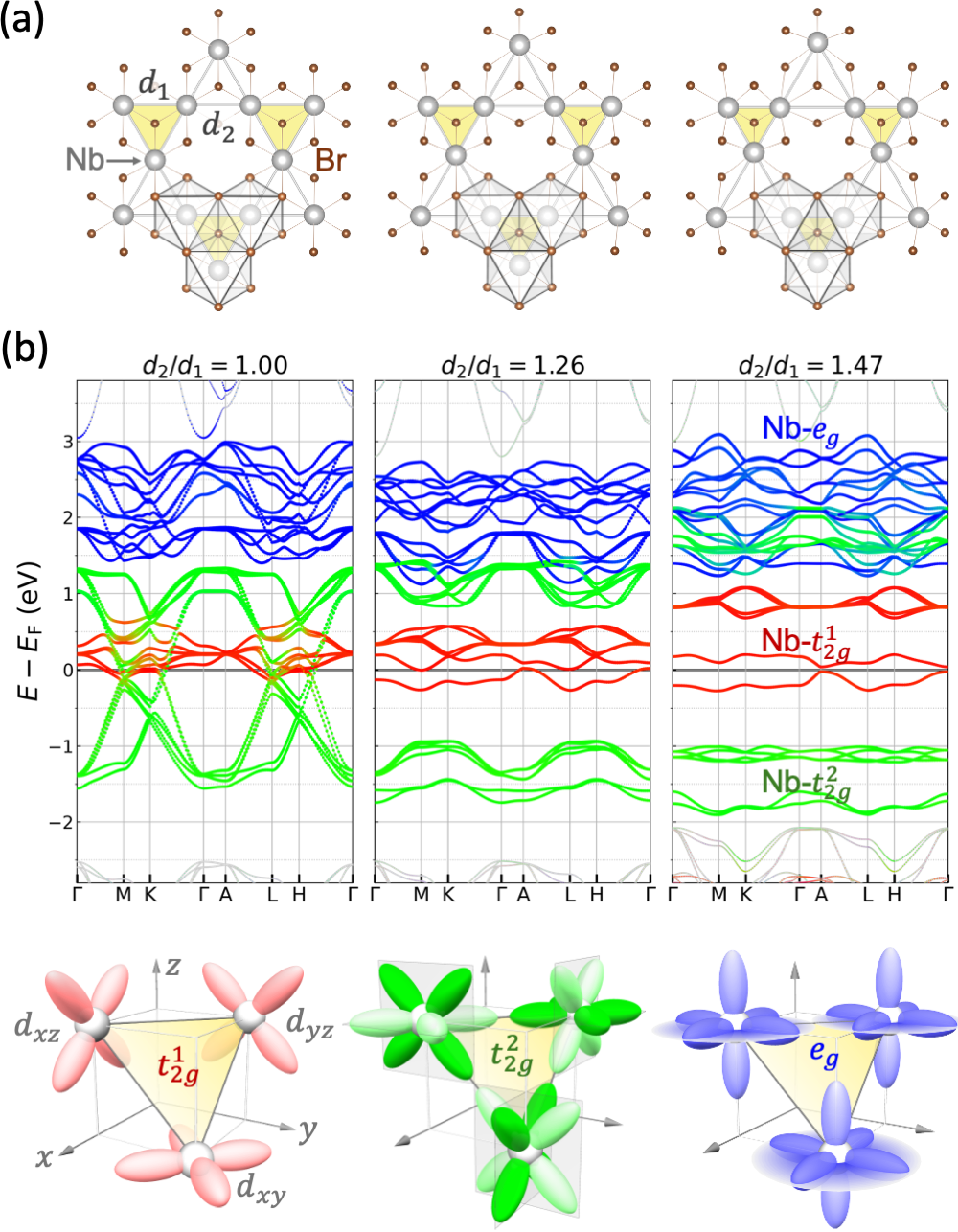}
    \caption{\textbf{Breathing Mode-Induced Flat Band Formation in Low-Temperature \NbX{Br}.} (a) and (b) illustrate the crystal structure (showing only one layer for clarity) and the electronic structure as a function of the size ratio of the small and large triangles $d_2/d_1$. The electronic structures in (b) are weighted between Nb$-t_{2g}^1$ (in red), Nb$-t_{2g}^2$ (in green), and Nb-$e_g$ (in blue) states, respectively. Black lines in (a) depict the ideal (left panel) and distorted (two following panels) octahedral environment surrounding the Nb atoms.}
    \label{fig:dftflatbands}
\end{figure}

\begin{figure*}[ht!]
    \centering
    \includegraphics[width=0.99\textwidth]{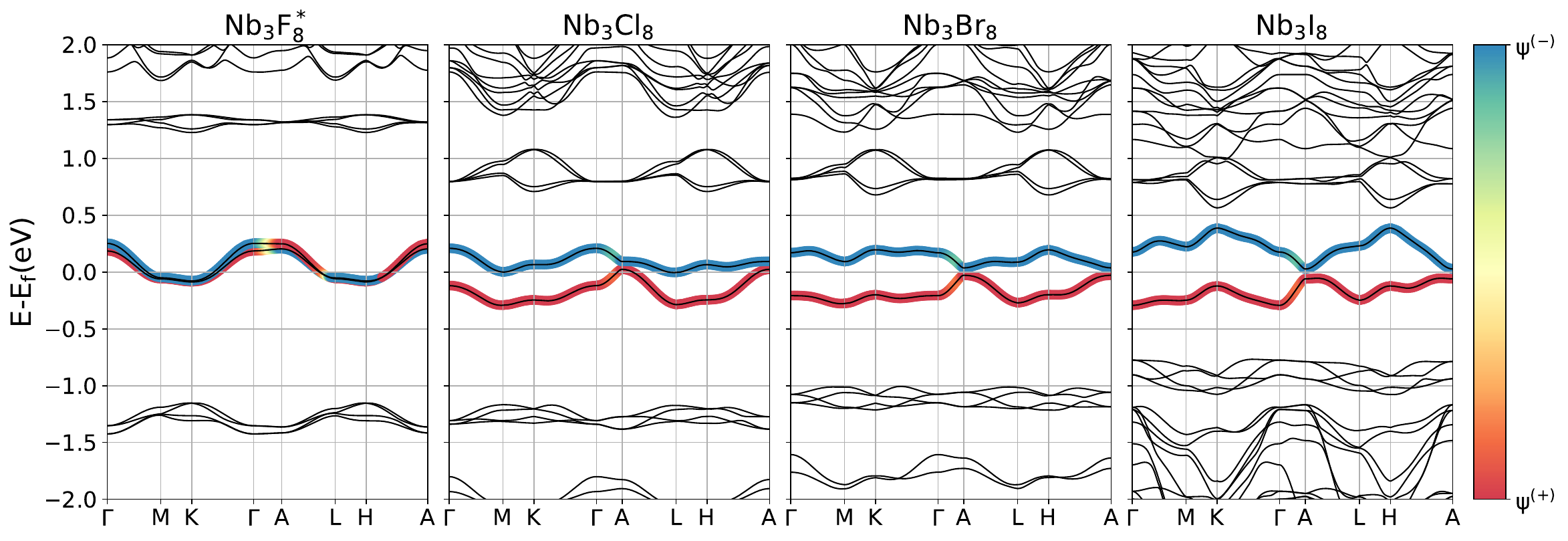}
    \caption{\textbf{Nb$_3$X$_8$ DFT Bandstructures.} Colors indicate the weight of the bonding ($\psi^{(+)}$) and anti-bonding ($\psi^{(-)}$) orbitals in the Wannier basis. $\psi^{(\pm)} = |\langle\psi^{(\pm)}|\psi\rangle|^2$, where $|\psi\rangle$ is the eigenfunction in the Wannier basis.}
    \label{fig:dft}
\end{figure*}

The formation of flat bands in \NbX{X} plays a crucial role in their electronic and magnetic properties. As the underlying lattice structure is of kagome type, it has been suspected that the flat bands originate from the topologically non-trivial flat bands of the well-known generic kagome band structure as described, e.g., by tight-binding models with isotropic hoppings; see for example Refs.~\onlinecite{meng_2024,Sun_2022}. However, as the low energetic orbitals of all \NbX{X} compounds are of predominant Nb $d$ character, this simplified picture has to be reconsidered~\cite{zhou_orbital_2023}. 
Therefore, we calculated the electronic structure of LT \NbX{Br} at different levels of the breathing mode perturbation and thus degree of in-plane Nb trimerization, which is measured by the ratio $d_2/d_1$, where $d_1$ and $d_2$ represent the sizes of the small and large triangles, respectively, cf. Fig.~\ref{fig:dftflatbands}a. When the structure is a conventional kagome lattice ($d_1 = d_2$), the octahedral crystal field causes the $d$ orbitals of the Nb ions to split into the lower-energy $t_{2g}$ ($d_{xy}$, $d_{yz}$, and $d_{xz}$) and the higher-energy $e_g$ ($d_{x^2-y^2}$ and $d_{z^2}$) states. In the actual breathing-mode ground state ($d_2/d_1 \approx 1.47$), the octahedral crystal field is distorted, and the $t_{2g}$ states are further split into two sublevels labeled $t_{2g}^1$ and $t_{2g}^2$.

In Fig.~\ref{fig:dftflatbands}b we show the corresponding DFT band structures, starting from the unperturbed conventional kagome lattice ($d_2/d_1=1$) and indicate Nb $t_{2g}^1$ orbitals in red, Nb $t_{2g}^2$ orbitals in green, and Nb $e_{g}$ orbitals in blue. From this, we see that in the putative conventional kagome structure, \NbX{Br} would be a metal with various Nb $d$ bands crossing the Fermi level. Upon trimerization, i.e. upon introducing the breathing mode perturbation, we see that the Nb $t_{2g}^2$ bands open up a full gap, while the Nb $t_{2g}^1$ bands stay behind, nearly pinned to the Fermi level. 
Finally, in the actual full breathing mode structure ($d_2/d_1 \approx 1.47$), we see that the Nb $t_{2g}^1$ bands further split into a set of two very flat bands located directly around the Fermi level and a set of four unoccupied bands. Next to this, we also find that the occupied Nb $t_{2g}^2$ bands split, yielding two sets of very flat bands, while their unoccupied counterparts only mildly change in their dispersion and now overlap in energy with the Nb $e_{g}$ bands.

The flat band formation in \NbX{X} is thus intimately connected to the Nb trimerization within the breathing mode distortion and independent of the LT or HT stacking. In the monolayer limit, we already showed that the Nb $t_{1g}$ band at the Fermi level can be best understood as being formed by a molecular orbital centered at the breathing-mode induced Nb trimer~\cite{grytsiuk2024nb3cl8}. 
As such, the origin of the two well-separated low-energy flat bands around the Fermi level in \NbX{X} is conceptually very similar to the formation of flat bands in the star-of-David phase of TaS$_2$~\cite{wilson1975charge} or twisted bilayer graphene~\cite{bistritzer2011moire}: due to new, larger structures (induced by the CDW in TaS$_2$, the moir\'e potential in twisted bilayer graphene, or here the breathing-mode distortion), new effective orbitals are formed, which are spatially rather far apart from each other. This large spatial separation yields a small effective hopping of electrons between these new effective CDW, moir\'e, or here molecular orbitals, such that their resulting bands are only weakly dispersive. Simultaneously, gaps in the rest of the electronic structures are opened, leaving (nearly) isolated flat bands behind. All of this is lifted upon destroying the larger structure, i.e. by melting the CDW in TaS$_2$, tuning away from the magic angle in twisted bilayer graphene, or suppressing the breathing mode distortion in \NbX{X}.

The key advantage of \NbX{X} is that its breathing mode distortion is a stable ground state (at all reported temperatures and doping levels) that does not require a precisely tuned twist angle or temperature-induced stabilization. The formation of flat bands in \NbX{X} is thus very robust, so that resulting correlation effects can be studied without additional complications.

\subsubsection{Out-of-Plane Dimerization as Origin of Flat Band Gapping}
Having established the in-plane breathing mode distortion and its accompanying trimerization as the mechanism producing the isolated flat bands, we are now ready to study their evolution throughout the \NbX{X} family.
In Fig.~\ref{fig:dft}, we show the bulk DFT band structures in the LT phases. As expected, we find two relatively flat bands around the Fermi level in all cases. 
In the case of \NbX{F}$^*$, both bands cross the Fermi level, while for the others, we consistently find one fully occupied and one completely unoccupied band. For in-plane momenta ($k_z = 0$), the two bands are clearly split apart, while for the out-of-plane momentum direction $\Gamma-A$, we observe a stronger dispersion, and the gap nearly closes. As will become clear from the subsequent Wannierization discussed below, this $k_z$ behaviour is a result of the interlayer dimerization of trimers within the strongly hybridized bilayers in the LT structure. Thus, in $k_z$, the low-energetic electronic structure is reminiscent of the behaviour of the Su-Schrieffer-Heeger (SSH) model~\cite{ssh1979}, which consequently yields a gap in this direction, due to the mismatch of the alternating out-of-plane hopping matrix elements in the LT phase. From Cl to Br and I, we already see the first clear trend in the material class: the in-plane gap between these two bands steadily increases. In the HT phase, this out-of-plane dimerization is strongly suppressed yielding negligible gaps, rendering the low-energy band structures of all HT structures qualitatively similar to the LT \NbX{F}* one.

\subsubsection{Molecular Orbital Basis}
\begin{figure}[h]
    \centering
    \includegraphics[width=0.99\columnwidth]{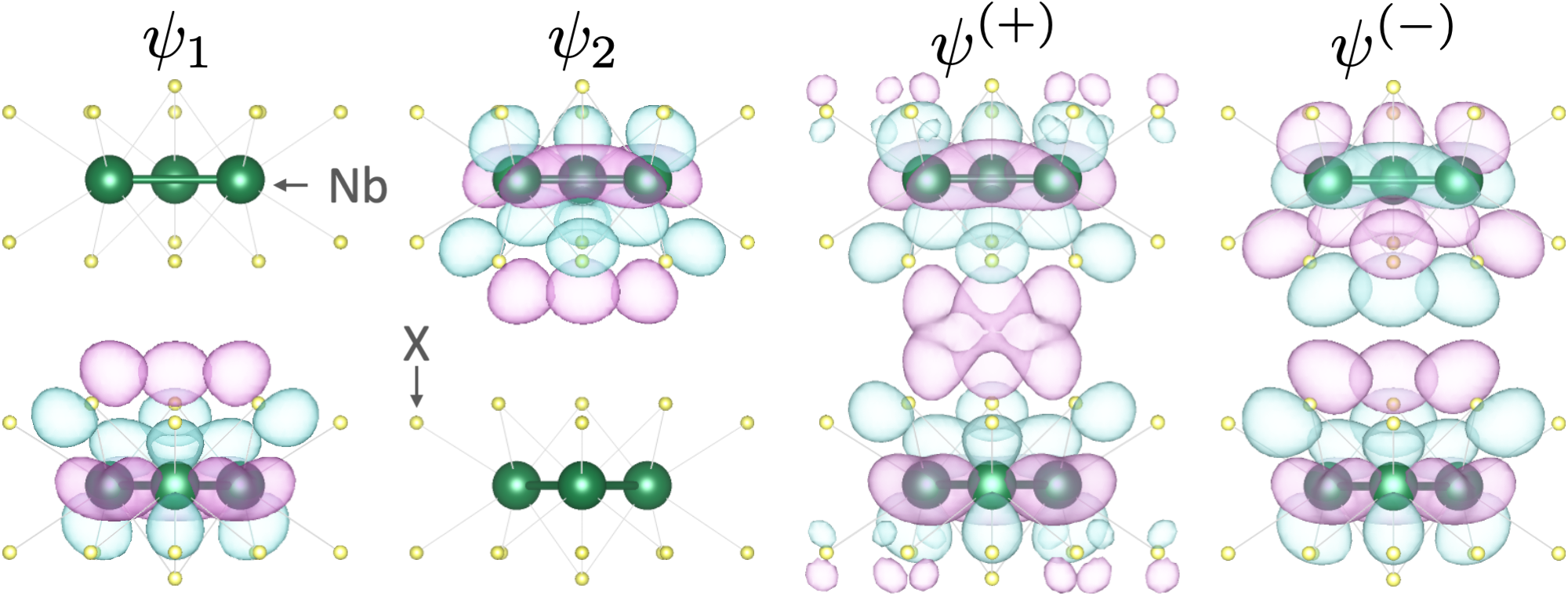}
    \caption{\textbf{Wannier Molecular Orbitals of \NbX{Br}.} $\psi_1$ and $\psi_2$, along with their 
    bonding ($+$) and anti-bonding ($-$) combinations, $\psi^{(\pm)} = ( \psi_1 \pm \psi_2 ) / \sqrt{2}$.}
    \label{fig:wan_orb}
\end{figure}
To prepare for the derivation of material-specific many-body Hamiltonians using ab initio downfolding and to further analyze the low-energetic band structure, we construct maximally localized Wannier functions for the two low-energy bands. In all four cases, we find that the bands can be perfectly Wannier-interpolated within a basis spanned by two molecular orbitals $\psi_1$ and $\psi_2$, which are centered on trimers of adjacent layers and which are directly above each other, see Fig.~\ref{fig:wan_orb}. As shown in Refs.~\cite{grytsiuk2024nb3cl8,hu_correlated_2023}, these molecular orbitals arise from a direct linear combination of three Nb $d$ orbitals from the local trimers. The original atomistic multi-orbital basis has, however, very little influence due to the large energy separation between the trimer orbital and all other states, which are therefore either fully occupied or empty.
In this molecular orbital basis, we find similar contributions of both orbitals to both low-energy bands. However, upon diagonalization of the single-particle Wannier Hamiltonian, we find that the lower (occupied) band is well described by a bonding orbital $\psi^{(+)}$ formed by the direct sum of the trimer-centered molecular orbitals $\psi^{(+)} = (\psi_1 + \psi_2)/\sqrt{2}$, while the upper (unoccupied) band is well described by the corresponding anti-bonding orbital $\psi^{(-)} = (\psi_1 - \psi_2)/\sqrt{2}$, as illustrated in Fig.~\ref{fig:wan_orb}. For the subsequent analysis of correlation effects this molecular orbital basis is most efficient, as it compresses six atomistic basis orbitals to just two molecular ones. Furthermore, it will be helpful to switch back and forth between the molecular orbital basis ($\psi_i$) and the bonding/anti-bonding basis ($\psi^{(\pm)}$), as the self-energy in the LT phase becomes diagonal in the bonding/anti-bonding basis.

Upon projecting the Kohn-Sham states to the $\psi^{(\pm)}$ basis, we see that this interpretation holds especially for in-plane momenta. 
For out-of-plane $k_z$ momenta, however, we observe that $\psi^{(\pm)}$ states mix, which is a result of the small, but finite wavefunction overlap between the dimerized layers. This admixture thus verifies the SSH-like behaviour in the $k_z$ direction and is consistent with Ref.~\cite{xu2024fillingobstructedatomicinsulator} within which \NbX{Br} and \NbX{I} are classified as obstructed atomic insulators~\footnote{See \url{https://www.topologicalquantumchemistry.fr/#/detail/25767} and \url{https://www.topologicalquantumchemistry.fr/#/detail/421609}.}. The latter are characterized by wave-function centers that are not located at
a atomic position, which holds in our case for the bonding combination of the trimer orbitals, cf. $\psi^{(+)}$ in Fig.~\ref{fig:wan_orb}. Although the Cl, Br, and I compounds all share the classification as obstructed atomic insulators, the details and specifically the size of the DFT gap change from material to material, as reflected in the Wannier hopping parameters in Tab.~\ref{tab:parametersfull} and as discussed in more detail in the following section.

\subsection{Downfolded Many-Body Hamiltonians from Ab Initio Constrained Random Phase Approximation Calculations}
{\renewcommand{\arraystretch}{1.2}
\begin{table*}[ht!]\centering
\caption{{\bf Structural and Ab Initio Downfolded Many-Body Hamiltonian Parameters for LT \NbX{X}.} In-plane ($a$) and out-of-plane ($c$) lattice parameters, molecular orbital spread ($\Omega$), nearest neighbor in-plane ($t_\parallel$) and out-of-plane ($t_\perp$) hopping parameters and corresponding cRPA screened Coulomb ($U_\parallel$ and $U_\perp$) interactions, on-site bare ($V_0$) and screened ($U_0$) Coulomb interactions including their ratio $\varepsilon_\text{eff} = V_0/U_0$.}
\begin{tabular}{c R{1.2cm} | R{0.9cm} R{0.9cm} R{0.9cm} |R{1.2cm}  R{1.2cm}  R{1.2cm} | R{1.2cm}  R{1.2cm}  R{0.9cm} | R{1.2cm}  R{1.2cm}  R{1.2cm}   }
\midrule
    \multicolumn{1}{c}{X} 
 &  \multicolumn{1}{c|}{} 
 & \multicolumn{1}{r}{$a$}
 & \multicolumn{1}{r}{$c$}
 & \multicolumn{1}{r|}{$\Omega$}
 & \multicolumn{1}{r}{$t_\parallel$}  & \multicolumn{1}{r}{$t_\perp^\text{s}$}  & \multicolumn{1}{r|}{$t_\perp^\text{w}$}  
 & \multicolumn{1}{r}{$V_0$} & \multicolumn{1}{r}{$U_0$} & \multicolumn{1}{r|}{$\varepsilon_\text{eff}$}
 & \multicolumn{1}{r}{$U_\parallel$}  & \multicolumn{1}{r}{$U_\perp^\text{s}$}  & \multicolumn{1}{r}{$U_\perp^\text{w}$}    
   \\\midrule
%1_NbF
\multirow{ 3}{*}{F} & \multicolumn{1}{l|}{Bulk}&
5.74&31.16 &4.12 & 34.3 &     -4.9 &     -6.5   &     7486.4 &     2590.5 & 2.89 &    804.5 &    714.6 &    572.5    \\
{} & \multicolumn{1}{l|}{BL}&       
&&& 33.8 &     -5.0 &       --   &     7496.2 &     3988.8 & 1.88 &   {1977.2} &   {1987.7} &       --     \\
{} & \multicolumn{1}{l|}{ML}&       
&&& 33.6 &       -- &       --   &     7497.2 &     4021.8 & 1.86 &   {2016.7} &       -- &       --     \\ \hline
%
%2_NbCl
\multirow{ 3}{*}{Cl} & \multicolumn{1}{l|}{Bulk}&      
6.75 &36.75&7.08 & 20.1 &   -136.0 &    -16.1   &     6126.4 &     1451.4 & 4.22 &    475.6 &    400.1 &    313.5     \\
{} & \multicolumn{1}{l|}{BL}&       
&&& 18.5 &   -136.2 &       --   &     6150.0 &     2697.6 & 2.28 &    {1598.5} &    {1570.9} &       --     \\
{} & \multicolumn{1}{l|}{ML}&       
&&& 18.6 &       -- &       --   &     6180.3 &     2771.3 & 2.23 &   {1670.1} &       -- &       --     \\ \hline
%
%3_NbBr
\multirow{ 3}{*}{Br} & \multicolumn{1}{l|}{Bulk}&       
7.08&40.84&8.54 & 5.9 &   -169.4 &    -20.4   &     5787.4  &     1186.6 & 4.88 &    381.5 &    342.0 &    262.4    \\
{} & \multicolumn{1}{l|}{BL}&        
&&& 3.4 &   -169.2 &       --   &     5819.4  &     2396.0 & 2.43 &    {1483.8} &    {1482.3} &       --     \\
{} & \multicolumn{1}{l|}{ML}&        
&&& 4.5 &       -- &       --   &     5887.7  &     2475.3 & 2.38 &    {1554.3} &       -- &       --    \\ \hline
%
%4_NbI
\multirow{ 3}{*}{I} & \multicolumn{1}{l|}{Bulk}&      
7.70&43.24&13.15& -8.3 &   -218.2 &    -24.6   &     5052.7 &      787.0 & 6.42 &    258.3 &    258.5 &    183.8     \\
{} & \multicolumn{1}{l|}{BL}&      
&&& -13.2 &   -218.4 &      --   &     5101.8 &     1928.7 & 2.65 &    {1323.3} &    {1349.0} &       --    \\
{} & \multicolumn{1}{l|}{ML}&      
&&& -11.5 &       -- &      --   &     5321.1 &     2052.4 & 2.59 &    {1397.3} &       -- &       --     \\ \hline
\bottomrule
\end{tabular}
\label{tab:parametersfull}
\end{table*}}

Using the molecular orbital basis, we proceed with deriving generalized Hubbard Hamiltonians of the form
\begin{align}
    H = \sum_{i,j} t_{ij} \, c^\dagger_i c_j + \frac{1}{2} \sum_{i,j,k,l} U_{ijkl} \, c^\dagger_i c^\dagger_j c_k c_l \label{eq:Hubbard}
\end{align}
for all compounds. Here $c_i^\dagger$ and $c_j$ are electron creation and annihilation operators in molecular orbitals at sites $i$ and $j$, respectively, $t_{ij}$ are single-particle (hopping) and $U_{ijkl}$ partially screened two-particle (Coulomb) matrix elements. The latter are evaluated within the constrained random phase approximation (cRPA)~\cite{cRPA}, which accounts for all RPA-level screening processes except those taking place within the molecular orbitals $\psi_i$. In this way, all intra- and interlayer screening properties are included. As a first-principles method, the cRPA therefore captures how screening differs across the class of materials, both due to changes in the chemical composition as well as due to differences in the screening environment between monolayers, bilayers, and bulk structures. 
This way, our material-specific localized molecular orbitals render adequately the different local chemical environments in each compound and in each stacking, while the screening captures global crystal effects including polarizations from higher and deeper lying electronic states.

The cRPA scheme gives access to the full tensor $U_{ijkl}$. 
In Tab.~\ref{tab:parametersfull}, we show a subset of these matrix elements focusing on the local and nearest-neighbor density-density Coulomb ($U_{ijji}$) as well as on intra- and inter-layer hopping matrix elements. For a complete overview and details about the calculations, see Section~\ref{sec_meth_dft}.

We start by discussing the molecular orbital spreads ($\Omega$) of the bulk structures, which we find to increase from F to I. This trend follows the lattice constants, which increase in the in- and out-of-plane direction from F to I. This shows how the molecular orbitals and thus the local orbital basis reflects the chemical changes in the structure. The tightest-bound lattice (\NbX{F}*) hosts highly localized molecular orbitals, while the weakest-bound lattice (\NbX{I}) has rather delocalized molecular orbitals.

These trends are further reflected in the hopping matrix elements. Overall, we find that the in-plane nearest-neighbor hopping ($t_\parallel$) shows a decreasing trend from F to I. Its magnitude is relatively small in all compounds, which reflects the flat in-plane (constant $k_z$) dispersion.
The strong out-of-plane nearest-neighbor hopping ($t^{(s)}_\perp$), in turn, grows significantly in magnitude from F to I, which is in line with a decrease in the orbital spread. The ``weak link'' in the out-of-plane directions ($t^{(w)}_\perp$) is in all compounds about an order of magnitude smaller than $t^{(s)}_\perp$ and accordingly increases in magnitude slightly from F to I. This reflects the increasing bilayer hybridization between the inter-layer dimerized trimers, leading to the growth of the bonding/anti-bonding single particle splitting, as observed around $\Gamma$ in Fig.~\ref{fig:dft}. 
In \NbX{F}*, the in- and out-of-plane hoppings are not very different, such that there is no opening of a full dimerization gap throughout the Brillouin zone and only a small splitting at $\Gamma$ is observed.
To analyze the gap at the high symmetry point $A$, we need to take into account that there are three adjacent layer neighbors in the ``weak-link'' directions, such that the SSH gap of a nearest-neighbor model would be given by $\tau_\perp = 2\left|t^{(s)}_\perp-\sqrt{3}t^{(w)}_\perp \right|$. This increases from about 216 to 351 \,meV going from Cl to I, indicating an increase of the dimerization strength (in \NbX{F}*, it is only $4\,$meV). The DFT gaps at $A$ (where the SSH gap opens) are, however of similar size in \NbX{Cl} and \NbX{Br}. We understand this as a result of long-range hoppings suppressing the dimerization, particularly hopping to next nearest and next-next-nearest neighbouring trimers in the adjacent van der Waals layer. These hoppings become slightly more pronounced from Cl to I.

The most important interaction parameter is the partially screened local density-density interaction $U_0$ (Hubbard's $U$), which depends on the size of the orbital and on the screening. We can disentangle these effects by also looking at the bare local density-density Coulomb interaction $V_0$, which depends only on the size of the orbital. It is largest in \NbX{F}* and smallest in \NbX{I}, inversely following the molecular orbital spread. The impact of screening can then be quantified by the ratio $\varepsilon_\text{eff} = V_0/U_0$, which is seen to increase drastically from F to I. This screening parameter is approximately inversely proportional to the ``background'' electronic gap (i.e., the gap between the filled and occupied bands above and below the flat bands). Thus, the screened local density-density Coulomb interactions are not just suppressed by the enhanced molecular orbital spread, but also by stronger screening going from F to I.

Next to this, our calculations show that the partially screened Coulomb interactions are dominated by density-density interactions. This is a result of having just a single molecular orbital being centered in each of the adjacent, strongly hybridized monolayers. Non-density-density interactions, such as Hund's exchange interactions, are thus non-local, which significantly suppresses their strengths (to a few meV). Due to reduced screening in the layered vdW structures, the density-density Coulomb interactions are also non-local and show a significant long-ranged tail. The latter is visible in Tab.~\ref{tab:parametersfull} by comparing the local interactions $U_0$ to the nearest-neighbor ones in both in- and out-of-plane direction.
 In contrast to the out-of-plane hopping matrix elements, we do not see a significant bilayer hybridization effect in the non-local density-density Coulomb interactions. Comparing $U^{(s)}_{1,\perp}$ and $U^{(w)}_{1,\perp}$, we find only small differences, which can be explained by the slightly different distances between the molecular orbital centers in these two directions.

Comparing the bulk parameter with the mono- and bilayer ones, we do not see sizable differences in the hopping parameter and only mild enhancements of the bare interactions, but strong enhancements of the \emph{screened} Coulomb matrix elements. The latter is mostly driven by the \emph{reduced} screening in the mono- and bilayer limit as compared to the bulk structures, as clearly visible from the reduced $\varepsilon_\text{eff}$ in these structures. Thus, while the bulk and bilayer structures behave rather similar in terms of their non-interacting electronic structure, they could be vastly different as a result of different screening environments.

Taken together, our ab initio downfolding calculations show that the local Coulomb interactions $U_0$ are very large in comparison to in-plane hopping terms with $U_0/t_\parallel$ ratios always larger than $80$. In comparison to the strong out-of-plane hopping matrix elements, we find $U_0/t_\perp$ ratios that decrease from about $12$ in \NbX{Cl} to around $4$ in \NbX{I}. These ratios point towards possible strong Coulomb-interaction induced correlation effects, which might change from system to system. To investigate this and quantify the correlation strengths, we proceed with solving the generalized Hubbard models using cluster dynamical mean-field theory.

\subsection{Generalized Phase Diagram from Cluster Dynamical Mean Field Theory at Half-Filling}
\subsubsection{Self-Energies}

\begin{figure}[t]
    \centering
    \includegraphics[width=\columnwidth]{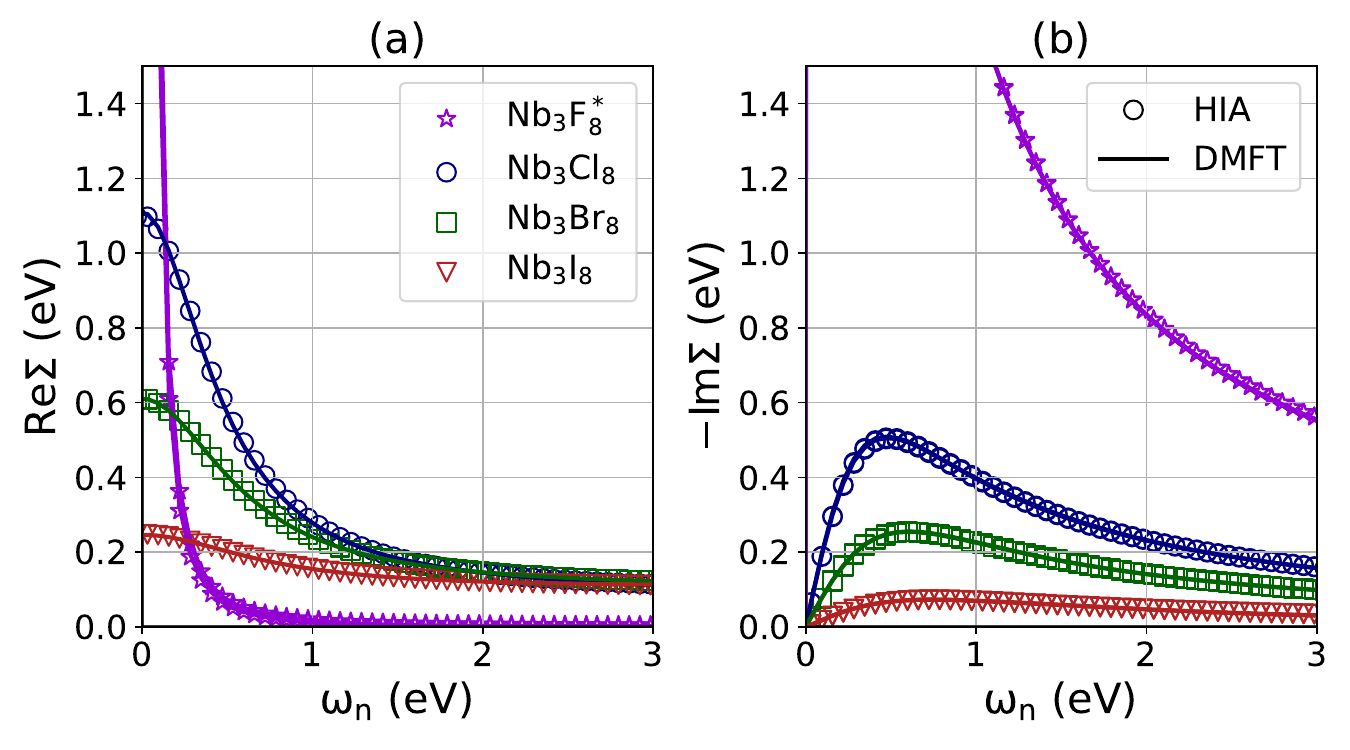}
    \caption{\textbf{\NbX{X} Matsubara Self-Energies.} (a) Real and (b) imaginary parts of the self energies in Matsubara frequencies in the bonding/anti-bonding basis. Markers represent Hubbard-I results, lines represent cluster DMFT results.}  
    \label{fig:DMFT}
\end{figure}

Using our cRPA-based downfolded molecular-orbital models, we calculate the self-energies at half-filling within dynamical mean-field theory (DMFT) and its Hubbard-I approximation \cite{Lichtenstein_1998,Kotliar_2006}. Specifically, we define the impurity problem as the two inter-layer dimerized molecular (trimer) orbitals within each bilayer and include the local $U_0$, the strong inter-layer hybridization $t^{(s)}_\perp$ as well as all inter-layer Coulomb interaction elements $U_{ijkl}$ (among which $U^{(s)}_{\perp}$ is by far the largest, see Tab.~\ref{tab:parametersfull}). Accordingly, we deal with a cluster DMFT problem  \cite{Lichtenstein_2000,Kotliar_2001,Maier_2005}. Fig.~\ref{fig:DMFT} shows the resulting cluster DMFT Matsubara frequency self-energies in the diagonal $\psi^{(\pm)}$ basis for all compounds calculated using a CT-HYB quantum Monte Carlo solver at $\beta=100\,$eV$^{-1}$ (see \ref{sec_meth_dft} for all details). In this basis at half-filling, the imaginary parts of the self-energy are the same for both channels ($\mathrm{Im}\Sigma_{(+)} = \mathrm{Im}\Sigma_{(-)}$), while the real parts differ only in their signs ($\mathrm{Re}\Sigma_{(+)} = -\mathrm{Re}\Sigma_{(-)}$) in the case of Cl, Br, and I. For \NbX{F}* the bonding and anti-bonding self-energy lie on top of each other in Fig. \ref{fig:DMFT}.

The real part, as depicted in Fig.~\ref{fig:DMFT}~(a), shows that correlation effects enhance the bonding/anti-bonding splitting, similar to the Stoner enhancement of the exchange splitting in magnetic systems. In I, the real part is almost frequency-independent, while it becomes strongly frequency-dependent moving towards Cl. This is a sign of the increasing importance of dynamic correlation effects, which necessitate (cluster) DMFT calculations. In the imaginary part depicted in Fig~\ref{fig:DMFT}~(b), we again find a clear trend: the overall amplitudes of the self-energy gradually increase from I to Br, Cl, and F, which is another sign of the increasing degree of correlation. Specifically, we see that the imaginary part of the self-energy of \NbX{I} is barely changing with frequency, while it clearly shows retardation effects in \NbX{Cl} and \NbX{Br}. Finally, \NbX{F}* shows a divergent imaginary self-energy for small Matsubara frequencies.

Since in \NbX{I} retardation effects are vanishingly small, while its $\operatorname{Re}[\Sigma(i\omega_n)]>0$, we can classify \NbX{I} as a weakly correlated dimerized (or obstructed atomic) band insulator. \NbX{Br} and \NbX{Cl} in turn show signatures of strongly correlated insulators, while \NbX{F}* behaves like a conventional Mott insulator~\cite{date2024mott}.

\subsubsection{Phase Classification Based on Local Molecular Dimer Properties}
\begin{figure}
    \centering
    \includegraphics[width=\columnwidth]{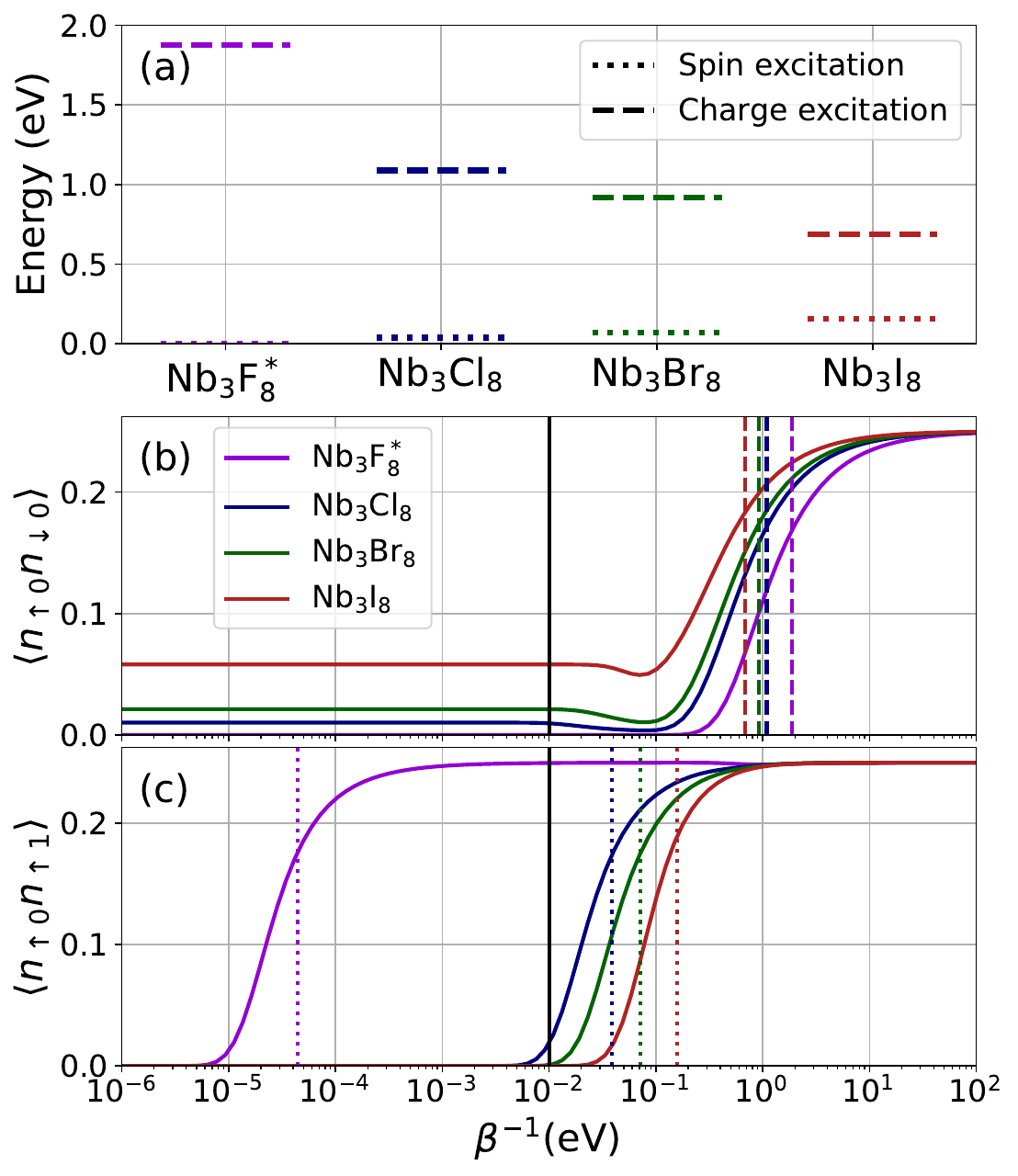}
    \caption{\textbf{Local Dimer Analysis.} (a) shows the charge neutral excitation spectrum. (b) and (c) show the temperature dependence of expectation values of $\braket{n_{0,\uparrow} n_{0,\downarrow}}$ double occupation and inter-layer spin alignment $\braket{n_{0,\uparrow} n_{1,\uparrow}}$. These can be interpreted as measures for the thermal occupation of the first charge and spin excitation respectively. This can be seen by clear transitions in behavior at the characteristic temperatures $T_c=E_{exc}/k_b$, plotted as dotted/dashed lines. The black line is the temperature at which we perform our DMFT study $\beta=100 eV^{-1}$.}
    \label{fig:dimer}
\end{figure}

To further substantiate this classification of LT bulk \NbX{X} into Mott, strongly and weakly correlated insulators, we proceed with analyzing the ``local impurity'', i.e. the dimer formed by the hybridized molecular orbitals from adjacent monolayers, in more detail. This is similar to the analysis by Zhang et al.~\cite{Zhang_2023_PRB}. However, we use here our material-specific model parameters and, importantly, the full Coulomb tensor including non-local interactions between the molecular orbitals sites (which is important for charge-neutral excitations~\cite{Schuler13}) to not only calculate $T=0$ properties, but also finite temperature expectation values.

The dimer ground-state is characterized by a multi-reference Slater determinant of the form $\alpha (\ket{\uparrow,\downarrow}-\ket{\downarrow,\uparrow}) + \beta (\ket{\uparrow\downarrow,.}-\ket{.,\downarrow\uparrow})$, while the first charge-neutral excitations can be described by the triplet $\{\ket{\uparrow,\uparrow}$, $\ket{\downarrow,\downarrow}$, $\ket{\uparrow,\downarrow}+\ket{\downarrow,\uparrow}\}$ and a charge-excitation of the form $\ket{\uparrow\downarrow,.}-\ket{.,\downarrow\uparrow}$. 

In Fig.~\ref{fig:dimer} we analyze and compare the ground and excited states among all compounds. At low temperatures, we find that the double occupation $\braket{n_{\uparrow0} n_{\downarrow0}}$ of \NbX{F}* is nearly fully suppressed, while it is small but finite in \NbX{Cl} and \NbX{Br} and finally rather large in \NbX{I}. This results from the large $U_0/\tau_\perp$ ratio in \NbX{F}*, which basically prohibits two electrons of different spins to be in the same molecular orbital. In turn, for \NbX{I} this ratio is strongly reduced, such that a double occupation is possible.
This trend continues in the charge-neutral excitation energies. We see that the spin-excitations gap of the \NbX{F}* dimer is vanishingly small, while it is small but finite for \NbX{Cl} and \NbX{Br}, and finally rather large in \NbX{I}. The charge-excitation energy in turn decreases from \NbX{F}* to \NbX{I}, such that the excitations are energetically orders of magnitude apart in \NbX{F}* and of the same order in \NbX{I}.

This zero temperature analysis thus clearly separates \NbX{F}* from the others, which underlines its special status as a conventional Mott insulator. The difference between \NbX{Cl}/\NbX{Br} and \NbX{I} is more quantitative, but due to their vastly different material parameters still obvious. In fact, \NbX{Cl}/\NbX{Br} and \NbX{I} are that much different that we find for \NbX{I} a perturbative $GW$-like description to be perfectly adequate, while it  breaks down in \NbX{Cl} and \NbX{Br}, as discussed in more detail in Appendix~\ref{app:ED}. Taking this and the self-energy property discussion from above together, this justifies differentiating \NbX{Cl}/\NbX{Br} and \NbX{I} into strongly and weakly correlated insulators.

Finally, as a result of the similar spin and charge excitation energies, we find the spin and charge excited states to get partially occupied in \NbX{I} at rather similar temperatures, while these temperature scales are (very) different in the other compounds, cf. Fig.~\ref{fig:dimer} (b) and (c). Moreover, we find the (lower) spin-excitation to be thermally activated in \NbX{Br} and \NbX{Cl} only at temperatures in which these compounds are already in their high-temperature phases. As such the temperature response of \NbX{F}*/\NbX{Br}/\NbX{Cl} and \NbX{I} are vastly different.

\subsubsection{Momentum-Resolved Spectral Functions}

\begin{figure*}[t]
    \centering
    \includegraphics[width=\textwidth]{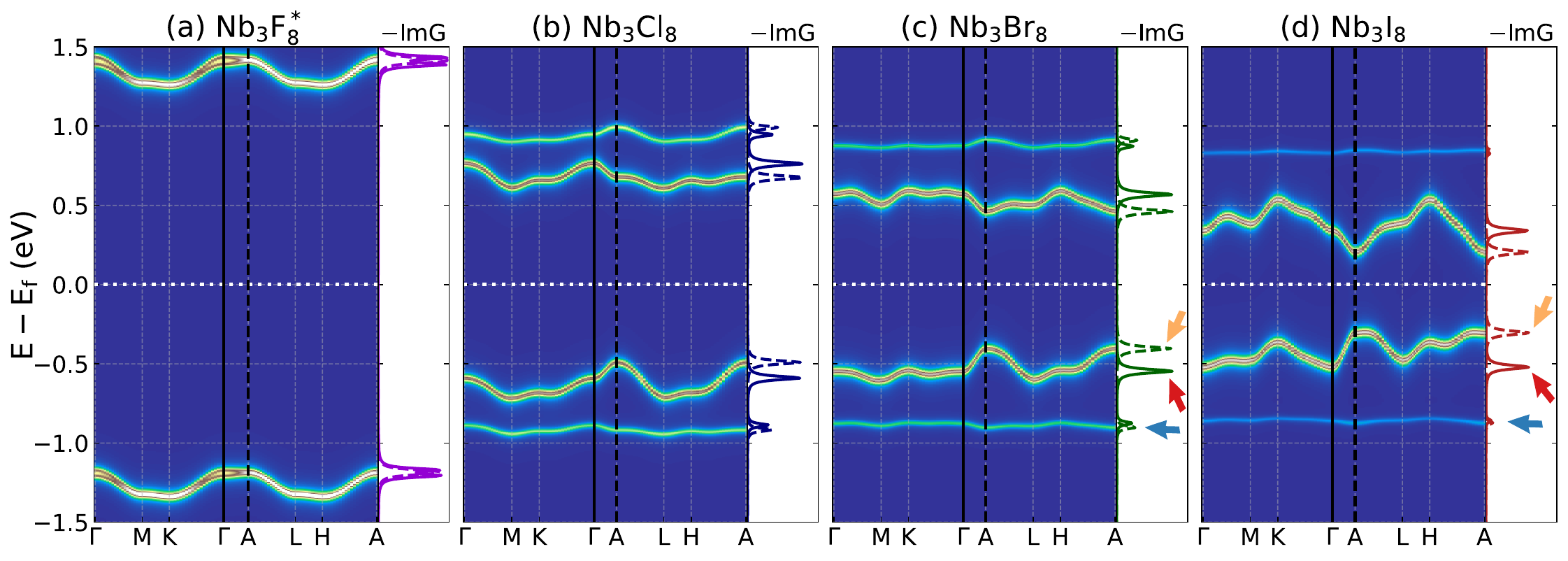}
    \includegraphics[width=\textwidth]{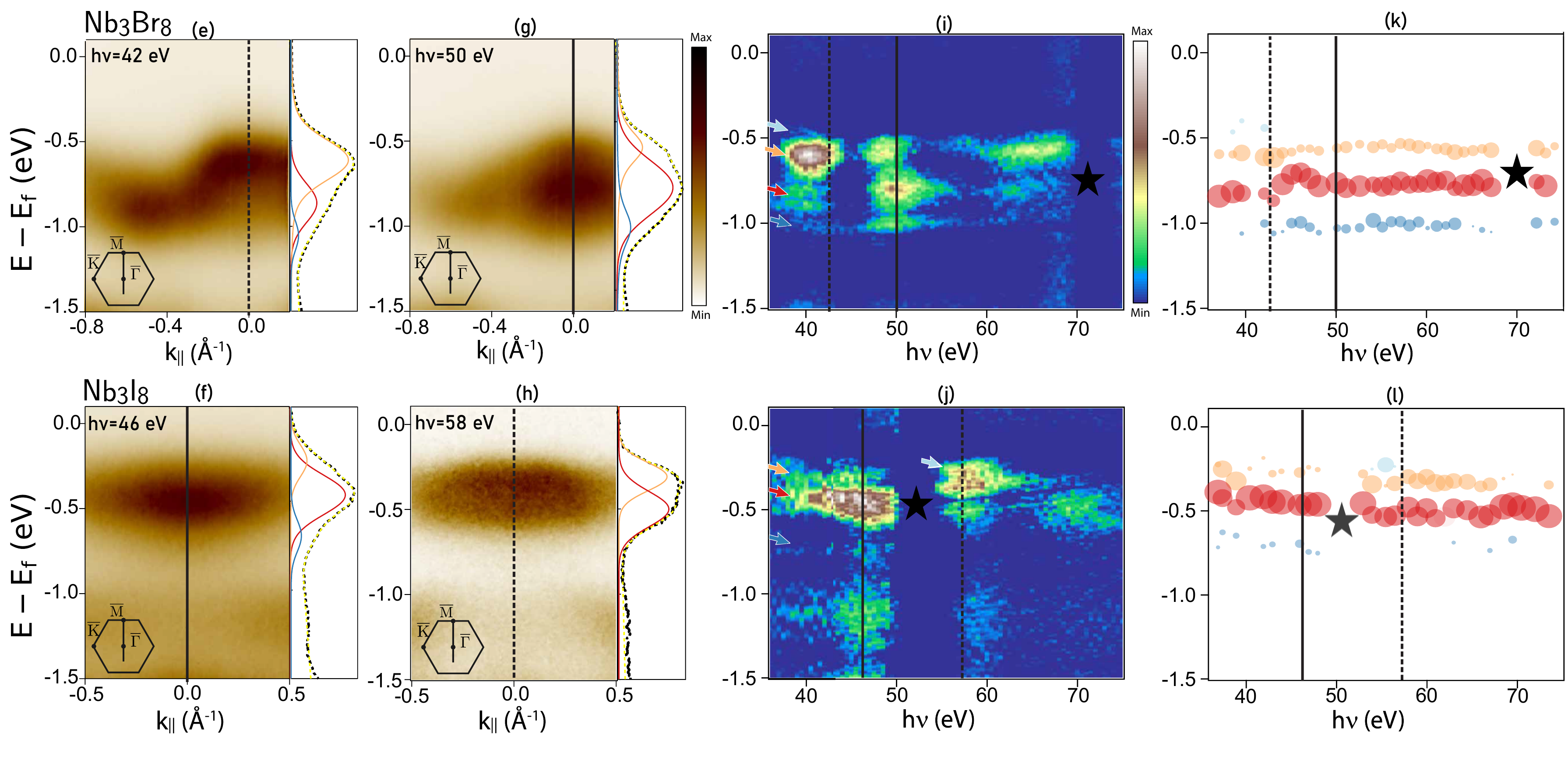}
    \caption{\textbf{Theoretical and Experimental Spectral Functions at Half-Filling ($\mathbf{N_e=2}$).} (a-d) Momentum resolved spectral functions of Nb$_3$X$_8$ within Hubbard-I approximation at half-filling. Line cuts at $\Gamma$ \& $A$ are shown in full and dashed lines respectively. (e-h) Raw ARPES data along $\overline{\Gamma}-\overline{\mathrm{M}}$ alongside energy distribution curves (EDCs) taken at $\overline{\Gamma}$. (i-j) 2D curvature plots of EDCs plotted against photon energy. (k-l) EDC position and normalized area from curve fits using Voigt function at each photon energy. Dot sizes represent the normalized area under the curve. Dashed and solid lines in (i-k) and (j-l) connect to photon energies indicated in (e-g) and (f-h), respectively. Arrows in (i-j) indicate features matching curve fits in (k-l) and are associated with features marked by the same arrows in (c) and (d).}
    \label{fig:Hubbard-I}
\end{figure*}

Next to the dimer-local charge-neutral excitation spectra, charged excitations, such as measured by angle-resolved photoemission spectroscopy, can give further insights into the differences between the four compounds. To this end we study in the following the full lattice spectral functions $A_k(\omega)$, which we can easily access via the so-called Hubbard-I approximation~\cite{Lichtenstein_1998}. This approximation should be in good agreement with our full cDMFT results at integer fillings due to the very large $U_0$ compared to the hopping to the bath (ie. $t_\perp^{(w)}/t_\parallel$) for all systems. This assumption is confirmed by the self-energy results shown in Fig.~\ref{fig:DMFT}. Importantly, the Hubbard-I approximation gives us direct access to retarded spectral functions without analytic continuation of Matsubara data. 

The fully interacting spectral functions $A_k(\omega)$ for all four compounds at half-filling ($N_e=2$ reflecting the number of electrons in the two flat bands formed by the molecular orbitals) are shown in Fig.~\ref{fig:Hubbard-I}~(a-d). In all cases, we see that the gap is widened in comparison to the mean-field DFT solutions shown in Fig.~\ref{fig:dft}. For \NbX{Cl}, \NbX{Br}, and \NbX{I}, we see additional flat features at lower and higher energies [indicated by blue arrows in Fig.~\ref{fig:Hubbard-I}~(c-d)], which are energetically separated the strongest in \NbX{I} and less and less in \NbX{Br} and \NbX{Cl} until they are fully merged with the main features in the case of \NbX{F}*. 

Upon analyzing the exact diagonalization results of the generalized Hubbard dimer, which serves as the origin of the Hubbard-I self-energies (see Appendix~\ref{app:ED} for details), we understand that, indeed, in \NbX{F}* a conventional Mott gap is formed. \NbX{F}* is thus a half-filled two-band Mott insulator, see Fig.~\ref{fig:Hubbard-I}~(a). In \NbX{Cl} and \NbX{Br}, the gap is formed between a lower Hubbard band (LHB) of the bonding orbital and the upper Hubbard band (UHB) of the anti-bonding orbital, which is further accompanied by a secondary LHB of the anti-bonding orbital (below the LHB of the bonding orbital) and a secondary UHB of the bonding orbital (above the UHB of the anti-bonding orbital), see Fig.~\ref{fig:Hubbard-I}~(b-c). This is not in line with the conventional Mott insulator characterization, but shares strongly retarded self-energies as a common origin as shown in Appendix~\ref{app:graphicalsol}. As such, classifying both \NbX{Cl} and \NbX{Br} as strongly correlated insulators is appropriate. In \NbX{I} all strong correlation (retardation) effects are significantly suppressed, such that the gap is best understood as being formed between the original bonding/anti-bonding bands, but slightly enhanced by Coulomb interactions, see Fig.~\ref{fig:Hubbard-I}~(d). \NbX{I} is thus indeed best classified as a weakly correlated band insulator. 
This is in qualitative agreement with the bilayer studies by Hu \textit{et al.}~\cite{Hu2024} and Zhang \textit{et al.}~\cite{Zhang_2023_PRB}. 
However, due to the lack of consistent material-specific model parameters for bilayer Cl and Br in their LT phases, Zhang et al., came to the conclusion that Cl is a Mott insulator, while and Br has been labeled a band insulator.
Altogether, we again find clear trends going from \NbX{F}* to \NbX{I}: the fundamental gap decreases and the energetic separation of the secondary LHB/UHB features increases, while their spectral weights vanish towards \NbX{I}.

Experimentally, the fundamental gaps of \NbX{X} have not yet been studied in great detail. To our knowledge, only \NbX{Cl} has previously been studied in absorption, photoluminescence and transport measurements, indicating a gap size of about $1.1\,$eV~\cite{Sun_2022,Gao_2023,meng_2024,yang2025gatetunableroomtemperaturemott}, which is in good agreement with our theoretical prediction of an indirect gap of about 1.11~eV [see Fig.~\ref{fig:Hubbard-I}~(b)]. 

\subsubsection{Strong Correlation Features Uncovered by Angle-Resolved Photoemission Spectroscopy}

With respect to the secondary LHB/UHBs in the LT bulk structures, which are most prominent in \NbX{Cl} and \NbX{Br}, we note that they are only captured by \emph{cluster} DMFT calculations. Using a \emph{single-site} DMFT approximation, as we previously applied for bulk \NbX{Cl}~\cite{grytsiuk2024nb3cl8} and as applied in other studies \cite{Gao_2023}, these features are much more coherent and even above/below the primary LHB/UHBs, as further detailed in Appendix~\ref{app:CDMFT}. This underlines the importance of two-site cluster DMFT calculations for the LT structures, which take the full matrix structure of the self-energy (from the two orbitals depicted in Fig.~\ref{fig:wan_orb}) into account.
We further note that our fully momentum dependent highly-resolved low-temperature real-frequency Hubbard-I spectral functions $A_k(\omega)$ uncovers an intriguing difference between the primary (bonding) and the secondary (anti-bonding) lower Hubbard bands: the primary (bonding) one still disperses (most prominently between $\Gamma$ and $A$), while the secondary (anti-bonding) one is nearly dispersion-less. 

Thus, from ARPES measurements performed in normal emission geometry, which measure $A_k(\omega)$ along $\Gamma$-$A$, we should naively expect a feature with a clear $k_z$ dispersion, representing the primary (bonding) LHB between $\Gamma$ and $A$, as well as a minor and less dispersive structure stemming from the secondary (anti-bonding) LHB. Note, however, that ARPES is a surface sensitive technique and the localization of the photohole can lead to a strong smearing in $k_z$, rendering the actual observation less clear.

To explore the manifestation of the many-body effects in the spectral function experimentally, we have performed ARPES experiments on \NbX{Br} and \NbX{I} (see Methods \ref{meth:ARPES} and Appendix~\ref{app:ARPES} for more details). Consistent with the literature~\cite{regmi_spectroscopic_2022,Regmi_2023_prb,date2024mott,Sun_2022,Gao_2023}, we see that the upmost occupied band consistently shows a splitting for these compounds, which is especially visible around $\bar{\Gamma}$, see Figs.~\ref{fig:Hubbard-I}~(e-l). We argue that this splitting actually reflects the expected three-dimensional character of the primary (bonding) LHB band and we also find spectroscopic evidence for the less dispersive and weaker secondary (anti-bonding) LHB.

We focus our analysis on data collected in normal emission (at $\bar{\Gamma}$), corresponding to the bulk A--$\Gamma$--A high-symmetry direction. This is accomplished by measuring the photoemission intensity as a function of photon energy, which corresponds to a variation in $k_z$.
Fig.~\ref{fig:Hubbard-I}~(e-h) show the photoemission intensity for \NbX{Br} and \NbX{I} along the $\bar{\Gamma}-\bar{\mathrm{M}}$ direction of the surface Brillouin zone at different photon energies. Also shown are energy distribution curves (EDCs) at $k_\parallel = 0$ (normal emission). These EDCs form the basis of the photon energy-dependent photoemission intensity which is shown in Figs.~\ref{fig:Hubbard-I}~(i-j) as a curvature plot. The main features in the EDCs and in the photon energy-dependent intensity are two intense peaks. For \NbX{Br}, a third minor peak is strictly required to obtain a good fit to the data; for \NbX{I} we also include such a third peak in EDC fits but it is found to be very weak throughout the range of investigated photon energies. The resulting fits to three Voigt lines is also shown in Fig.~\ref{fig:Hubbard-I}~(e-h). For now, we focus on the two intense peaks. The intensity of lower (higher) peak reaches a maximum at the photon energies corresponding to the bulk $\Gamma$ (A) points, as marked by red (orange) arrows in Figs.~\ref{fig:Hubbard-I}~(i-j) and the corresponding dots in Figs.~\ref{fig:Hubbard-I}~(k-l). We interpret these peaks not as two separate bands but rather as the extrema reached in the dispersion of the primary (bonding) LHB between the bulk $\Gamma$ and $A$ points. Such interpretation is consistent with the fact that the high density of states at the extrema of a one-dimensional dispersion probed in normal emission can be expected to dominate the photoemission intensity in the presence of $k_z$ broadening. There are many examples of such behavior in the literature, e.g., in the case of of the non-correlated topological insulator Bi$_2$Se$_3$, see Fig. 6 in Ref.~\cite{bianchi_electronic_2012}. In this interpretation, the separation between the peaks of 0.25~eV and 0.19~eV for \NbX{Br} and \NbX{I}, respectively, can be interpreted as the total band-width along the $\Gamma$-A direction. This is in good qualitative agreement with our theoretical estimates for these $k_z$ band widths, as indicated by the spectra taken at $\Gamma$ and $A$ and shown in the side-panels of Fig.~\ref{fig:Hubbard-I}~(c) and (d). We note that these band widths are mostly defined by the underlying PBE functional based on which our cluster DMFT calculations are performed. Correlation-driven renormalizations are on the order of $30$ to $50\,$meV and thus play only a minor role.

We now return to the fact that a good fit for the EDCs from \NbX{Br} requires taking a third band at lower energies into account.
Figs.~\ref{fig:Hubbard-I}~(k-l) show how the fitted peak positions and relative areas develop vs. photon energy. The third minor peak in the fits [marked by the blue arrow in Figs.~\ref{fig:Hubbard-I}~(i) and blue dots in Figs.~\ref{fig:Hubbard-I}~(k)] is strictly needed to obtain good fits for \NbX{Br} and the resulting band has a strongly suppressed dispersion and spectral weight. Based on these features we interpret this as the secondary LHB of the anti-bonding orbital. We note that this feature might also be present in the EDCs by Date \textit{et al.}~\cite{date2024mott}. In contrast, for \NbX{I} this third feature is not strictly needed to obtain a satisfactory fit to the EDCs. It is very weak in the 46~eV data and it is absent in the 58~eV data, see Fig.~\ref{fig:Hubbard-I}~(f,h). This is in agreement with our cluster DMFT calculations that predict a significant spectral weight reduction of the secondary LHB in \NbX{I} compared to \NbX{Br}.

We thus stress that the presence of two resonances, which has been argued to be a hallmark of (strong) correlation effects~\cite{Gao_2023,Zhang_2023_PRB}, is not sufficient to unambiguously differentiate between a conventional band, a strongly correlated band, and a Mott insulator based on ARPES data, as it results from the $k_z$ dispersion of the upmost occupied band, which is present even without correlation effects.  
While Date \textit{et al.}~\cite{date2024mott} argue that the details of this $k_z$ dispersion is altered by strong correlation effects, we instead suggest to search for the existence of the third resonance at slightly lower energies resulting from the secondary LHB. This is a clear and unambiguous feature of strong correlation effects in the strongly correlated insulators in their LT phases resulting from Coulomb driven retardation properties. 

Finally, we note that further comparison with the available ARPES results show that the lower $t^2_{2g}$ bands, which we neglect within our cluster DMFT calculations, are still separated from the upmost bands by about $0.5$ to $1\,$eV (increasing from I to Br), which is in agreement with the trends expected from comparison to DFT calculations, ruling out charge-transfer insulating states.

From this analysis we conclude that also charged excitations, as measured by photoemission spectroscopes, are adequate probes to observe the qualitative and quantitative differences between the four compounds.

\begin{figure*}[ht!]
    \centering
    \includegraphics[width=0.99\linewidth]{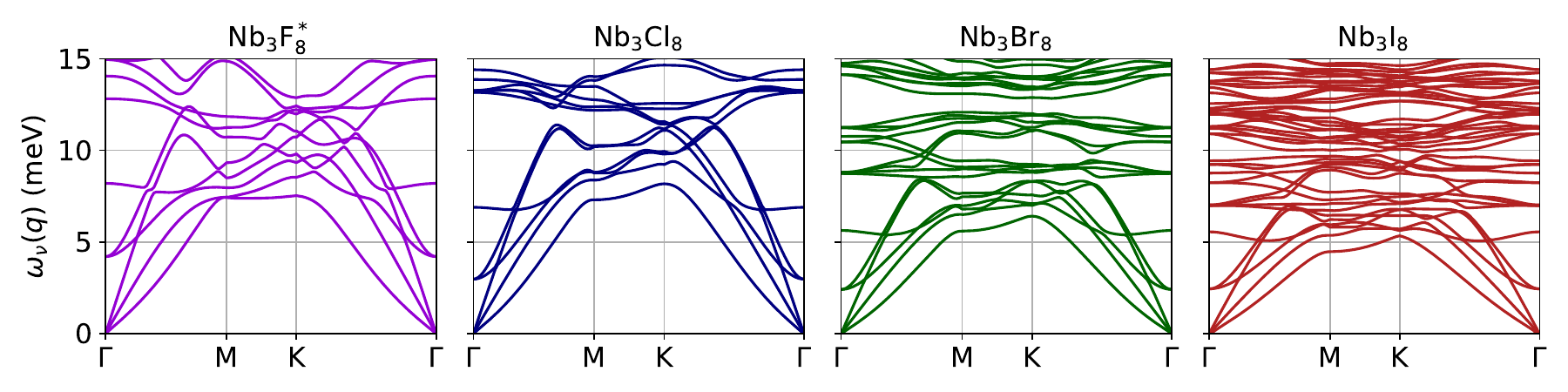}
    \caption{\textbf{Phonon Spectra of LT Bulk \NbX{X}.} From left to right \NbX{F}*, \NbX{Cl}, \NbX{Br}, and \NbX{I} calculated using density functional perturbation theory.}
    \label{fig_new:DFPT}
\end{figure*}

\subsubsection{Combined Phase Diagram}

To organize the four different compounds of the whole \NbX{X} material class, including their monolayer, bilayer, and bulk structures, in both low and high-temperature phases, we propose to position each system into a plane spanned by the inverse in- vs. out-of-plane correlation strengths. 
The inverse in-plane correlation strength can be measured by the ratio of the in-plane hopping $\tau_\parallel = t_\parallel$ to the local cRPA screened Coulomb interaction $U_0$, as $t_\parallel$ dominates the in-plane kinetics in all systems. 
The inverse out-of-plane correlation strength can be measured by the ratio of the out-of-plane dimerization strength to $U_0$. 
The dimerization strength can be conveniently characterized by the single-particle gap of the local dimer as approximated by $\tau_\perp = |t_\perp^{s} - \sqrt{3}t_\perp^w|$ in the LT phase. In the HT (monolayers) structures this is approximately (by definition) zero and for LT (HT) bilayers given by $\tau_\perp = |t_\perp^{s(w)}|$.
This extends and generalizes the schematic phase diagram by Date et al.~\cite{date2024mott} and the simplified one by Zhang et al.~\cite{Zhang_2023_PRB}. Importantly, our proposal uses appropriately normed (unitless) dimensions and is based on all relevant material properties including the local Coulomb strengths $U_0$ as well as the out-of- and in-plane hoppings $t_\perp^s$, $t_\perp^w$, and $t_\parallel$. This way, we can consistently organize all materials in all structures in one and the same phase diagram.

The result is presented in Fig.~\ref{fig-new:phasediagram} for which we used our material-specific parameters from Tab.~\ref{tab:parametersfull}. This way, the pure (single or two band) Mott insulators are all positioned on the bottom of the phase diagram. Their positions on the horizontal in-plane correlation strength axis changes, with monolayer \NbX{Br} being the strongest Mott insulator (due to the weakly screened $U_0$ and small $t_\parallel$) and LT bulk \NbX{F}* the weakest Mott insulator (due to its strongly screened $U_0$). Most of the LT bilayer and all thicker \NbX{Cl} and \NbX{Br} systems lie above the boundary in the correlated insulator phase, mostly driven by rather large out-of-plane dimerization strengths as a result of small dimerization gaps and large Coulomb repulsions. LT bulk \NbX{I} is positioned in the band insulator phase as result of a large dimerization strength and rather weak (strongly screened) local Coulomb interaction. Upon thinning LT \NbX{I} down from the bulk to a bilayer structure, which barely affects the in-plane hopping or the out-of-plane dimerization strengths, the LT bilayer \NbX{I} nevertheless behaves fundamentally different (as a correlated insulator) as a result of the drastically enhanced local Coulomb interaction due to the reduced screening in the bilayer structure.

Based on our phase classification discussion from above, we note that we expect the transition from a Mott to a correlated insulator to happen rather abruptly (as it is accompanied by clear qualitative differences, e.g., in the self-energies), while the transition from correlated to band insulators is more continuous. We furthermore expect that the phase boundaries are not only controlled by the out-of-plane correlation strengths, but also by the in-plane one. This is reflected by taking the critical $t_\parallel/U_0$ ratio for the Mott transition in triangular lattices as found by Ref.~\cite{Wietek21} into account here.

Most importantly, this generalized phase diagram gives a clear guidance into which direction a system changes. For example, upon applying pressure the in- or out-of-plane hopping likely increase such that a Mott insulator might be tuned to a correlated insulator or a correlated one into a conventional band insulator. Similarly, decreasing the Coulomb interaction, e.g. in bilayer systems via external screening, can tune the material through the same phases, but affecting simultaneously the in- and out-of-plane correlation strengths. This way this phase diagram clearly shows that uniaxial in- or out-of-plane strain or stress affects the material at hand differently than external screening. Also, from this it becomes clear why all monolayer as well as HT bulk \NbX{X} systems will be Mott insulators: in all of them the out-of-plane correlation strengths is zero (or strongly suppressed), while the in-plane one is still large as result of small $t_\parallel$ and (rather) large $U_0$

\subsection{Dynamical Lattice Stability and Electron-Phonon Coupling to Low-Energy Subspace}

So far, we showed that the low-energy electronic structure of all compounds will be renormalized by the (partially strong) Coulomb interactions. Together with the inclusion of the so-far only putative \NbX{F}* structure, this raises questions about the dynamical stability of the underlying LT bulk crystal structures. To address this, we performed density functional perturbation theory (DFPT) calculations for all compounds (see Methods for computational details).

The resulting low-energy phonon dispersions for bulk LT \NbX{X} are shown in Fig.~\ref{fig_new:DFPT} and full dispersions can be found in Appendix~\ref{sec:fullphonons}. We see that within DFPT all compounds are dynamically stable, even the new putative \NbX{F}*. From \NbX{I} to \NbX{F}*, we see clearly how the phonon spectra get stiffer as a result of the decreasing lattice constant, cf. Tab.~\ref{tab:parametersfull}. This trend is in line with previous phonon calculations for the monolayer compounds~\cite{jiang_exploration_2017,mortazavi_first-principles_2022}. We also note that the monolayer and bilayer LT structures are dynamically stable in our calculations (see Appendix~\ref{sec:fullphonons}) and that the monolayer phonon dispersions are in good agreement with previous data~\cite{jiang_exploration_2017,mortazavi_first-principles_2022}.

While this is a promising result, we need to reflect on the applicability of DFPT in the presence of (strong) Coulomb correlation effects. Specifically, the corresponding renormalization effects to the electronic structures are not present on the DFT level, which has been used to relax all structures and which is used to perform the perturbation calculations within DFPT. Thus, it is a priori not clear how reliable the DFT optimized lattice structures and the presented phonon dispersions are. To investigate this, we additionally performed so-called constrained DFPT (cDFPT)~\cite{cDFPT,Berges_cdfpt}  calculations for \NbX{F}* and \NbX{Cl}, which allows to calculate the phonon spectra upon removing all electron-phonon couplings to the low-energy electronic structure. Therefore, the (most strongly) renormalized parts of the electronic structure are effectively disregarded within cDFPT.

Comparing the DFPT and cDFPT phonon spectra of \NbX{F}* and \NbX{Cl} in Appendix~\ref{sec:cdfpt}, we see that the coupling to the low-energy electronic states barely changes the phonon energies. This is an important result, as it shows that the phonons are weakly coupled to the low-energy electronic states. Therefore, it is reasonable to assume that beyond-DFT corrections to this electronic sub-space do not affect the dynamical stability, and that our relaxed structures and the resulting DFPT phonon spectra are trustworthy.

\subsection{Doping Effects}

\begin{figure*}[!]
    \centering
    \includegraphics[width=0.99\linewidth]{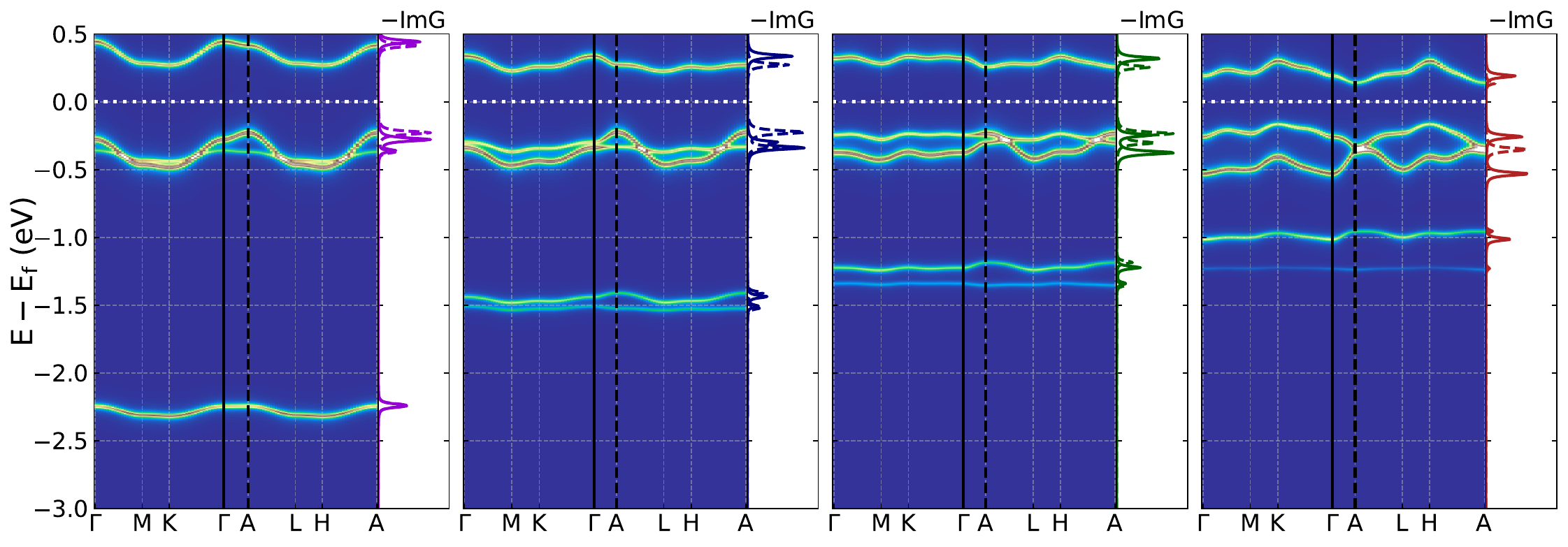}
    \caption{\textbf{Theoretical Spectral Functions at $\mathbf{N_e=3}$.} (a-d) Interacting spectral functions of doped compounds ($N_e=3$) calculated using the Hubbard-I approximation. Line cuts at $\Gamma$ \& $A$ are shown in full and dashed lines respectively.}
    \label{fig:Hubbard-I-N3}
\end{figure*}

The study of doping provides a way to differentiate between strongly and weakly correlated insulators. In conventional weakly correlated band insulators, electron or hole doping just yields a shift in the chemical potential while the band structure does not change. In contrast, strongly correlated insulators can show entirely new electronic features, such as Kondo resonances~\cite{Georges96}, upon doping.

\subsubsection{Integer Doping ($N_e=3$)}

We begin with doping the system by exactly one electron per dimerized inter-layer trimer. 
In this case we still deal with integer filling of the two flat bands, but with $N_e=3$ instead of half filling with $N_e=2$, such that we can still use the Hubbard-I approximation. The resulting doped spectral functions are shown in Fig.~\ref{fig:Hubbard-I-N3}. From a pure band-structure perspective, we would expect that doping the systems with one electron (or equivalently one hole), would just half-occupy the non-interacting anti-bonding (bonding) band, making all systems good metals. Instead, we find that \emph{all systems} have a gap of around $0.5\,$eV, which we understand as clear evidence for strong correlation effects developing in all systems, including \NbX{I}. In detail, by doping with only one electron (hole), \NbX{F}* and \NbX{Cl} are now strongly correlated charge transfer insulators (see overlap between the new uncorrelated bonding bands and the lower Hubbard band of the anti-bonding band), while \NbX{Br} and \NbX{I} are now Mott insulators. In all cases the gaps are opened by strong retardation effects of the self-energy as we explain in more detail in Appendix~\ref{app:graphicalsol}. Taking the approximate particle-hole symmetry of the flat non-interacting bonding/anti-bonding bands into account, we expect that doping in the other direction, to $N_e=1$, yields similar correlated electronic structures. Experimentally, this could be achieved by replacing one halide with one chalcogen atom per dimerized unit cell, e.g., within Nb$_3$X$_{7.5}$Te$_{0.5}$. We note that Refs.~\cite{Zhang_2023} and \cite{Nb3TeI7_2025} find Nb$_3$Cl$_7$Te and Nb$_3$I$_7$Te still displaying the breathing mode kagome lattice, respectively, indicating the stability of the lattice even under higher doping.

We further note that for $N_e=3$, the non-local (density-density) Coulomb interaction is now playing a quantitatively important role by strongly increasing the gap. If we neglect this contribution, the strongly correlated gaps are much smaller (see Appendix \ref{app:localU}). This once again stresses the relevance of having access to all material-specific parameters, including the non-local Coulomb interactions.

\begin{figure*}[t]
    \centering
    \includegraphics[width=0.99\textwidth]{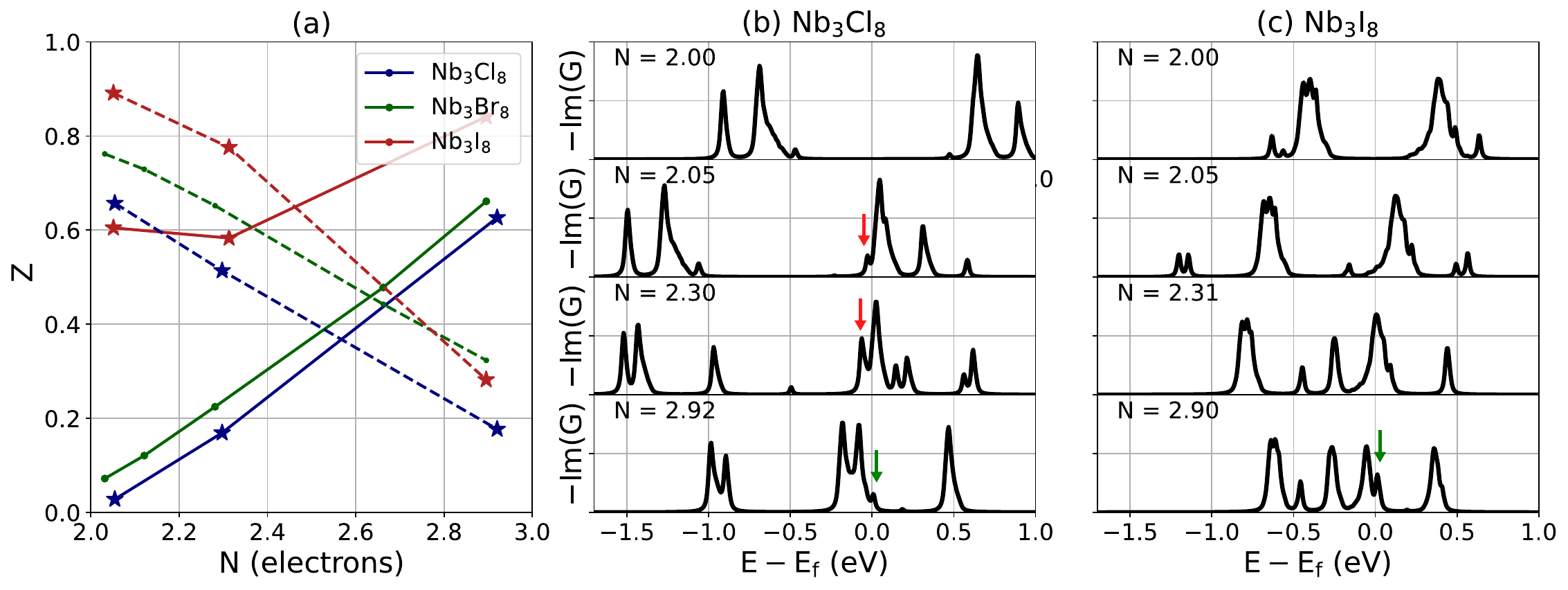}
    \caption{\textbf{Doping Dependence.} Behavior under doping of (a) the quasiparticle weight of the bonding (solid) and anti-bonding (dashed) components and the analytically continued local spectral functions of (b) \NbX{Cl} and (c) \NbX{I}. Stars in (a) indicate doping levels, which are analyzed in (b) and (c). \label{fig:doping}}
\end{figure*}

\subsubsection{Continuous Doping ($N_e=2+\delta$)}

To analyze the details of the changes from $N_e=2$ to $N_e=3$, we performed additional calculations for the Cl, Br, and I compounds at various doping levels in between denoted by $N_e = 2 + \delta$. In these cases Hubbard-I is not a good approximation any more, such that we are restricted to imaginary time/frequency cluster DMFT and corresponding quantum Monte Carlo calculations (see \hyperref[sec:methods]{Methods} for details). The resulting imaginary frequency self-energies are discussed in Appendix \ref{app:Sigma_matsubara}. In the bonding orbital channel we find for small dopings $\delta$ stronger dynamic correlation effects in \NbX{Cl} and \NbX{Br} compared to \NbX{I}, which become rather weak for all materials toward $N_e=3$. The anti-bonding self-energy behaves oppositely for all systems. Upon increasing the doping from $N_e=2$ to $N_e=3$, dynamic correlations generally increase. This is also reflected in the orbital-dependent quasiparticle weights, which we extracted from the imaginary frequency self-energies and show in Fig.~\ref{fig:doping}~(a). The bonding quasiparticle weights of \NbX{Cl} and \NbX{Br} start close to zero in the vicinity of $N_e=2$, indicative of strong correlation effects, and steadily increase with doping, while for \NbX{I} it is larger than $0.5$ at all fillings, indicative of moderate correlation effects. On the other hand, the anti-bonding orbital quasiparticle weight drastically decreases towards $N_e=3$ for all three compounds. This can be understood from the band perspective, where $N_e=3$ corresponds to a half-filled anti-bonding band. The large Coulomb interaction compared to the kinetic energy thus renders the $N_e=3$ systems more or less conventional single-band Mott insulators, but with a lower Hubbard band that is close to the fully-occupied bonding band.

To gain further spectroscopic insights into the continuous doping responses of the systems, we continued our imaginary frequency data to the real axis. To this end, we apply the recently proposed minimal pole representation method ~\cite{prony1,prony2} to analytically continue the matrix-valued local self-energies and calculate from this the interacting local spectra for \NbX{Cl} and \NbX{I}. We chose the latter two to highlight the differences between the doping dependence of a strongly and a weakly correlated system. The results are shown in Fig.~\ref{fig:doping}~(b) and (c). For \NbX{Cl} starting with a small doping of $\delta=0.05$, we see that the main features shift and, importantly, a small resonance at $\omega=0\,$eV arises, which grows in intensity towards $\delta=0.3$, as indicated by the red arrows in Fig.~\ref{fig:doping}~(b). This is reminiscent of a Kondo resonance, which is a hall-mark of doped strongly correlated insulators. Simultaneously, we see that at higher energies new resonances appear, which form at large dopings ($\delta=0.92$) a new upper Hubbard band. In fact, we can understand the \NbX{Cl} $\delta=0.92$ data as a slightly hole-doped variant of the $N_e=3$ situations, which already shows the single-orbital lower and upper Hubbard bands, but here, due to finite hole doping, accompanied by a Kondo resonance, as indicated by the green arrow in Fig.~\ref{fig:doping}~(b).

In contrast, the \NbX{I} spectra more or less just shift in energy until around $\delta=0.3$, without introducing any new features and especially without showing any evidence of a Kondo resonance at zero energy. This is the doping dependence we expect from conventional weakly correlated band insulators. However, for larger dopings, we start to see new features, which are in line with our interpretation of \NbX{I} as a single-band Mott insulator at $N_e=3$. In detail, at $\delta=0.9$ we should understand \NbX{I} as a hole-doped single-band Mott insulator showing a Kondo resonance, cf. green arrow in Fig.~\ref{fig:doping}~(c).

\subsection{Magnetic Properties of Monolayer and Bulk \NbX{X}}

Magnetism is an ubiquitous concept in correlated materials and has correspondingly received considerable attention in \NbX{X} as well. 
For the \NbX{Cl} monolayer, our previous work~\cite{grytsiuk2024nb3cl8} and recent spin-spiral DFT~\cite{mangeri_2024} calculations both show a magnetically ordered state with 120$^\circ$ angles between trimer-centered moments. 
For the bulk LT phase, on the other hand, out-of-plane dimerization can lead to singlet-formation~\cite{NbCl_structC2, Pasco_2019, Gao_2023, Zhang_2023_PRB}, which is common in bilayer Hubbard models at large $U$ and large $t_\perp/t_\parallel$~\cite{Kancharla07,Hafermann_2009,Hu2024}.

Although a detailed account of the quantum magnetism in \NbX{X} is beyond the scope of this work, it is insightful to study the connection between the magnetic, electronic and lattice structure that arise from our downfolded models. 
Given the dominance of the Coulomb interaction over the hopping matrix elements [e.g., for \NbX{Br} $U^\text{eff}$ is approximately five (hundred) times larger than the largest out-of-plane (in-plane) hopping, see Tab.~\ref{tab:parametersfull}], it is appropriate to describe the magnetic properties of \NbX{X} using a Heisenberg model that incorporates trimer-localized magnetic moments.
Accordingly, we approximate the exchange interactions parameters $J_{ij} = -2 (t_{ij})^2\slash U^{\text{eff}}$ using the strong coupling expansion that incorporates the effective Coulomb interaction (see Section~\ref{sec_jij} for details).
This formulation indicates that the exchange interaction is strongest when the hopping matrix element between two sites is large, leading to antiferromagnetic (AFM) exchange interactions ($J_{ij} < 0$). This behavior is characteristic of materials with large $U$, such as all Nb halides, and explains the differences to DFT estimates that do not account for large values of $U$~\cite{feng_enabling_2023,mangeri_2024}. 

Using our downfolded material-specific parameters from Tab.~\ref{tab:parametersfull} and~\ref{tab:ht_ab_papram}, which includes appropriately screened non-local Coulomb matrix elements, we can thus calculate all (long-ranged) magnetic exchange interactions $J_{ij}$ for all monolayer and bulk structures, see Tab.~\ref{tab:jijparameters_ML} and~\ref{tab:jijparameters}, respectively. Afterwards, using either atomistic spin dynamics  or classical Monte Carlo simulations (see Section~\ref{sec_meth_spirit}), we solve the resulting Heisenberg Hamiltonian 
\begin{equation}
H = -\sum_{i,j} J_{ij} {\bf S}_i \cdot {\bf S}_j - \sum_i {\bf B} \cdot {\bf S}_i\, ,
\label{eq:Heisenberg}
\end{equation}
where the first term represents isotropic exchange interactions and the second term accounts for the Zeeman energy in the presence of an external magnetic field ${\bf B}$.

\subsubsection{Spin Spirals in Monolayer \NbX{X}}

{\renewcommand{\arraystretch}{1.2}
\begin{table}[h!]\centering
\caption{
Effective exchange parameters $J^{(n)}$ (in meV) and ground state spin spiral periods $\lambda^{(n)}$ (in units of in-plane lattice parameter $a$) for ML \NbX{X}.
Here, $n=1,2,3$ denotes the first, second, and third shells of nearest neighbors, with 
$n \leq 2$ and $n \leq 3$ indicating truncations for exchange interactions $J_{ij}^{(n)}$ to two and three shells, respectively.
\label{tab:jijparameters_ML}}
\begin{tabular}{l  R{1.2cm} R{1.2cm} R{1.2cm} R{1.2cm} 
}\Xhline{2\arrayrulewidth}
{} & \NbX{F} &  \NbX{Cl} & \NbX{Br} &  \NbX{I} \\ \hline\hline
$J^{(1)}$ & -1.12	 &  -0.63  & -0.04  &  -0.40 \\
$J^{(2)}$ &  0.00	 &  -0.04  & -0.07  &  -0.16 \\
$J^{(3)}$ &  0.00  &  -0.03  & -0.08  &  -0.26 \\\hline
$\lambda^{(n \leq 2)}$  & $1.5 a$ & $1.5 a$ & $2.1 a$ & $1.7 a$\\
$\lambda^{(n \leq 3)}$& $1.5 a$ & $1.5 a$ & $2.8 a$ & $2.3 a$\\
\bottomrule
\end{tabular}
\end{table}}

We first investigate the magnetic ground-state properties of \NbX{X} monolayers. Using the exchange interaction parameters provided in Tab.~\ref{tab:jijparameters_ML}, 
we solve Eq.~\ref{eq:Heisenberg} at $|{\bf B}| = 0$ using atomistic spin dynamics simulations. The results indicate that the ground states display spin spiral order across all \NbX{X}. However, due to the finite size of the simulated structures, the resulting configurations yield commensurate spin spiral ordering. To address this, we analytically minimize the total energy given by the Hamiltonian from Eq.~\ref{eq:Heisenberg} with respect to the spin-spiral wave vector ${\bf q}$ assuming flat spirals (see  Appendix~\ref{sec:ML_mag}). The ground state periods of the spin spirals given by $\lambda = 2\pi/|{\bf q}|$ for various truncation distances between interacting effective spin pairs are given in Tab.~\ref{tab:jijparameters_ML} and further details can be found in Appendix~\ref{sec:ML_mag}.

\NbX{F}* and \NbX{Cl} exhibit identical spin spirals with a periodicity of $1.5a$, corresponding to a $120^\circ$ anti-ferromagnetic (AFM) order, independent of taking the third nearest neighbor interaction into account or not. This changes for \NbX{Br} and \NbX{I}. Notably, the relatively large $J^{(2)}$ for these compounds gives rise to incommensurate spin-spiral configurations, and when $J^{(3)}$ is taken into account, the modulation period increases further.

\subsubsection{Out-of-plane Magnetic Frustrations as the Origin of the Lattice Phase Transition in Bulk \NbX{X}}

{\renewcommand{\arraystretch}{1.2}
\begin{table}[h!]\centering
\caption{Effective exchange parameters for bulk Nb$_3$X$_8$ in the LT and HT structure in meV.  The labels (s) and (w) denote  strongly and weakly bound neighbors in the out-of-plane ($\perp$) directions, respectively, while $n=1,2,3$  corresponds to first, second, and third nearest neighbors in the in-plane ($\parallel$) direction. 
The last row presents the energies (in meV) associated with the magnetic ground state. For more details, see Appendix~\ref{sec:ML_mag}.
\label{tab:jijparameters}}
\begin{tabular}{R{0.5cm}  R{1.2cm} R{1.2cm} R{1.2cm} R{1.2cm} | R{1.2cm} R{1.2cm}
}\Xhline{2\arrayrulewidth}
        & \multicolumn{4}{c|}{LT phase} & \multicolumn{2}{c}{HT phase}\\\cline{2-7}
        & Nb$_3$F$_8^*$
        & \NbX{Cl}
        & \NbX{Br} 
        & \NbX{I} 
        & \NbX{Cl}
        & \NbX{Br}
        \\ \hline\hline
\multicolumn{1}{l}{$J^{(\text{s})}_\perp $} & -0.03 & -35.21 & -67.98&  -180.21 & -0.55 & -0.99\\
\multicolumn{1}{l}{$J^{(\text{w})}_\perp $} & -0.05 &  -0.49 &  -0.98 &   -2.29 & -0.48 & -0.96\\ \hline
\multicolumn{1}{l}{$J^{(1)}_\parallel    $} & -1.32 &  -0.83 &  -0.09 &   -0.26 & -1.18 & -0.30\\
\multicolumn{1}{l}{$J^{(2)}_\parallel    $} & -0.00 &  -0.05 &  -0.07 &   -0.16 & -0.05 & -0.07\\
\multicolumn{1}{l}{$J^{(3)}_\parallel    $} &  0.00 &  -0.04 &  -0.10 &   -0.33 & -0.04 & -0.10\\ \hline
\multicolumn{1}{l}{$E_\text{ex}$} &-3.99 & -37.61& -70.38&-185.98 &-3.62 & -3.94\\
\bottomrule
\end{tabular}
\end{table}}

We proceed with the discussion the magnetic properties in both low- and high-temperature bulk structures as resulting from the magnetic interaction parameters given in Tab.~\ref{tab:jijparameters}.

In bulk, The main difference between the LT and HT phases lies in the connectivity of the layers, as illustrated in Fig.~\ref{fig:structure}~(b). In the HT phase, every trimer is magnetically connected to three trimers in the layer above and three in the layer below by (rather small) $J_\perp^{(s/w)}$. As a result of the anti-ferromagnetic inter-layer coupling, this leads to magnetic frustration in the HT structure: the trimer-moment cannot align consistently with all three moments in the upper and lower neighboring layers. 
On the other hand, in the LT phase, the dimerization in the vertical direction lifts this frustration and every trimer moment has another trimer moment to which it is coupled strongly by $J_\perp^{(s)}$, supporting singlet formation. Overall, this lowers the total energy of the system.

\begin{figure}[!t]
    \centering
    \includegraphics[width=0.4\textwidth]{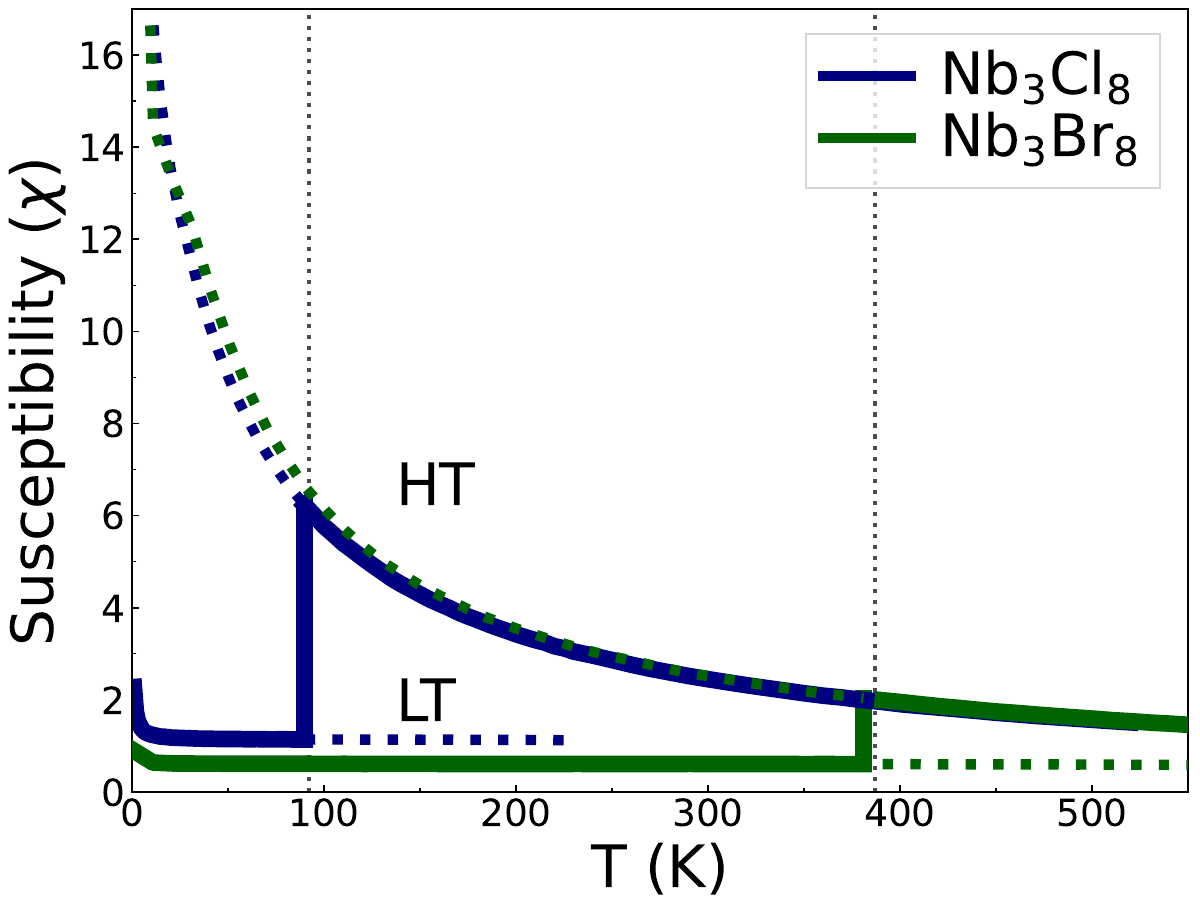}
    \caption{\textbf{Magnetic Susceptibility of \NbX{Cl} and \NbX{Br}.} Classical Monte Carlo simulations of the magnetic susceptibility for both HT and LT structures as function of temperature $T$ at magnetic field $B=5\,$T. The dotted vertical lines mark the transition temperatures at $T = 92$ and $387\,$K.}
    \label{fig:Nb3X8_Mag}
\end{figure}

This is reflected in the total (magnetic) energies of \NbX{Cl} and \NbX{Br}, which are reduced by about $34\,$meV and $66\,$meV in the LT phases, respectively (see Tab.~\ref{tab:jijparameters}).
This trend clearly correlates with the critical transition temperatures of \NbX{Cl} and \NbX{Br}~\cite{Pasco_2019} and offers a simple explanation of why \NbX{I}, that has the largest exchange interaction energy among other LT \NbX{X} compounds, has so far been only observed in the LT phase. 
These observations made possible by our material-specific matrix elements are in line with the suggestion by Sheckleton et al.~\cite{NbCl_structC2}, that the inter-layer magnetic frustration in the HT phase combined with energy gain due to singlet formation in LT phase drives the structural transition.

We note that in the case of LT \NbX{F}*, singlet formation is not energetically favorable. This is due to 
$J^\text{(s)}_\perp < J^\text{(w)}_\perp < J^\text{(1)}_\parallel$, which indicates that the primary contribution to the ground state arises from the in-plane $120^\circ$-AFM ordering.

This interpretation also aligns with the local magnetic susceptibilities, which we compute using classical Monte Carlo simulations (for computational details see Section~\ref{sec_meth_spirit}) for \NbX{Cl} and \NbX{Br} in their LT and HT structures, solving Eq.~\ref{eq:Heisenberg} and considering a finite magnetic field. Indeed, as shown in Fig.~\ref{fig:Nb3X8_Mag}, the computed magnetic susceptibilities for the HT structures of both \NbX{Cl} and \NbX{Br} match and follow a Curie-Weiss behavior. A concomitant collapse of the magnetic susceptibility accompanies the structural phase transition to the LT phase for both compounds. The reduction in magnetic susceptibility of LT \NbX{Br} compared to LT  \NbX{Cl} can be attributed to a larger $J_{\perp}^{\text{(s)}}$, indicating that LT \NbX{Br} is less responsive to magnetic fields. This seems to be perfectly consistent with available experimental data~\cite{NbCl_expmagsusc,NbCl_structC2,Pasco_2019,Jeff_2023,Kim_2023,Liu_2024}.
However, we note that for fully compensated singlets, the susceptibilities in the LT phases should be completely suppressed, while our calculations show small but finite responses accompanied by small tails at low temperatures. This aspect thus necessitates a full quantum mechanical treatment in the future.

\subsection{Symmetry Breaking in Few-Layer Stacks and Connections to the Field Free Josephson Diode Effect}

\begin{figure}
    \centering
    \includegraphics[width=0.40\textwidth]{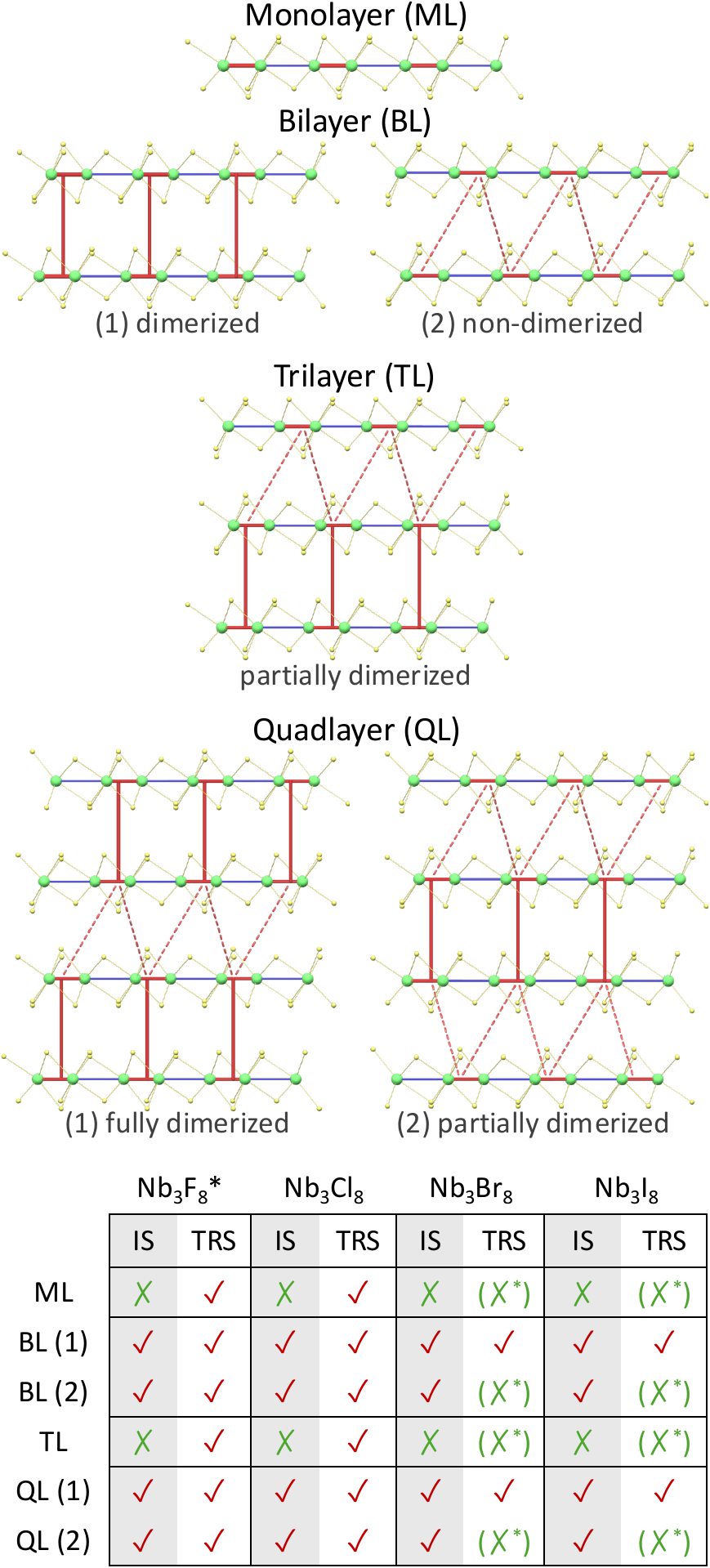}
    \caption{\textbf{Symmetries of Finite Stacks of \NbX{X}.} Inversion and time-reversal symmetry in monolayer, bilayer, trilayer, and quadlayer of \NbX{X}, taking into account various surface ("edge") terminations. 
    The solid dark red line represents strong interlayer coupling between the nearest trimers, while the dashed line indicates weak coupling. 
    Crosses (in brackets) indicate (possible but not necessary) symmetry-breaking. Red checkmarks denote the preservation of symmetry.}
    \label{fig:symmetries}
\end{figure}

We now turn our focus to finite stacks of a few \NbX{X} layers and their potential role for the field-free Josephson diode effect. The latter, as observed in NbSe$_2$/\NbX{Br}/NbSe$_2$~\cite{wu2022field}, requires simultaneous time-reversal (TRS) and inversion symmetry (IS) breaking~\cite{PhysRevB.103.245302,zhang_JDiode_2022}.

The IS breaking can be straightforwardly discussed: \NbX{X} stacks with an even number of layers are inversion symmetric, while stacks with an odd number of layers break inversion symmetry. In particular, inversion symmetry is broken in the monolayer.
The TRS breaking requires more attention, especially with respect to the impact of the dimerization (alternating hybridization) of bilayers, which depends on the number of involved layers. Specifically, there can be top and/or bottom surface layers that only weakly hybridize with their neighbor and can therefore approximately behave like freestanding monolayers. 
A trilayer, for example, consists of a fully dimerized bilayer accompanied by a nearly-free monolayer, as depicted in Fig.~\ref{fig:symmetries}.
Taking the strong correlation effects into account, these surface monolayers realize (approximately) conventional single-band Mott insulators~\footnote{
    In the monolayer limit, all four compounds host a single half-filled flat band around the Fermi level and, as a result of reduced screening, local Coulomb interaction matrix elements are larger than in the bulk (e.g., $U=1.9\,$eV for monolayer \NbX{Cl}~\cite{grytsiuk2024nb3cl8} versus $1.5\,$eV in bulk according to Tab.~\ref{tab:parameters}). For a surface layer in a finite stack, the expectation is that $U$ is between the latter values, since there is screening from one side. 
}, which are only weakly coupled to the neighboring dimerized bilayer, as we show in more detail in Appendix~\ref{app:trilayer}. 
In contrast, mean field DFT calculations, wrongly predict the trilayer to behave like a metallic monolayer in contact with a band-insulating bilayer, cf.\ Appendix~\ref{app:trilayer} and Ref.~\cite{zhang_JDiode_2022}. Thus, in the context of symmetry breakings and the Josephson diode effect, correlation properties must be considered consistently throughout the whole structure with a special focus on unpaired surface monolayers.

As previously discussed for \NbX{Cl}~\cite{grytsiuk2024nb3cl8,mangeri_2024} and mentioned above and in Appendix~\ref{sec:ML_mag}, these effective monolayers likely host variants of magnetic spin-spiral states. 
Thus, all stacks with at least one unpaired surface layer can be classified as magnetic, which potentially breaks TRS. 
In contrast, stacks consisting of fully dimerized bilayers are non-magnetic as a result of the singlet formation, such that TRS stays intact.

However, whether a magnetic spin-spiral state and its potential TRS breaking also has a global effect on the electronic properties might depend on the electron coherence length~\cite{Spiral_sup_cond}. If the latter is smaller than the spin-spiral wavelength, there might be a global effect. For example, the 120$^\circ$ Néel state of monolayer \NbX{Cl} seems to leave the TRS after backfolding to the primitive unitcell intact~\cite{mangeri_2024}. We find the spin-spiral wavelengths to be ranging from $10\,$\AA\, to $18.5\,$\AA\, in \NbX{Cl} and \NbX{I}, respectively (see Appendix \ref{sec:ML_mag}), which is on the same order as the electronic coherence length of other Mott insulators~\cite{Coherence_length,kokin1975metal,kokin1976low}. Based on this, we assume a possible global effect of the TRS breaking in the surface monolayers of \NbX{Br} and \NbX{I}, but possibly not in \NbX{F}* and \NbX{Cl}.
Additionally, we note that the interfaces to the superconducting leads in NbSe$_2$/\NbX{Br}/NbSe$_2$, as well as magnetic impurities, might further contribute to the TRS breaking.

Taken together and as summarized in Fig.~\ref{fig:symmetries}, this suggests that odd stacks of \NbX{Br} and \NbX{I} might intrinsically allow for the field-free Josephson diode effect, while \NbX{Cl} and \NbX{F}* or in the case of fully dimerized bilayers in even stacks, interface effects to the encapsulating superconductors or impurities might be additionally required. 

\section{Conclusion and Discussion}

Our analysis shows that the low-energetic electronic structure as well as the magnetic properties of the whole class of \NbX{X} are dominated by the formation of flat bands as a result of the breathing-mode kagome distortion. Since the resulting low-energetic bands are well separated from those formed by all other Nb and halide orbitals, low energy generalized Hubbard models can be reliably obtained by ab intio downfolding using the constrained random phase approximation in combination with Wannier orbitals representing molecular orbitals centered at the trimerized Nb clusters.

The resulting models properly render the local chemistry of the effective molecular orbitals as well as the global screening properties. This allow us to quantify the strengths of in- and out-of-plane hopping as well as local and non-local Coulomb matrix elements in a material-specific manner. 
Our ab initio model database further shows that the whole material family is in the regime of strong local interaction ($U_0/t_\parallel > 60$). Since the ratios between these kinetic and local interaction terms vary from monolayer to bulk structures and from \NbX{F}* to \NbX{I} as a result of modifications to the global screening, as well as from varying local confinements of the molecular orbitals, the degree of Coulomb correlations gradually decrease from \NbX{F}* to \NbX{I} and from ML to bulk structures. Based on a strong coupling expansion we can further derive effective magnetic models, which show that the high-temperature crystal structures host inter-layer magnetic frustrations, which are lifted by singlet formation in the low-temperature structures. As these frustrations are the strongest in \NbX{I} and gradually diminish in \NbX{Br} and \NbX{Cl} accompanied by an decreasing energy gain from singlet formation, the accompanying critical temperatures of the crystallographic phase transition also gradually decrease from \NbX{I} to \NbX{Cl}. 

As a result, the low-temperature bulk materials are best understood as dimerized bilayers, which only couple weakly among each other through vdW interactions. The magnetic ground states are thus most likely singlets formed between inter-layer neighboring spins localized on the strongly hybridized molecular orbitals. On the electronic side, our combined cRPA and cluster DMFT calculations show that the halide atoms control the ratios between the local Coulomb interactions and the in- and out-of-plane hopping terms and thereby drive the low-temperature bulk materials from a conventional two-band Mott insulator (\NbX{F}*) to strongly correlated insulators (\NbX{Cl} and \NbX{Br}) and finally to a weakly correlated band insulator (\NbX{I}). Moreover, since the screening is significantly reduced in the bilayer structures, we find that even the \NbX{I} bilayer becomes a strongly correlated insulator. In turn, due to the missing out-of-plane dimerization, all monolayer, as well as high-temperature structure variants of \NbX{X}, are single- or two-band Mott insulators, respectively. Thus, already in the undoped case the class of \NbX{X} host several correlated ground states of different characters. To differentiate between these variants of strongly and weakly correlated insulators, we showed how the charge-neutral and temperature dependent excitations spectra as well as the charged excitation spectra of the local (interlayer molecular) dimer differ and how this is reflected in experimentally accessible dispersive bonding and nearly non-dispersive anti-bonding lower Hubbard bands.

These correlated states can further be tuned either by doping or by the thickness of finite stacks. We showed that upon electron doping \NbX{X} can be driven through intermediate states of correlated metals to charge transfer or conventional Mott insulators. Furthermore, as a result of the alternating strong and weak hybridziation between adjacent layers, finite stacks of \NbX{X} can host weakly hybridized surface monolayers, which host on their own already at half-filling Mott insulating states.

In terms of the in-plane trimerization, similar compounds, such as Mo$_3$O$_8$, are also accompanied by strong correlations~\cite{nikolaev2021quantum}, but do not allow for single molecular orbital Hubbard model descriptions. 
In terms of the out-of-plane dimerization, \NbX{X} shares many similarities with bulk VO$_2$, one of the most well-studied correlated systems. It also hosts a crystal structure transition from a high temperature to a low temperature phase, which is accompanied by a dimerization of the vanadium atoms into pairs resulting in singlet formation~\cite{PhysRevB.10.1801,PhysRevLett.94.026404}. The latter yields a correlated gapped state, which is driven by strong Coulomb correlation effects, the dimerization, and the accompanying singlet formation~\cite{PhysRevLett.95.196404,PhysRevLett.97.116402,Tomczak_2007,PhysRevB.78.115103}, which can also be understood using a molecular orbital bonding/anti-bonding Wannier basis~\cite{GOODENOUGH1971490,PhysRevB.11.4383}.  However, the vdW material class of \NbX{X} comes with the significant advantage of a rather simple low-energy band structure, which allows for a clean disentanglement and thus for a trustworthy downfolding to minimal correlated models. 

Thus, the combination of in-plane trimerization and out-of-plane dimerization renders \NbX{X} an extremely robust, versatile, but also simple platform to study and engineer many-body phenomena.
In combination with the various and tunable degree of correlations, \NbX{X} can thus be seen as a promising \textit{drosophila melanogaster} of layered correlated (Mott) insulators, which offers the exciting possibility to precisely study numerous correlation phenomena both theoretically and experimentally. The models which we derived here yield an optimal theoretical framework for this purpose, and the complete model specifications for the entire \NbX{X} family are made freely available via a public repository, see Ref.~\cite{Nb3X8ModelDB}.

\section{Outlook}\label{Outlook}

An important difference to other prototypical bulk strongly correlated insulators is that as a result of the layered vdW structure screening is generically suppressed in \NbX{X}, such that significantly long-ranged Coulomb interactions are present, cf. Tab.~\ref{tab:parametersfull}. These could yield a non-local coupling of holons and doublons forming Mott-excitons~\cite{MottEx2020} in the strongly correlated insulators \NbX{Br} and \NbX{Cl} or conventional excitons in \NbX{I}, which could be influenced by the magnetic properties of the system. We can thus expect in-gap features in optical spectroscopy experiments, which might have been experimentally seen already in the case of \NbX{Cl}~\cite{Gao_2023,meng_2024}. Furthermore, in few- or monolayer systems, these non-local Coulomb interactions are tunable by means of environmental screening as, e.g., resulting from substrate materials. Together with first indications that long-range Coulomb interactions in combination with doping could yield novel charge ice states~\cite{stepanov2024signatureschargeicestate}, this renders \NbX{X} a promising platform for strong-correlation engineering of unconventional magnetic, charge, or superconducting order but also of optical effects via non-local Coulomb interactions~\cite{van_loon_coulomb_2023}.

Another intriguing many-body engineering pathway is via chemical modification, such as already realized in admixtures of the form Nb$_3$X$_{8-x}$Y$_x$ with X and Y being both halide atoms~\cite{Pasco_2019,Gao_2023}. As the halide controls the degree of correlation in this system class, such admixtures might be best understood as disordered (strongly) correlated (Mott or charge transfer) insulators. Alternatively, Nb$_3$X$_{8-x}$Y$_x$ with Y being a chalcogen atom~\cite{Zhang_2023,Nb3TeI7_2025} might allow for precise chemical doping to access the predicted strongly correlated metal regime and to realize the recently suggested Sordi transition~\cite{sordi2024}.

Finally, the vdW nature of \NbX{X} offers the possibility to mix and stack them into layered vdW heterostructures. Experimentally, this has already been done in the form of NbSe$_2$/\NbX{Br}/NbSe$_2$ heterostructures, which host non-reciprocal superconducting properties. Taking furthermore into account the different degrees of correlations already present within the class of \NbX{X}, stacking them into \NbX{X}/\NbX{Y}/\NbX{Z} heterostructures and/or mixing them with other correlated layered materials opens up an incredibly rich new toolbox for designing strongly correlated vdW heterostructures, e.g., with spatially varying local Coulomb interaction strengths. So far, there is only 4Hb-TaS$_2$, which comes with similar possibilities~\cite{silber_two-component_2024,crippa_heavy_2024}. The electronic structure of 4Hb-TaS$_2$ is, however, drastically more involved, requiring additional approximations, which are not necessary within the class of \NbX{X}. This allows, e.g., for implementing the proposal of enhanced critical temperatures in superconducting heterostructures~\cite{Maza_2021} or to realize Mott field effect transistors or correlated Schottky barriers.

Taking all these chemically, environmentally, and/or doping tunable correlation phenomena in \NbX{X} into account renders this material class an exciting framework for many-body functionalization concepts. For these and other functionalization schemes, we however stress that the interfaces to other (substrate) materials as well as among different variants of \NbX{X} need to be studied in detail.

\acknowledgments

The authors acknowledge useful discussions with Alberto Carta, Claude Ederer, Dmytro Afanasiev, Dieter Vollhardt, Giorgio Sangiovanni, Mikhail Titov, Maurits Haverkort, Roser Valenti, Antoine Georges, Andrew J. Millis, and Silke Biermann. The authors especially thank Emanuel Gull and Lei Zhang for supporting the analytical continuation using Prony as well as Jan Berges for helpful discussions on the phonon calculations.

We acknowledge the National Academic Infrastructure for Supercomputing in Sweden (NAISS), partially funded by the Swedish Research Council through Grant Agreement No. 2022-06725, for awarding this project access to the LUMI supercomputer, owned by the EuroHPC Joint Undertaking and hosted by CSC IT Center for Science (Finland) and the LUMI consortium, as well as computer resources hosted by the PDC Center for High Performance Computing and the National Supercomputer Centre (Projects Nos.\ NAISS 2023/1-44, 2023/6-129, 2023/8-9, 2024/1-18, 2024/6-127, and 2024/8-15) where the computational pipeline was developed and tested. The material specific computations were performed at the Dutch National Supercomputer Snellius under Projects No. EINF-11321 \& EINF-7490.
XL acknowledges Guangzhou Elite Project scholarship (S. J. [2022] No. 1). 
ZD acknowledges the fellowship from the Chinese Scholarship Council (No.202206750016). 
MW acknowledges financial support from the Punjab Educational Endowment Fund (PEEF). 
AGC acknowledges the financial support of the Zernike Institute for Advanced Materials.  
PH acknowledges support from the Independent Research Fund Denmark  (Grant No. 4258-00002B).
MIK and HURS acknowledge funding from the European Research Council (ERC) under the European Union’s Horizon 2020 research and innovation programme (Grant Agreement No. 854843-FASTCORR).
EGCPvL acknowledges support from the Swedish Research Council (Vetenskapsrådet, VR) under grant 2022-03090, from the Royal Physiographic Society in Lund and by eSSENCE, a strategic research area for e-Science, grant number eSSENCE@LU 9:1.
MIK and MR acknowledge support from the Dutch Research Council (NWO) via the “TOPCORE” consortium. JA, CK, MW, MA, MIK, AGC, and MR acknowledge support from the research program “Materials for the Quantum Age” (QuMat). This program (registration number 024.005.006) is part of the Gravitation program financed by the Dutch Ministry of Education, Culture and Science (OCW). M.R. acknowledges support from the Vidi ENW research programme of the Dutch Research Council (NWO) under the grant https://doi.org/10.61686/YDRHT18202 with file number VI.Vidi.233.077.

\section{Methods}
\label{sec:methods}

\subsection{Ab Initio Modeling}
\label{sec_meth_dft}

\subsubsection{Density functional Theory Calculations and Lattice Structure Relaxations}

 We investigate the material-specific electronic structure starting from DFT calculations using the Perdew-Burke-Ernzerhof GGA exchange-correlation functional~\cite{PBE} within a basis of projector augmented waves \cite{PAW} as implemented in the Vienna Ab Initio Simulation Package (VASP) \cite{VASP,VASPPAW}. We apply a Gaussian smearing of $0.05\,$eV and a plane-wave cut-offs of $550\,$eV for \NbX{F} and \NbX{Cl} and $350\,$eV for \NbX{Br} and \NbX{I} and use pseudo-potentials from VASP 6.4. The electronic self-consistency criterion was set to $10^{-7}\,\text{eV}$ and the $k$-points were sampled on $6\times6\times6$ $k$ grids.
 
 We conducted lattice and structural optimizations for all compounds in their LT (also for \NbX{Cl} and \NbX{Br} in HT) bulk phases, keeping the space group fixed at R$\bar{3}$m (at P$\bar{3}$m1). 
 Atomic positions were allowed to relax until the forces on each atom fell below $0.01\,$eV\AA$^{-1}$. Mono- and bilayer structures were derived from the relaxed bulk phases by introducing a vacuum space that is three times larger than the interlayer spacing between the nearest Nb layers.
 The details of the relaxed lattice parameters are summarized in Tab.~\ref{tab:parametersfull} and \ref{tab:ht_ab_papram}. Additionally, all atomic positions and calculation inputs are accessible through our public repository~\cite{Nb3X8ModelDB}.

{\renewcommand{\arraystretch}{1.5}
\begin{table}[t!]\centering
\caption{Structural and Ab Initio Downfolded Many-Body Hamiltonian Parameters for HT \NbX{Cl} and HT \NbX{Br}. In-plane ($a$) and out-of-plane ($c$) lattice parameters, molecular orbital spread ($\Omega$), nearest neighbor in-plane ($t_\parallel$) and out-of-plane ($t_\perp$) hopping parameters and corresponding cRPA screened Coulomb ($U_\parallel$ and $U_\perp$) interactions, on-site bare ($V_0$) and screened ($U_0$) Coulomb interactions including their ratio $\varepsilon_\text{eff} = V_0/U_0$.}
\label{tab:ht_ab_papram}
\begin{tabular}{l R{2.5cm} R{2.5cm}   }
\Xhline{2\arrayrulewidth}
                        & \NbX{Cl} & \NbX{Br} \\\hline\hline
$a$                     & 5.84     & 6.14\\
$c$                     & 12.28    & 12.94\\
$d_2/d_1$               & 1.40     & 1.47\\
$\Omega$                &  7.26    &  8.69\\\hline
$t_\parallel$           &  23.54   & 10.69\\
$t_\perp^\text{(s)}$    & -17.11   & -20.56\\
$t_\perp^\text{(w)}$    & -15.92   & -20.25 \\\hline
$V_0$                   & 6088.7   & 5752.7\\
$U_0$                   & 1401.0   & 1129.1\\
$\epsilon_\text{eff}$   & 4.35 & 5.09\\\hline
$U_\parallel$           & 460.0    & 360.2 \\
$U_\perp^\text{(s)}$    & 336.8    & 276.5\\
$U_\perp^\text{(w)}$    & 299.2    & 242.0\\
\bottomrule
\end{tabular}
\end{table}}

\subsubsection{Wannier Function Basis, Constrained Random Phase Approximation Calculations, and Many-Body Model Matrix Elements}

  Following the recipe from Ref.~\cite{grytsiuk2024nb3cl8} and starting from the DFT calculations described above, we use Wannier90~\cite{Wannier90} to construct molecular orbitals describing the two low energy bands of our DFT calculation. To this end we start from $s$-orbital projections localized at the centers of the small trimers and proceed with a maximal localization of the Wannier functions.

  Within this basis, we calculate all single-particle (hopping) and many-body (Coulomb) matrix elements according to $t_{ij} = \braket{\psi_i | \hat{H}_\text{DFT} | \psi_j}$ and $U_{ijkl} = \braket{ \psi_i \psi_j | \hat{U}_\text{cRPA}(\omega) | \psi_k \psi_l}$, respectively. For the latter we use the constrained Random Phase Approximation (cRPA) as implemented in VASP \cite{cRPA_vasp}. 
  To converge the bare Coulomb matrix elements as well as the cRPA polarization we vary the plane-wave cut-offs as well as the number of involved bands until changes are smaller than ca. $50\,$meV in the cRPA screened local Coulomb interaction $U_0$. This requires Coulomb matrix element cut-offs between $350$ and $550\,$eV and up to $256$ bands. For the extraction of monolayer and bilayer Coulomb matrix elements, we further apply an extrapolations scheme as described in Appendix~\ref{app:WFCEExtra}. Since the electronic gaps in the ``rest'' band structures are rather large and since the low-energy flat bands are not entangled with the rest-space, using the static limit of the cRPA screening is a good approximation for the \NbX{X} material class~\cite{grytsiuk2024nb3cl8}

  In Tab.~\ref{tab:parametersfull} and \ref{tab:ht_ab_papram} we show the most relevant matrix elements for the discussion of the LT and HT compounds, respectively. All other elements, including all necessary parameter files, are available via our \NbX{X} model parameter database~\cite{Nb3X8ModelDB}.

\subsubsection{DMFT}

  The resulting generalized Hubbard models are solved using dynamical mean-field theory using TRIQS~\cite{TRIQS2015}, applying two different impurity solvers. At integer filling and large $U/t$, the Hubbard-I approximation provides a good description of the physics while being computationally efficient and giving access to real-frequency data directly. We use the implementation by Schüler~\cite{HIA}. The Hubbard-I spectral functions in Figs.~\ref{fig:Hubbard-I}, \ref{fig:Hubbard-I-N3}, \ref{fig:localU} \& \ref{fig:graphicalN23} are calculated at $\beta =5000 \text{ eV}^{-1}$ for $\text{\NbX{F}}^*$,  $1000 \text{ eV}^{-1}$ for \NbX{Cl} and $100 \text{ eV}^{-1}$ for \NbX{Br} and \NbX{I}. This ensures that the contribution to the spectrum from the first excited state, which is less and less separated in energy from \NbX{I} to \NbX{F}*, is sufficiently small. The results are (Lorentzian) broadened with $\delta = 0.01\,$eV.    

  For the doped system, we use the continuous-time hybridization expansion~\cite{Werner:2006rt, Werner:2006qy, Haule:2007ys, Gull:2011lr} (CTHYB) solver implemented in w2dynamics~\cite{w2dyn} with worm sampling~\cite{PhysRevB.92.155102, PhysRevB.94.125153}. The Discrete Lehmann Representation~\cite{DLR2022, KAYE2022108458, Kaye2024} is used to represent the Green's functions efficiently. The DMFT calculations were carried out at $\beta =100 \text{ eV}^{-1}$.

We track the electron density $N = \text{Tr}(\hat{\rho})$, with $\hat{\rho} = G(\tau=0^-)$, upon changes in the chemical potential. To facilitate the analysis, the self-energy and related quantities are shown in the bonding/anti-bonding basis which nearly diagonalizes the self-energy and dispersion (off-diagonal components are at least two orders of magnitude smaller than diagonal). The approximate quasiparticle weight is then derived from the self-energy on the Matsubara axis
\begin{equation}
    Z = \left( 1-\left.\frac{\operatorname{Im}\left[\Sigma\left(i \omega_n\right)\right]}{\omega_n}\right|_{\omega_n \rightarrow 0}\right)^{-1},
\label{eq:Z}
\end{equation}
as discussed in Ref.~\cite{arsenault2012benchmark}. 

The Matsubara self-energies are analytically continued using the minimal pole representation method recently introduced by Zhang \textit{et al.}~\cite{prony1,prony2}, whereby all poles were restricted to lie on the real axis. Consequently the relative strength and energy of excitations are captured, but broadening effects due to interaction are disregarded.  

\subsection{Phonon Calculations}
\label{sec:phonons}

All DFPT calculations are performed within Quantum ESPRESSO~\cite{Giannozzi:2009vu,Giannozzi:2017tv} v7.3.1 and cDFPT calculations within v7.4.1 using v1.2 norm-conserving pseudopotentials from the Schlipf-Gygi library~\cite{schlipf_optimization_2015} as available via http://www.quantum-simulation.org. For the cDFPT calculations we used the patch provided by Berges via the elphmod package~\cite{berges_elphmod_2025}. All structures are re-relaxed with respect to the atomic positions using the lattice vectors from our VASP relaxations (as described above in section~\ref{sec_meth_dft}), $6\times6\times6$ $k$ meshes, energy cut offs of $100\,$Ry and $400\,$Ry for the wave functions and charge density, respectively, and a force convergence threshold of $10^{-6}\,$Ry/Bohr. (c)DFPT calculations are subsequently performed on $2\times2\times2$ $q$ meshes. If not stated otherwise, we applied acoustic sum rules.

The raw data as well as all necessary parameter files are available via our \NbX{X} model parameter database~\cite{Nb3X8ModelDB}.

\subsection{Effective exchange interaction parameters}
\label{sec_jij}

Given the on-site Coulomb interactions are significantly larger than the hopping matrix elements in all \NbX{X} compounds (see Tab.~\ref{tab:parametersfull} and \ref{tab:ht_ab_papram}), we apply a strong-coupling expansion to obtain exchange interactions of the form
$$
J_{\perp\slash\parallel} = -2 (t_{\perp\slash\parallel})^2\slash U^{\text{eff}}_{\perp\slash\parallel},
$$
where labels $\perp$ and $\parallel$ denote sites $i$ and $j$ in the out-of-plane and in-plane directions, respectively.
Instead of using $U_0$ in the equation above, we employ $U^\text{eff}_{\perp} = U_0 - U_{\perp}^{(s)}$ and $U^\text{eff}_{\parallel} = U_0 - U_{\parallel}^{(1)}$ for the out-of-plane and in-plane values, respectively. This includes the renormalization of the energy cost due to non-local Coulomb interactions~\cite{grytsiuk2024nb3cl8} and can be understood as the renormalization of the energy cost associated with virtual doublon-holon excitations due to non-local Coulomb interactions ~\cite{Schuler13}. Such renormalization is essential for accurately capturing charge-neutral excitations; however, it is not suitable for spectral functions, which describe excitations involving changes in total charge ~\cite{grytsiuk2024nb3cl8, PhysRevB.110.155120}. The resulting exchange interactions for both \NbX{X} monolayers and \NbX{X} bulk are presented in Tab.~\ref{tab:jijparameters_ML} and \ref{tab:jijparameters}, respectively.

\subsection{Atomistic Spin Dynamics Simulations}
\label{sec_meth_spirit}

To investigate the ground-state magnetic properties of the surface \NbX{X} monolayers, as well as the temperature and magnetic field dependence of the magnetic susceptibility in bulk \NbX{Cl}, we utilize spin dynamics simulations conducted with the SPIRIT code~\cite{Spirit_2019}. 
All simulations are based on a classical Heisenberg Hamiltonian given by Eq.~\ref{eq:Heisenberg}.
Our simulations are grounded in the Landau-Lifshitz-Gilbert (LLG) equation of motion to assess ground-state properties, while we employ classical Monte Carlo (MC) calculations to compute the magnetic susceptibility.
Our test simulations indicate that a lattice of $50 \times 50 \times 1$ spins is sufficient for the surface monolayer simulations (solving the LLG equations), while a $20 \times 20 \times 4$ spins suffices for the bulk simulations (using the MC method). To minimize boundary effects, for the surface monolayer case, we consider only in-plane periodic boundary conditions, whereas for the bulk case, periodic boundaries are applied in all dimensions.

In the Monte Carlo simulations, we employed $500$ thermalization steps, followed by $1,000,000$ samples in the Metropolis algorithm, and conducted five decorrelation steps. The isothermal susceptibility is evaluated according to
$$
\chi = \frac{1}{k_BT}(\langle {\bf M}^2\rangle - \langle {\bf M}\rangle ^2),
$$
where $ {\bf M} =\frac{1}{N} \sum_i^N {\bf S}_i$ represents the average magnetization of the sample.

\subsection{ARPES Measurements \label{meth:ARPES}}

  \NbX{Br} and \NbX{I} crystals were synthesized using chemical vapor transport~\cite{wu2022field}. To avoid air exposure, crystals were mounted in a glove box and transferred to the beamline in a vacuum suitcase. Samples were cleaved \textit{in situ} prior to ARPES measurements.
  The ARPES measurements were performed at the SGM3 beamline of the ASTRID2 synchrotron radiation facility at Aarhus University in Denmark \cite{SGM3_2004}. Measurements were carried out at 150~K with a base pressure of $1\times10^{-10}$ mbar. The ARPES data were acquired at photon energies (h$\nu$) ranging from 30.5 to 130.5~eV, with energy and angular resolutions better than 50~meV and 0.2$^\circ$, respectively. The beam spot size was 190~$\mu$m$\times$90~$\mu$m. The synchrotron radiation polarization and the sample-to-analyzer direction were both in the plane of incidence, and the analyzer slit was perpendicular to the plane of incidence. 
  The EDC fitting used a linear background and Voigt profiles, employing the minimum number of fitting components required for accurate fits.

\onecolumngrid
\newpage
\section{Appendix}
\appendix

\subsection{Vacuum Extrapolation of the Mono- and Bilayer cRPA Coulomb Matrix Elements \label{app:WFCEExtra}}

For the mono- and bilayer cRPA calculations we separate the mono-/bilayers by an effective vacuum of three times their individual monolayer heights. Since the true free-standing limit would only be achieved by further increasing this effective “vacuum height”, these cRPA calculation suffer from spurious screening from the periodic replicas of the systems in c direction. To extrapolate the Coulomb matrix elements to the true free-standing limit, we make use of our Wannier Function Continuum Electrostatic (WFCE) approach~\cite{WFCE} as applied for monolayer \NbX{Cl} in Ref.~\cite{grytsiuk2024nb3cl8}.

\begin{figure*}[h!]
    \centering
    \includegraphics[width=0.49\linewidth]{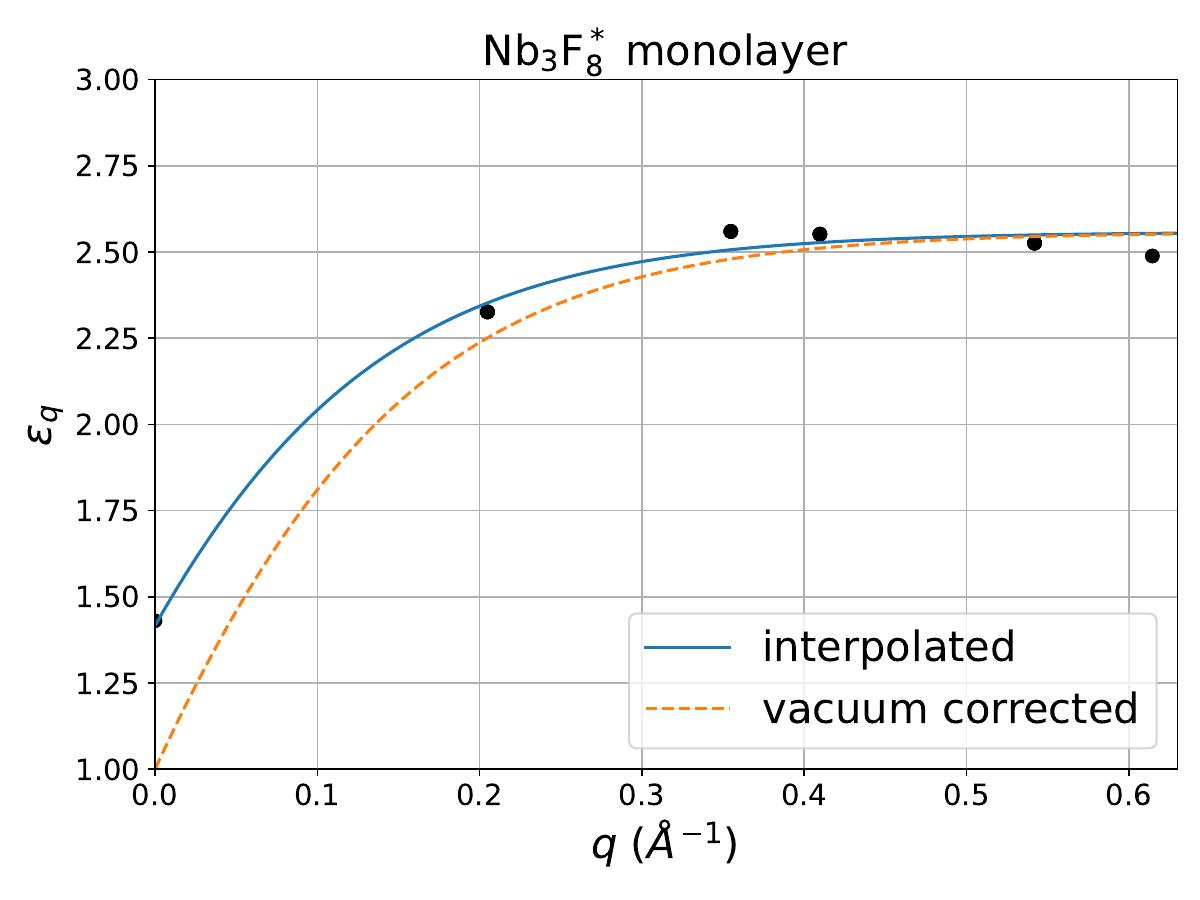}
    \includegraphics[width=0.49\linewidth]{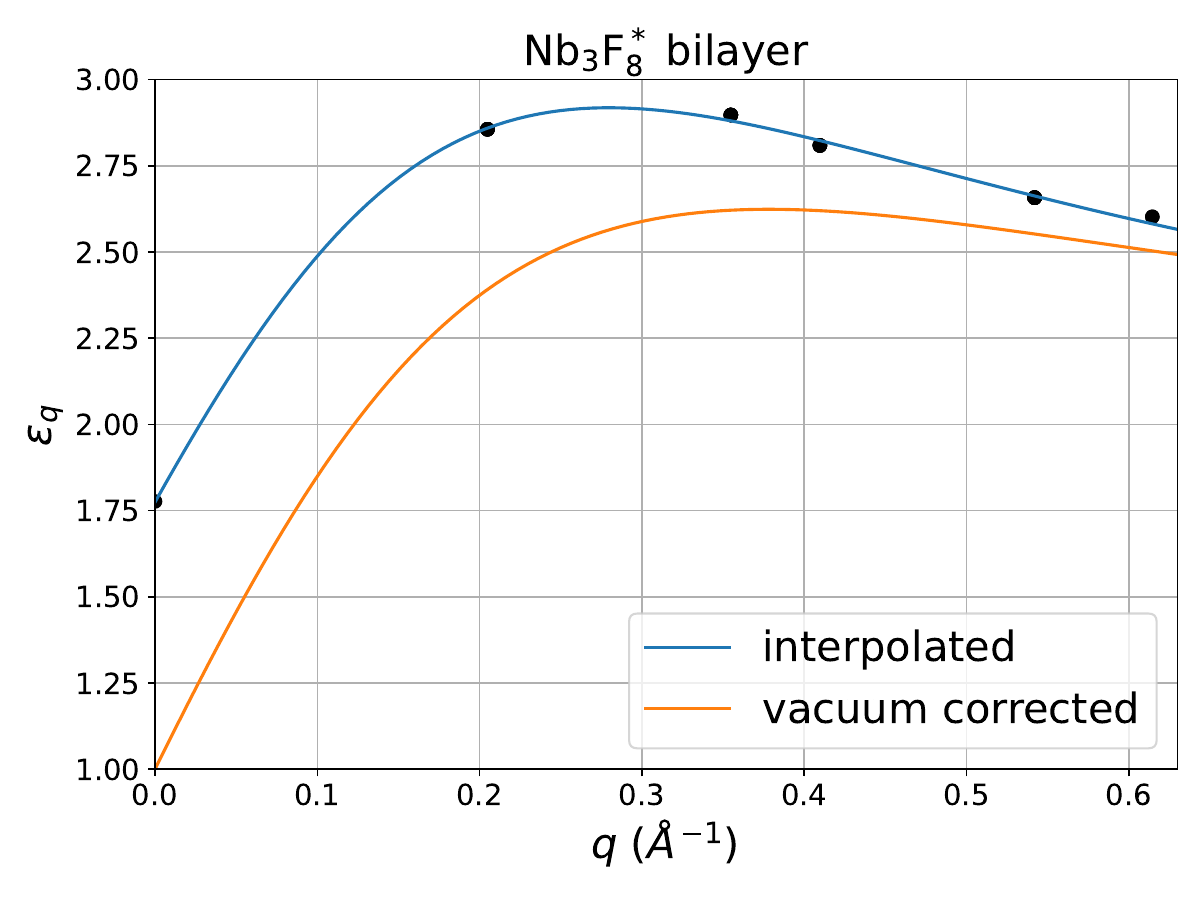}
    \caption{Leading cRPA dielectric functions of monolayer (left) and bilayer (right) \NbX{F}*. Markers indicate initial ab initio calculation results. Blue lines are analytic fits to the ab initio data. Orange lines represent the ``vacuum corrected'' behaviour. For the monolayer $\varepsilon_1^{(0)}$ was assumed to be a constant, while we used a Resta-like $q$-dependent model for the bilayer. \label{fig_new:WFCE}}
\end{figure*}

To this end, we start with the non-local bare Coulomb interaction of the mono-/bilayer as obtained from our cRPA calculations in momentum space. Within a matrix representation $v_{\alpha \beta}(q)$ using a product basis $\alpha, \beta = \{n,m\}$ we can diagonalize the Coulomb tensor $v(q) = \sum_\nu v_\nu(q) \ket{v_\nu(q)} \bra{v_\nu(q)}$ with $v_\nu(q)$ and $\ket{v_\nu(q)}$ being the corresponding eigenvalues and eigenvectors of the Coulomb tensors and $q = |\mathbf{q}|$. Assuming that the eigenbasis does not drastically change upon the effects of the cRPA screening, we can thus represent the full cRPA Coulomb tensor as $U(q) = \sum_\nu \frac{v_\nu(q)}{\varepsilon_\nu(q)} \ket{v_\nu(q)} \bra{v_\nu(q)},$
where $\varepsilon_\nu(q)$ are the corresponding pseudo-eigenvalues of the dielectric tensor describing the different screening channels. 
The leading eigenvalue $v_1(q)$ renders Coulomb penalties for mono-pole-like perturbations (all orbitally resolved electronic densities are in phase). The mono-pole-like screening as rendered by $\varepsilon_1(q)$ can be modeled by solving the Poisson equation for a dielectric slab of height $h$ embedded in some different dielectric environment yielding
\begin{align}
    \varepsilon_1(q) &= \frac{\varepsilon_1^{(0)} \left[ 1 - \tilde{\varepsilon}_0^{(1)} \tilde{\varepsilon}_0^{(2)} e^{-2q h} \right]}{1 + \left[ \tilde{\varepsilon}_0^{(1)} + \tilde{\varepsilon}_0^{(2)} \right] e^{-q h} + \tilde{\varepsilon}_0^{(1)} \tilde{\varepsilon}_0^{(2)} e^{-2q h}} 
    &\text{with}
    &&\tilde{\varepsilon}_0^{(1)} = \frac{\varepsilon_1^{(0)} - \varepsilon_\text{sub}^\text{below}}{\varepsilon_1^{(0)} + \varepsilon_\text{sub}^\text{below}}, \qquad
    \tilde{\varepsilon}_0^{(2)} = \frac{\varepsilon_1^{(0)} - \varepsilon_\text{sub}^\text{above}}{\varepsilon_1^{(0)} + \varepsilon_\text{sub}^\text{above}}.
\end{align}
Example results for $\varepsilon_1(q)$ of the inital cRPA calculations for mono- and bilayer \NbX{F}* are shown in Fig.~\ref{fig_new:WFCE}. For both structures we see that $\varepsilon_{q=0} > 1$ which is indicative for the spurious inter-layer screening due to the finite effective vacuum height. However, we also find that we can fit the raw initial cRPA data with the analytic $\varepsilon_1(q)$ function from above well, finding $\varepsilon_\text{sub}^\text{above} = \varepsilon_\text{sub}^\text{below} > 1$ (blue curves in Fig.~\ref{fig_new:WFCE}). By setting $\varepsilon_\text{sub}^\text{above} = \varepsilon_\text{sub}^\text{below} = 1$ we enforce $\varepsilon_{q=0} = 1$ (orange curves in Fig.~\ref{fig_new:WFCE}). This way we correct for the artificial spurious inter-layer screening. Using this corrected $\varepsilon_q$ and the bare $v_q$ we can reconstruct the vacuum corrected $U_q$ from which we calculate in a final step via Fourier transforms the real space $U_R$, which are listed in Tab.~\ref{tab:parametersfull}.

\newpage
\subsection{Exact Diagonalization of the Dimer Model and Relevance of Hedin Vertices \label{app:ED}}
\begin{figure*}[h!]
    \centering
    \includegraphics[width=0.99\textwidth]{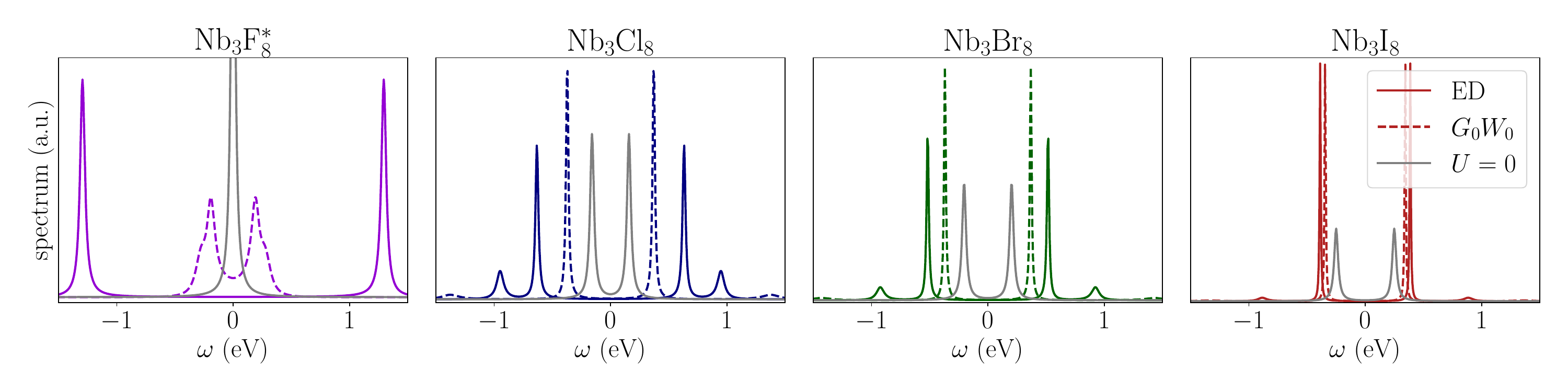}
    \caption{Dimer model results using parameters from cRPA. We compare bare bonding/anti-bonding splitting (grey solid line) with dimer $G_0W_0$ results (dashed lines) and exact diagonalization results (colored solid lines).}
    \label{fig:ED}
\end{figure*}

In order to establish the notion of strong and weak correlations we perform calculations using strong (Hubbard-I) and weak coupling (Hedin's GW) approximations. Comparing the results enables us to classify the compounds in the \NbX{X} family with respect to the degree of correlation effects.

We start this analysis by investigating the local (impurity) Green's functions in more detail. In the Hubbard-I approximation these are equivalent to the exact diagonalization (ED) results of a generalized Hubbard dimer using the full $U_{ijkl}$ tensor and an effective dimerization $\tilde{t}_\perp$, where $i,j,k,l$ belong to one of the two sites representing the two molecular orbitals of the strongly hybridized/dimerized planes. 

For the (simplified) Hubbard dimer with only local Coulomb interactions (i.e.\ only $U_{iiii}=U$ terms taken into account), the exact interacting Green's functions and the $G_0W_0$ approximation can both be derived analytically~\cite{tomczak_PhD,Zhang_2023_PRB}. The latter neglects all (Hedin) vertex contributions. The corresponding self-energies read
\begin{align}
    \Sigma_{ED}(\nu) &= \frac{U^2}{4} \frac{1}{\nu \mp 3 \tilde{t}_\perp}\\
    \Sigma_{G_0W_0}(\omega) &=  \frac{U^2 \tilde{t}_\perp}{c} \frac{1}{\nu \pm \tilde{t}_\perp \pm c}
\end{align}
with $c=\sqrt{4\tilde{t}_\perp^2+4U\tilde{t}_\perp}$. To connect this simple model to our ED calculations with the full $U_{ijkl}$, we fit the self-energy of the latter using the effective dimerization $\tilde{t}_\perp$ as the fit parameter. We find $\tilde{t}_\perp \approx 0.00, 0.15, 0.19$ and $0.25\,$eV for F, Cl, Br, and I, respectively. Setting $U$ to the cRPA downfolded values $U_{iiii}$, the analytic self-energies of the simplified Hubbard dimer fit the full ED ones very well.
By further comparing the Hubbard-I spectral functions obtained with and without the non-local Coulomb interactions between the two molecular orbitals of the dimer (see Appendix~\ref{app:localU}), we find the that non-local Coulomb interactions do not affect the single particle Green's functions at half-filling. 

We therefore compare in Fig.~\ref{fig:ED} the spectral functions of $G_{ED}(\omega)$ and $G_{G_0W_0}(\omega)$ obtained for the simplified Hubbard dimer allowing us to study how relevant Hedin vertex corrections are for each system. 
In the case of \NbX{Cl} we see that the ED and $G_0W_0$ show significant discrepancies. In detail, the ED result shows a gapped local spectral function with two main peaks forming the gap, which are accompanied by two side-peaks with reduced spectral weight. We can understand these four features as lower and upper Hubbard bands from the bonding and anti-bonding orbitals, whereby the gap is formed between the lower Hubbard band of the bonding orbital and the upper Hubbard band of the anti-bonding orbital. In the $G_0W_0$ approximation, we find a similar pole structure. However, the inner poles are at much smaller energies, while the outer poles are at much higher energies with suppressed spectral weight. Also, the nature of the $G_0W_0$ approximation does not allow us to identify the different poles as resulting from Hubbard bands. Instead, in the $G_0W_0$ approximation a more appropriate interpretation would be as shake-off bands resulting from charge fluctuations. Altogether, this clearly stresses the relevance of Hedin vertices for a qualitatively and quantitatively correct description of \NbX{Cl}. This allows us to identify \NbX{Cl} as a strongly correlated insulator. The same holds for \NbX{Br}, albeit with smaller deviations between ED and $G_0W_0$ results.
\NbX{F}* in turn starts from a metallic regime and is neither quantitatively nor qualitatively well described within the $G_0W_0$ approximation. The ED gap of \NbX{F}* can thus be best understood as a conventional Mott gap opened by a self-energy of the form $\frac{U^2}{4\nu}$. Finally, for \NbX{I} we see that the two results are qualitatively and quantitatively very similar such that we can understand \NbX{I} as a weakly correlated insulator, which is well approximated by the $G_0W_0$ scheme.

\newpage
\subsection{Single-Site vs. Cluster Dynamical Mean Field Theory}\label{app:CDMFT}
\begin{figure*}[h!]
    \centering
    \includegraphics[width=0.60\linewidth]{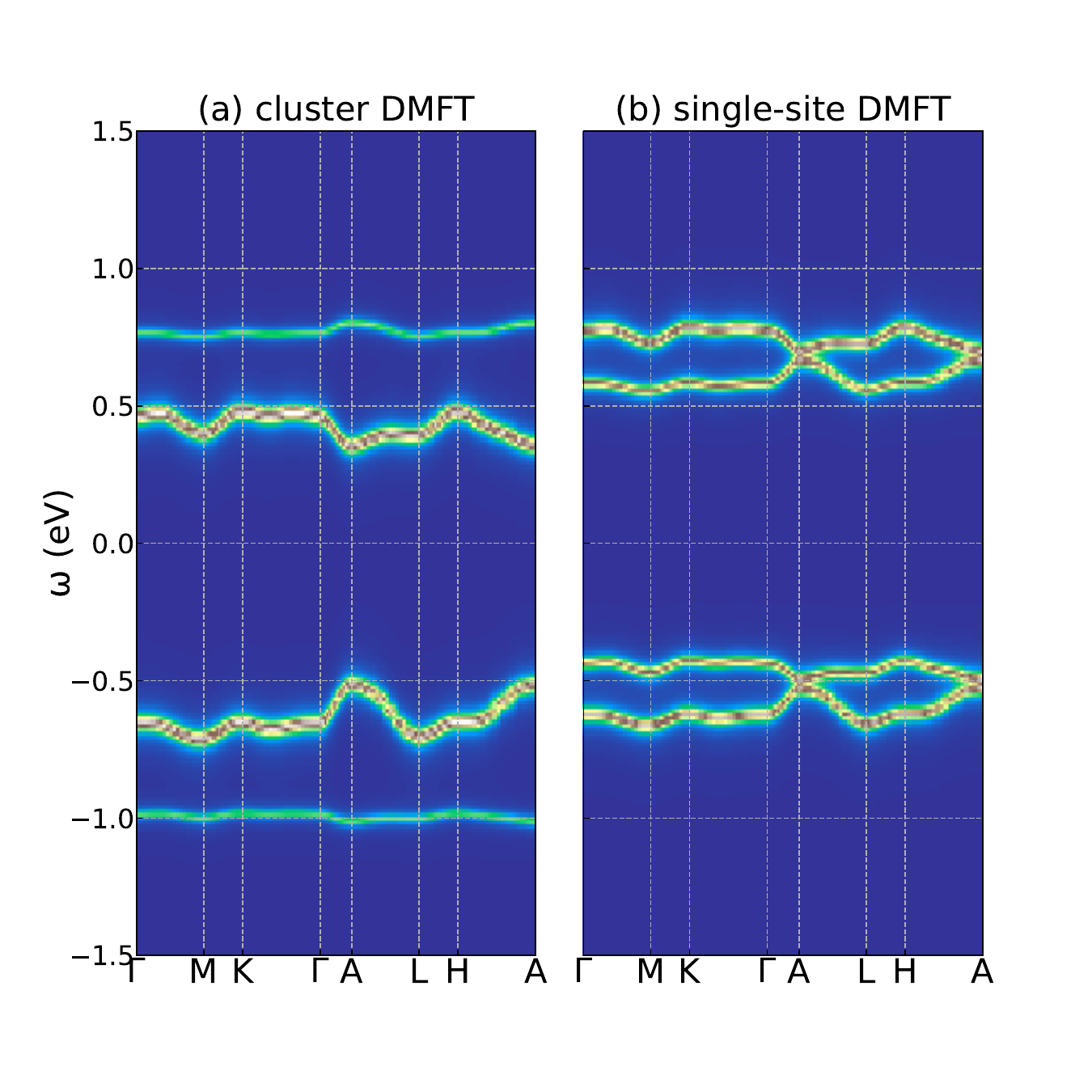}
    \caption{Bulk low-temperature \NbX{Br} in a cluster and single-site Hubbard-I calculation.}
    \label{fig:CDMFT}
\end{figure*}

In Fig.~\ref{fig:CDMFT} we show the differences between a cluster and a single-site DMFT calculation, which respectively include the full matrix structure of the self-energy or just its diagonal components in the molecular orbital basis, for the low-temperature bulk structure of \NbX{Br}. From this it is clear that to correctly describe the bonding and anti-boding Hubbard bands, cluster DMFT in a full two-molecular-orbital basis is necessary.

\newpage
\subsection{Hubbard-I Spectral Functions with only on-site Coulomb Interaction }
\label{app:localU}
\begin{figure*}[h!]
    \centering
    \includegraphics[width=0.99\linewidth]{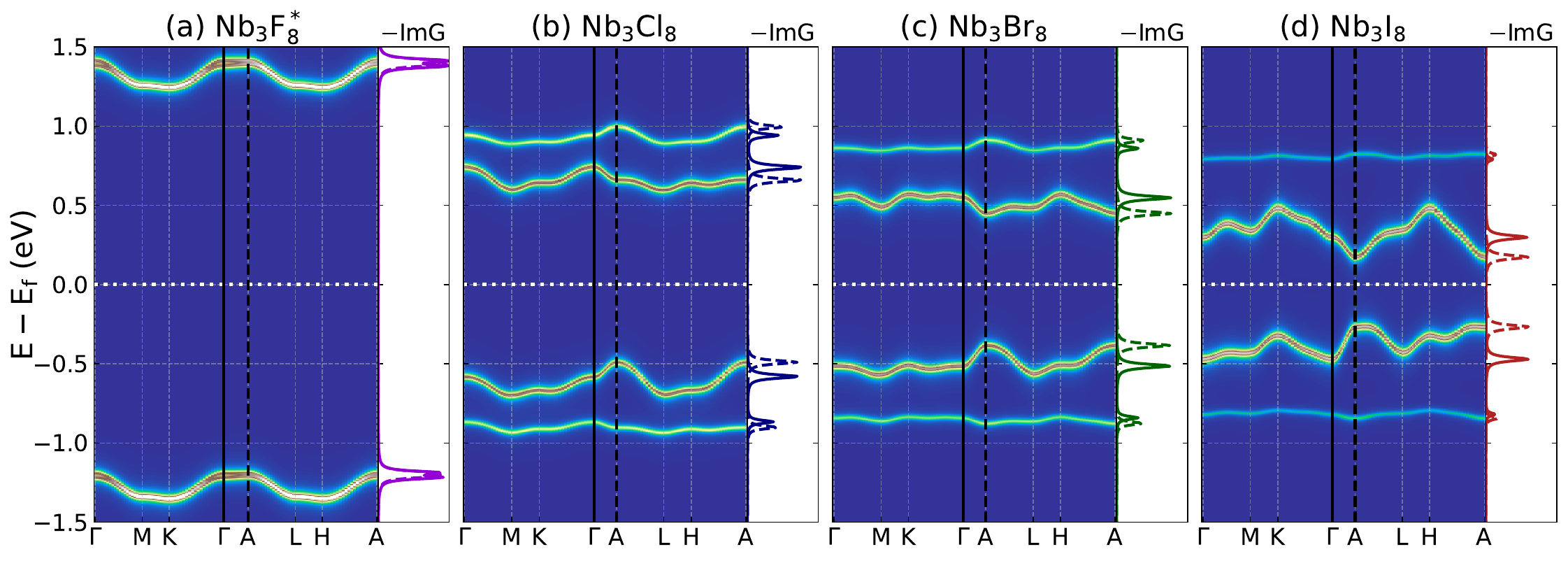}
    \includegraphics[width=0.99\linewidth]{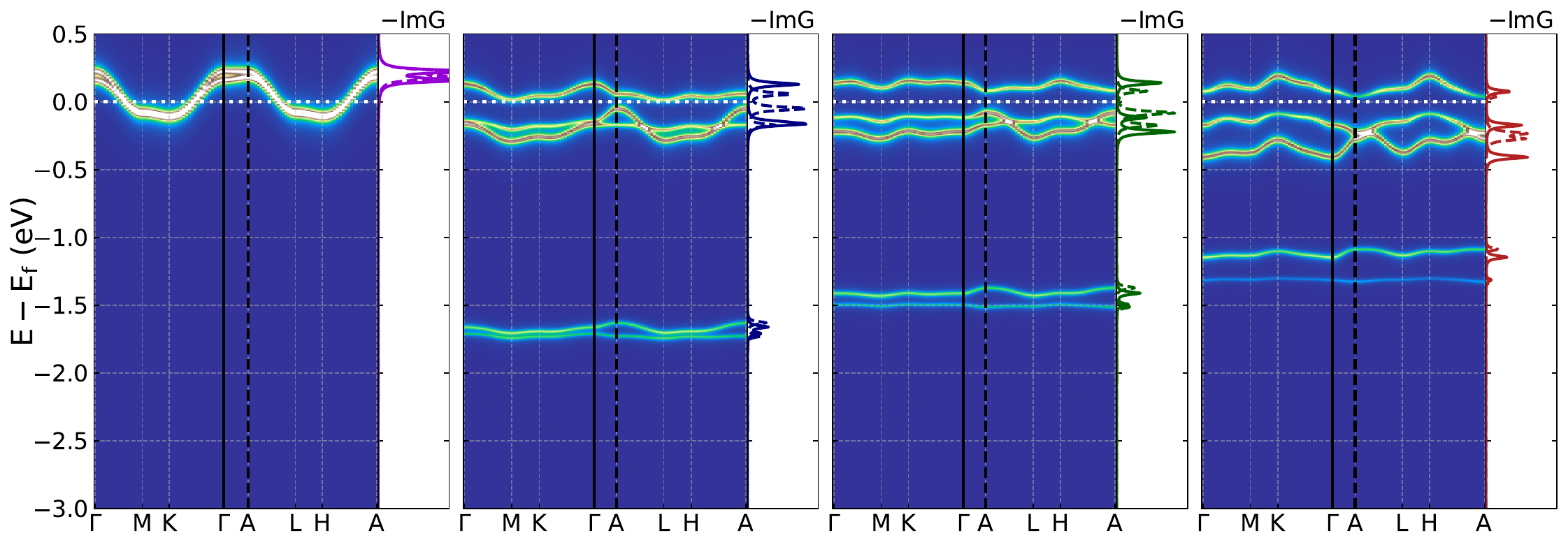}
    \caption{(a-d) Cluster DMFT spectral functions within Hubbard-I with only on-site coulomb interactions $U_0$ (but inter-cluster hybridization). For $N_e=2$ (top row) and $N_e=3$ (bottom row). Line cuts at $\Gamma$ \& $A$ are shown in full and dashed lines respectively.}
    \label{fig:localU}
\end{figure*}

In Fig.~\ref{fig:localU} we show the Hubbard-I spectral functions for $N_e=2$ and $N_e=3$ obtained using only local Coulomb interactions ($U_{iiii}$). Comparing these to Fig.~\ref{fig:Hubbard-I} and Fig.~\ref{fig:Hubbard-I-N3}, respectively, allows us to investigate the role of the interlayer dimer density-density Coulomb matrix elements (between the molecular orbitals $\psi_1$ and $\psi_2$) as all other elements of the full $U_{ijkl}$ tensor were already vanishingly small. 

For the undoped case ($N_e=2$), we find that the spectral functions do not change and the fundamental gaps are unaffected. This is different in the doped case ($N_e=3$). Here, removing the interlayer dimer density-density Coulomb interaction changes the spectral functions quantitatively in the case of Cl, Br, and I, and even qualitatively in the case of F. In the former cases, all gaps between sets of bands are modified when we neglect non-local Coulomb interactions. However, the qualitative properties are the same. \NbX{Cl} and \NbX{Br} are still charge-transfer insulators and \NbX{I} is still a (nearly) conventional Mott insulator. In the case of \NbX{F}* we however find a metallic solution as soon as we neglect the non-local Coulomb interaction at $N_e=3$. The \NbX{F}* result at three quarter filling with only on-site Coulomb interaction should be taken with care. As there is no actual dimerization in this system on might interpret it as two separate Mott insulators at non-integer filling and thus possibly away from the regime in which the Hubbard-I approximation is applicable.

\newpage
\subsection{Full Phonon Dispersions \label{sec:fullphonons}}

\begin{figure*}[h!]
    \centering
    \includegraphics[width=0.99\linewidth]{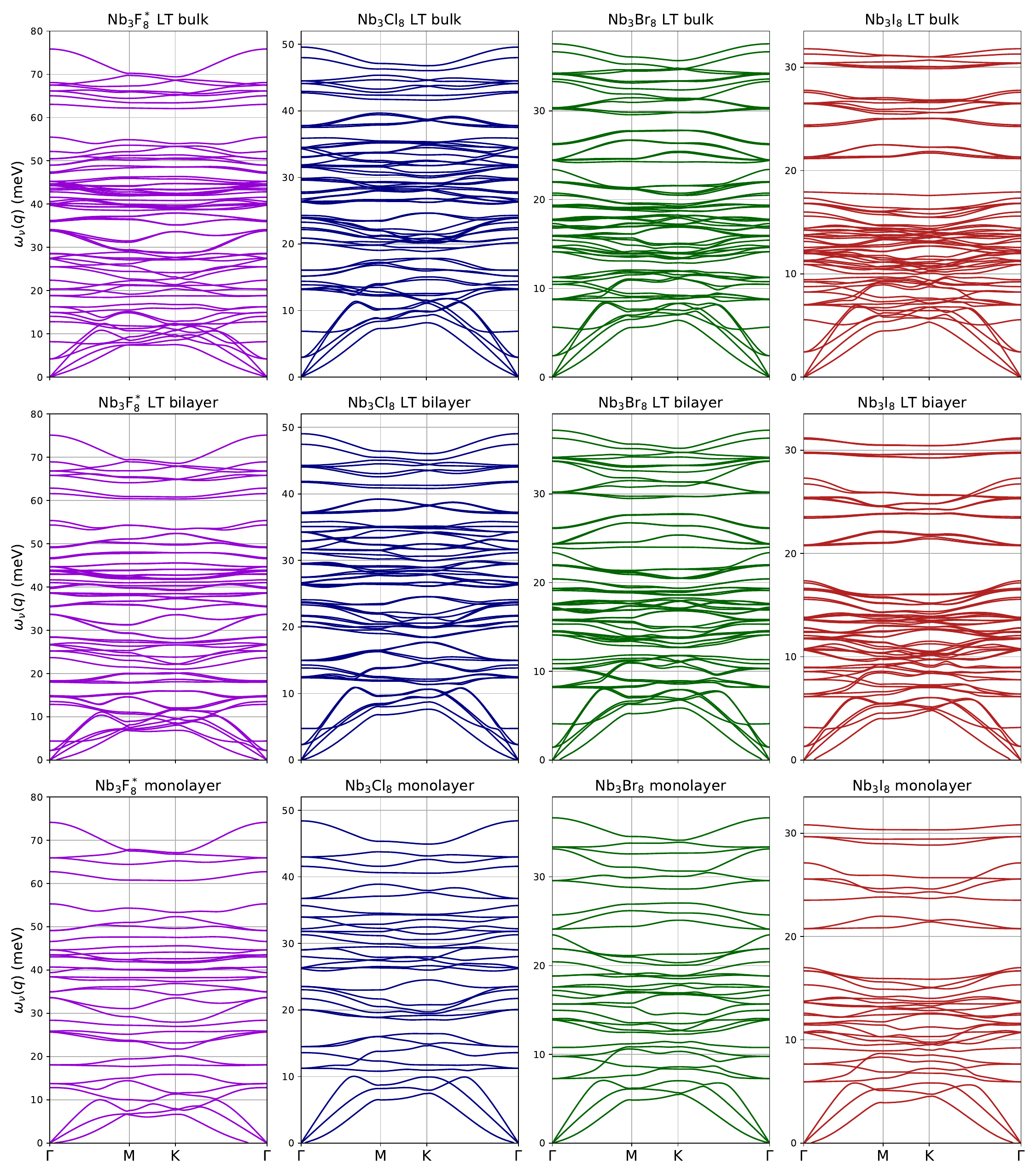}
    \caption{DFPT phonon dispersions for \NbX{X} in their LT bulk, bilayer, and monolayer structures.}
    \label{fig_new:DFPTfull}
\end{figure*}

In Fig.~\ref{fig_new:DFPTfull} we show the full phonon dispersion for all four compounds in their LT bulk, bilayer, and monolayer structures as calculated within DFPT.

\newpage
\subsection{Constrained Density Functional Perturbation Theory Calculations for \NbX{F}* and \NbX{Cl} \label{sec:cdfpt}}

\begin{figure*}[h!]
    \centering
    \includegraphics[width=0.49\linewidth]{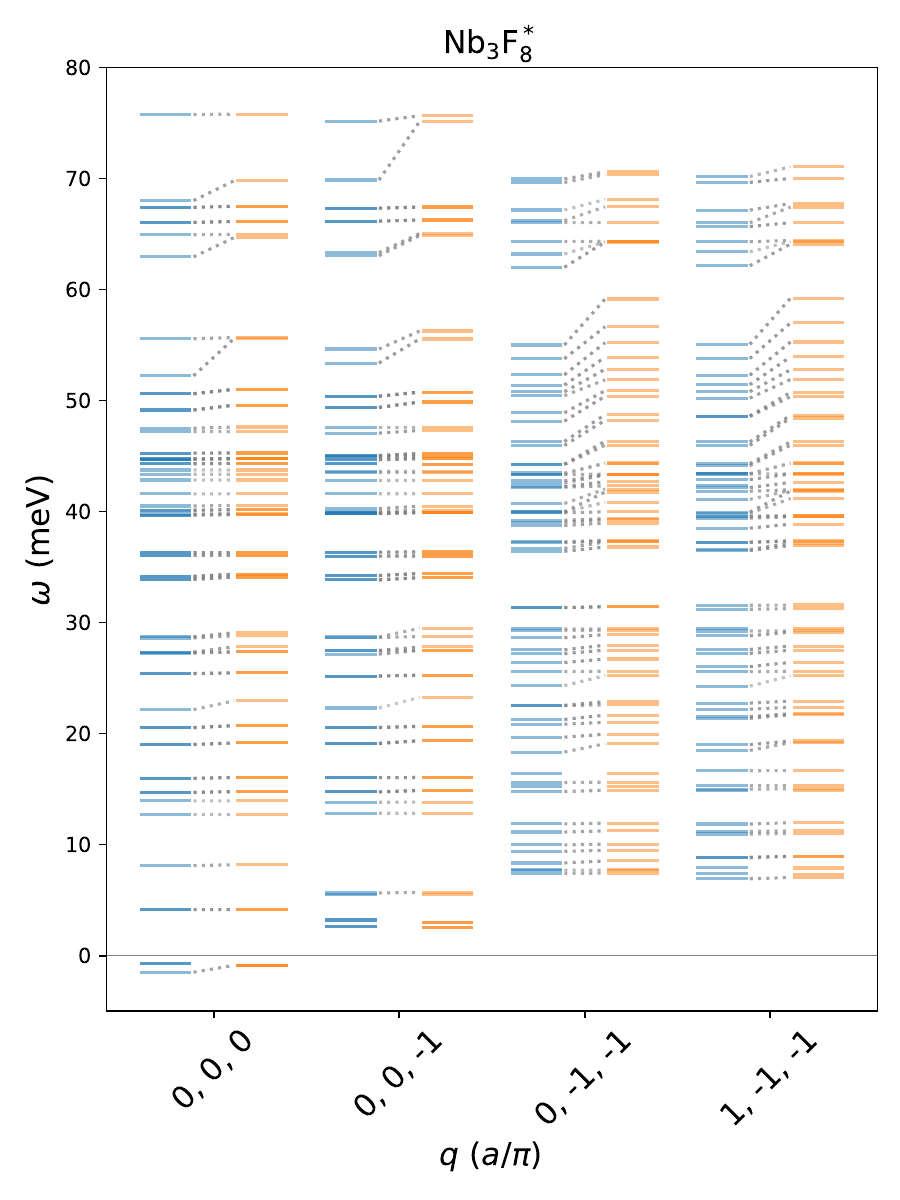}
    \includegraphics[width=0.49\linewidth]{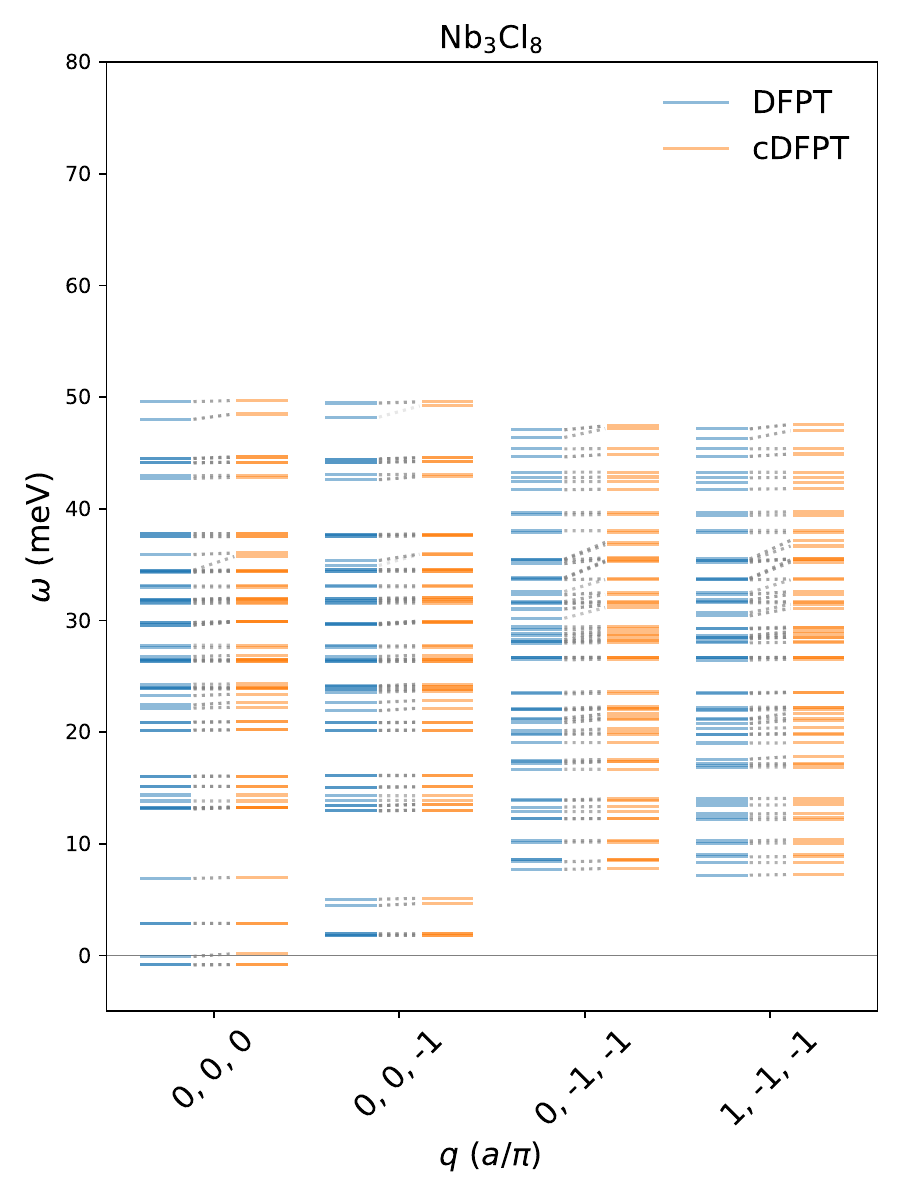}
    \caption{Comparison between DFPT (blue/left) and cDFPT (orange/right) phonon energies of LT bulk \NbX{F}* and \NbX{Cl} at selected $q$ points. Dashed lines serve as a guide to the eye and connect DFPT with corresponding cDFPT phonon energies. No acoustic sum rules are applied.}
    \label{fig_new:cDFPT}
\end{figure*}

Within the cDFPT calculations we removed the two lowest bands around the Fermi level from the Sternheimer equation. For detailed comparison between the DFPT and cDFPT results, we show in Fig.~\ref{fig_new:cDFPT} the phonon energies of LT bulk \NbX{F}* and \NbX{Cl} at the original (coarse) $q$ points without including any LO-TO splitting or applying any sum rules. The latter results in slightly imaginary frequencies at $\Gamma$ (indicated as negative values). This comparison shows that the DFPT and cDFPT results are very similar with only a few (mostly higher energy) modes being affected. As expected, those modes which change in cDFPT are shifted to higher energies, since the energy cost of atomic displacements is higher when (some) electronic states cannot move along~\cite{cdfpt_molecules}. Overall, cDFPT effects are small in \NbX{Cl} and only mildly larger in \NbX{F}*. The electron-phonon coupling to the slightly gaped low-energetic flat bands of \NbX{Cl} has thus a smaller effects to the phonons than the electron-phonon coupling to the metallic flat bands in \NbX{F}*.

\newpage
\subsection{Doping Dependence of DMFT Matsubara Self-Energies}\label{app:Sigma_matsubara}
\begin{figure}[h!]
    \centering
    \includegraphics[width=0.7\linewidth]{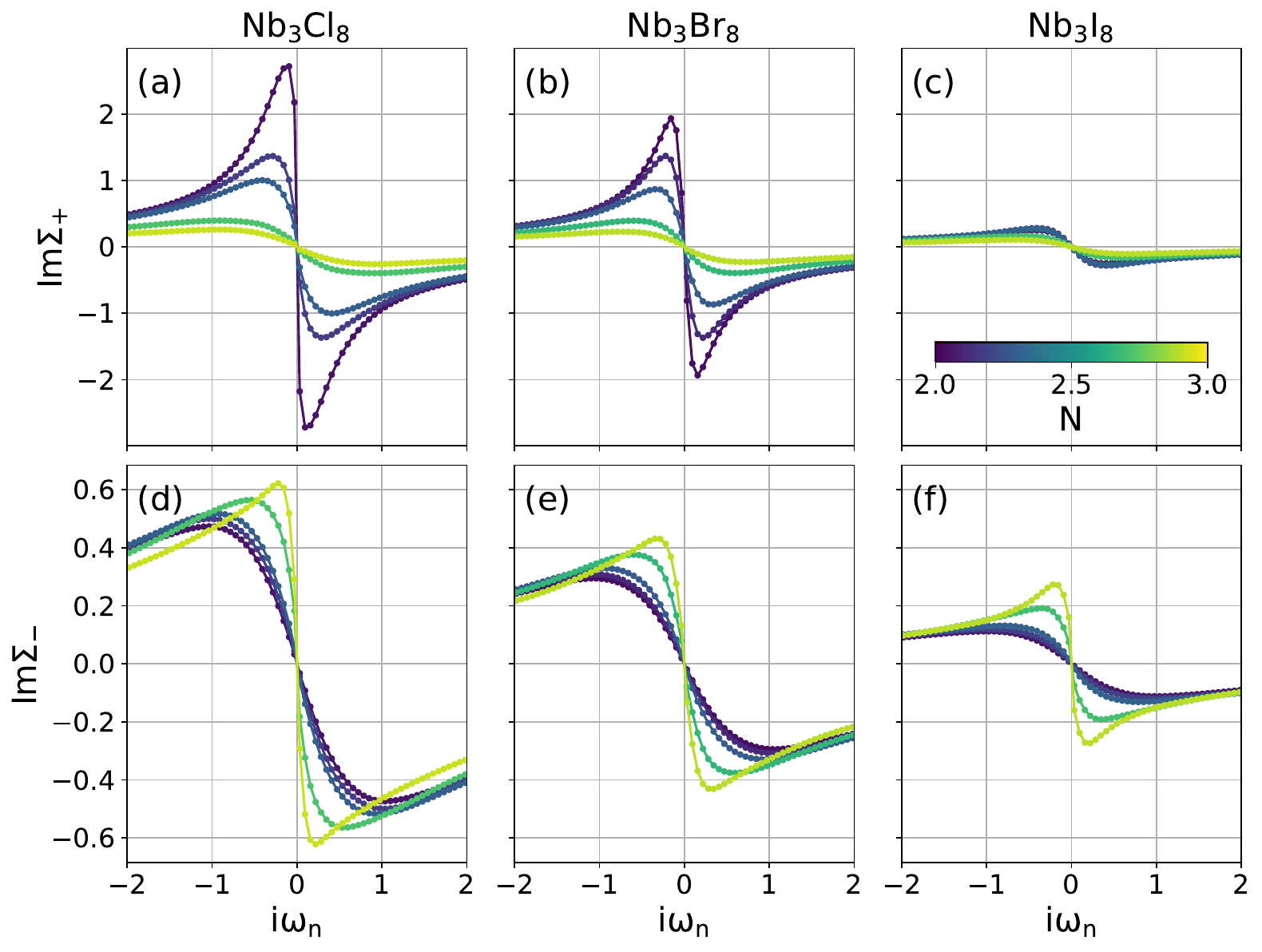}
    \caption{Doping-dependence of the imaginary Matsubara self-energy, for the bonding (a-c) and anti-bonding component (d-f).}
    \label{fig:ImSigma}
\end{figure}

Fig.~\ref{fig:doping} shows the doping-dependence of the approximate quasiparticle weights $Z$. $Z$ is calculated using Eq.~(\ref{eq:Z}) from the imaginary part of the Matsubara self-energy shown in Fig.~\ref{fig:ImSigma}. This self-energy is calculated in cluster DMFT using \texttt{w2dynamics}.

At small doping the bonding components of the \NbX{Cl} and \NbX{Br} self-energies have a significant frequency dependence around zero frequency, which results in the nearly vanishing quasiparticle weight shown in Fig.~\ref{fig:doping}~(a). In contrast, for \NbX{I} these retardation effects are strongly suppressed, resulting in a significant quasiparticle weight for \NbX{I} at small doping. This is in line with our identifications as strongly (\NbX{Cl} and \NbX{Br}) and weakly (\NbX{I}) correlated insulators at half-filling. 

As the doping level increases towards $N_e=3$ we see that the frequency response of the bonding component of the self-energies is suppressed in all compounds. As a consequence, the bonding channel becomes more and more coherent with doping.  In the anti-bonding component we observe a qualitatively similar behavior, i.e. \NbX{Cl} shows the strongest retardation effects and \NbX{I} the weakest. With increasing doping these retardation effects increase for all compounds. Thus, under (significant) electron doping even \NbX{I} shows correlation effects in the anti-bonding channel. In fact, close to $N_e=3$ all systems seem to behave as hole-doped Mott insulators, which is reflected in the significantly suppressed quasiparticle weights shown in Fig.~\ref{fig:doping}a.

\newpage
\subsection{Graphical Solution of the Dimer Models \label{app:graphicalsol}}
\begin{figure*}[h!]
    \centering
    \includegraphics[width=0.45\textwidth]{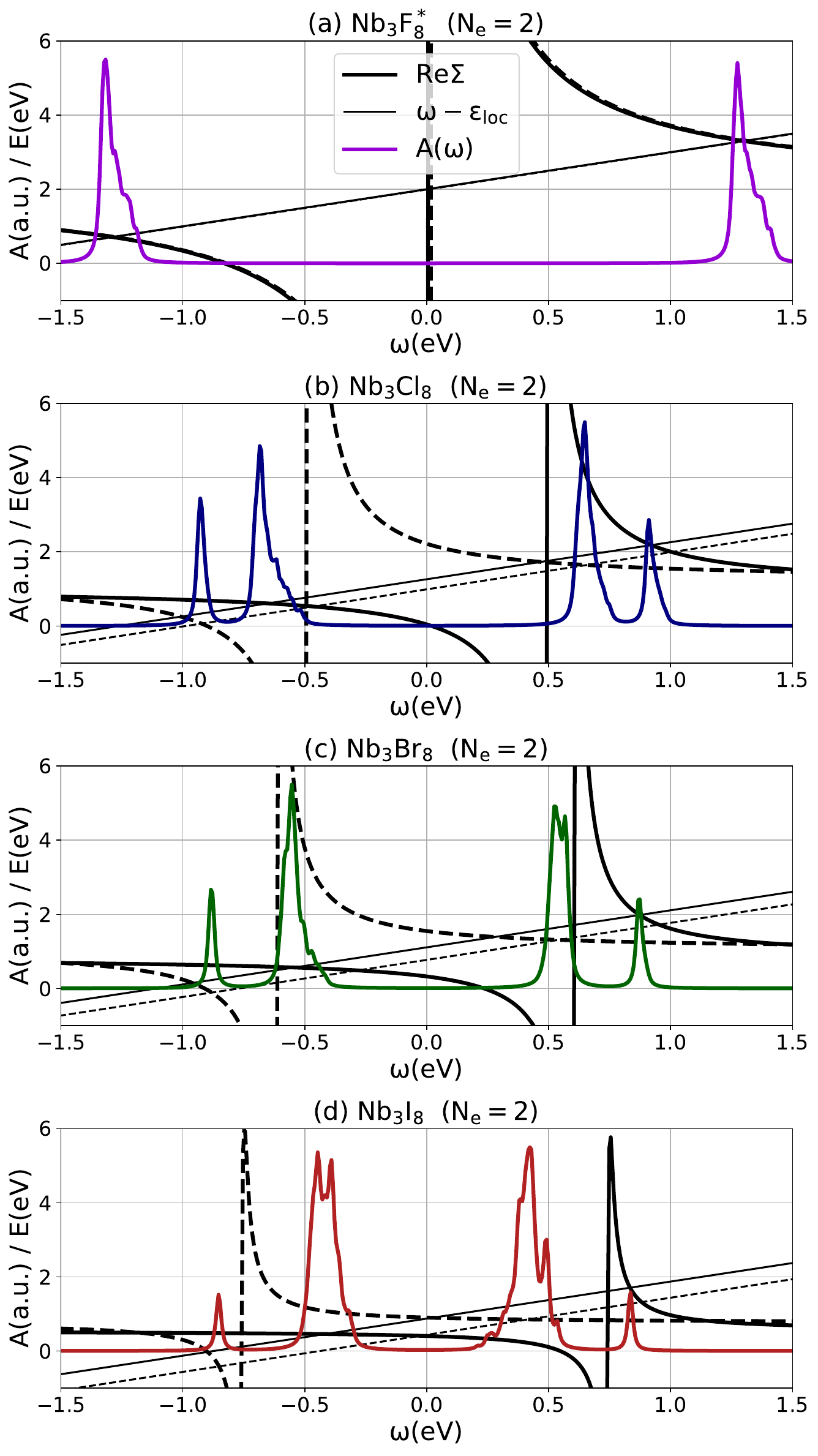}
    \includegraphics[width=0.45\textwidth]{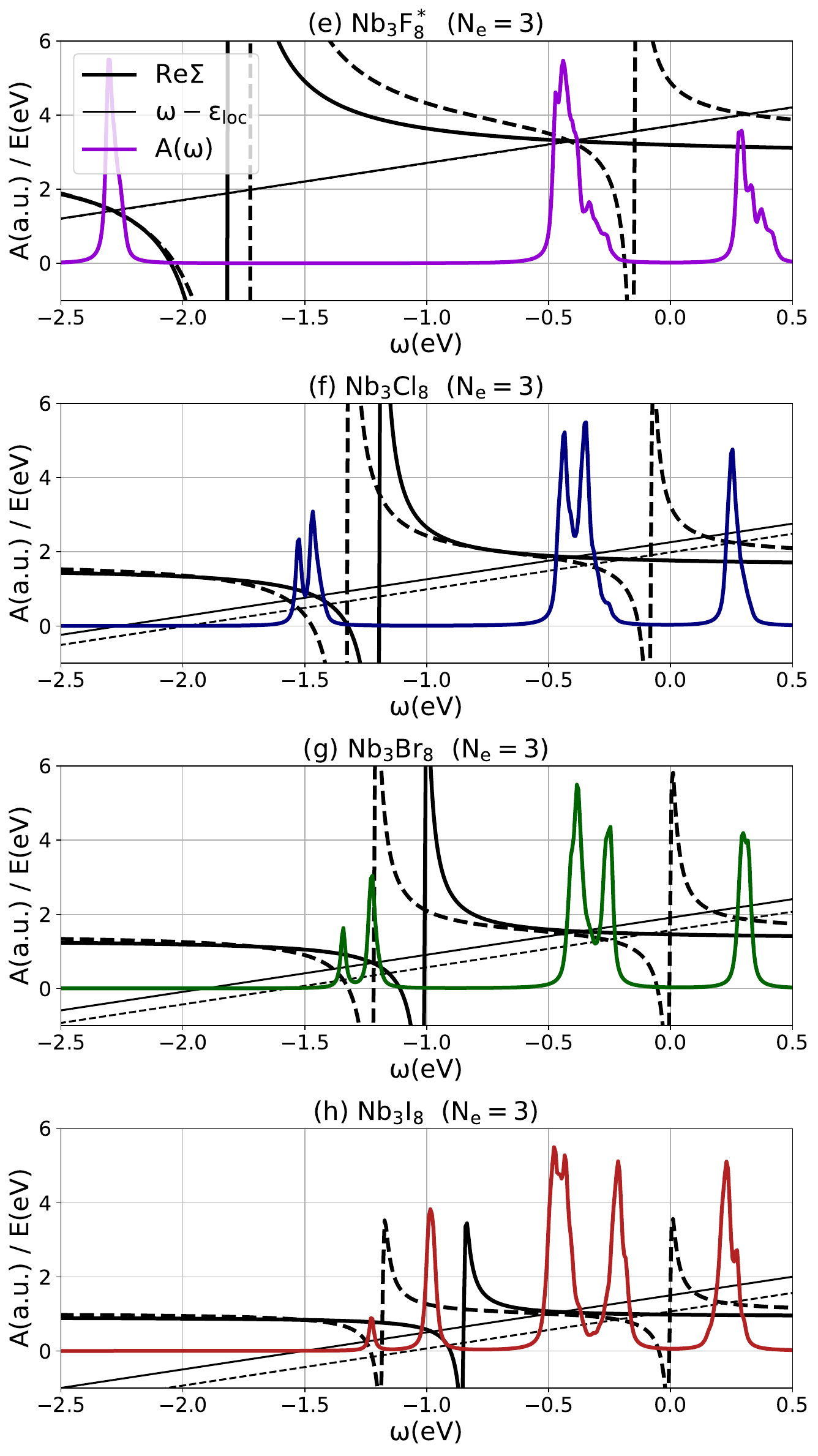}
    \caption{Graphical representation of the local Dyson equation for the Hubbard-I calculations. Solid lines represent bonding component and dashed represent anti-bonding component. Plots in the left column (a-d) show the graphical solutions at half-filling  (cf. Fig.~\ref{fig:Hubbard-I}). Plots in the right column (e-h) show graphical solutions at three-quarter filling: $N_e=3$ (cf. Fig.~\ref{fig:Hubbard-I-N3}).}
    \label{fig:graphicalN23}
\end{figure*}

To further clarify and visualize the qualitative differences between the strongly and weakly correlated states (with and without doping), we present in Fig.~\ref{fig:graphicalN23} the graphical solutions of the Dyson equations governing the solutions (local spectral functions) of the dimer models for all four compounds and for $N_e=2$ and $N_e=3$, respectively. In detail, we show for all cases $\mathrm{Re}\Sigma(\omega)$ and $\omega - \epsilon_{\text{loc}}$ for both, bonding and anti-bonding channels, with $\epsilon_{\text{loc}} = \sum_{BZ}\epsilon_k$. Therefore, the solutions to the equation
\begin{equation}
    \omega - \epsilon_{\text{loc}} = \mathrm{Re}[\Sigma(\omega)]
    \label{eq:qp-roots}
\end{equation}
define excitation energies (poles in the spectral functions).

For $N_e=2$ we see in the left column of Fig.~\ref{fig:graphicalN23} three qualitatively different solutions. For \NbX{F}* we find (nearly) degenerate bonding and anti-bonding self-energies, both of which have a strong pole around $\omega=0$. As a consequence of this (effectively) single-pole structure, we find two solutions to Eq.~(\ref{eq:qp-roots}) (around $-1.2$ and $+1.2\,$eV) yielding two features in the spectral function. This is the classic Mott-gap scenario resulting in a lower and upper Hubbard band with a gap defined by the local Coulomb interaction strength $U$. In contrast, for the other compounds, we see that the bonding and anti-bonding self-energies still exhibit single poles, which are, however, shifted with respect to each other. As as result we find in all cases four distinct solutions, with two main spectral features around the gap and two accompanying side features. Upon close inspection we see that the upmost occupied state is given by a solution resulting from the bonding channel, and the lowest unoccupied one by a solution resulting from the anti-bonding channel. The fundamental gap is thus always opened between the bonding and anti-boding channel. However, due to the large $U$ in the case of \NbX{Cl} and \NbX{Br} the corresponding $\mathrm{Re}\Sigma(\omega)$ at the solutions $\omega=\varepsilon^{(\pm)}$ still have a finite slope. The fundamental gaps are thus significantly affected by the Coulomb interaction and especially the retardation effects imprinted into $\mathrm{Re}\Sigma(\omega)$. At a first glance, the situation seems to be similar for \NbX{I}, but upon close inspection, we see that the gap is formed between $\omega=\varepsilon^{(\pm)}$ where the self-energies are nearly constant and very similar to their static values $\mathrm{Re}\Sigma(\omega=0) \approx \frac{U^2}{\mp 12 \tilde{t}_\perp}$ (for the bonding and anti-bonding orbital, respectively). For the reduced $U$ and enhanced $\tilde{t}_\perp$ of \NbX{I}, this is dominated by $\tilde{t}_\perp$. The gap of \NbX{I} is thus mostly controlled by static mean-field correlations.

In the doped case with $N_e=3$ (right column of Fig.~\ref{fig:graphicalN23}) the same analysis shows that the fundamental gap between highest occupied and lowest unoccupied states is always opened due to significant retardation effects, but solely in the anti-bonding channel and in (close) vicinity of a weakly correlated state resulting from the bonding orbital (see Fig. \ref{fig:graphicalN23} right column). This single orbital Mott-insulator behavior shows a gradual increase in gapsize from \NbX{I} to \NbX{F*}. In the cases of \NbX{F}* and \NbX{Cl} the lower Hubbard bands are so close in energy to the weakly correlated bonding-band that charge transfer occurs, rendering these compounds at these doping levels correlated charge-transfer insulators.

\newpage

\newpage

\subsection{Trilayer \NbX{Br} at half-filling \label{app:trilayer}}
\begin{figure}[h!]
    \centering
    \includegraphics[width=0.49\textwidth]{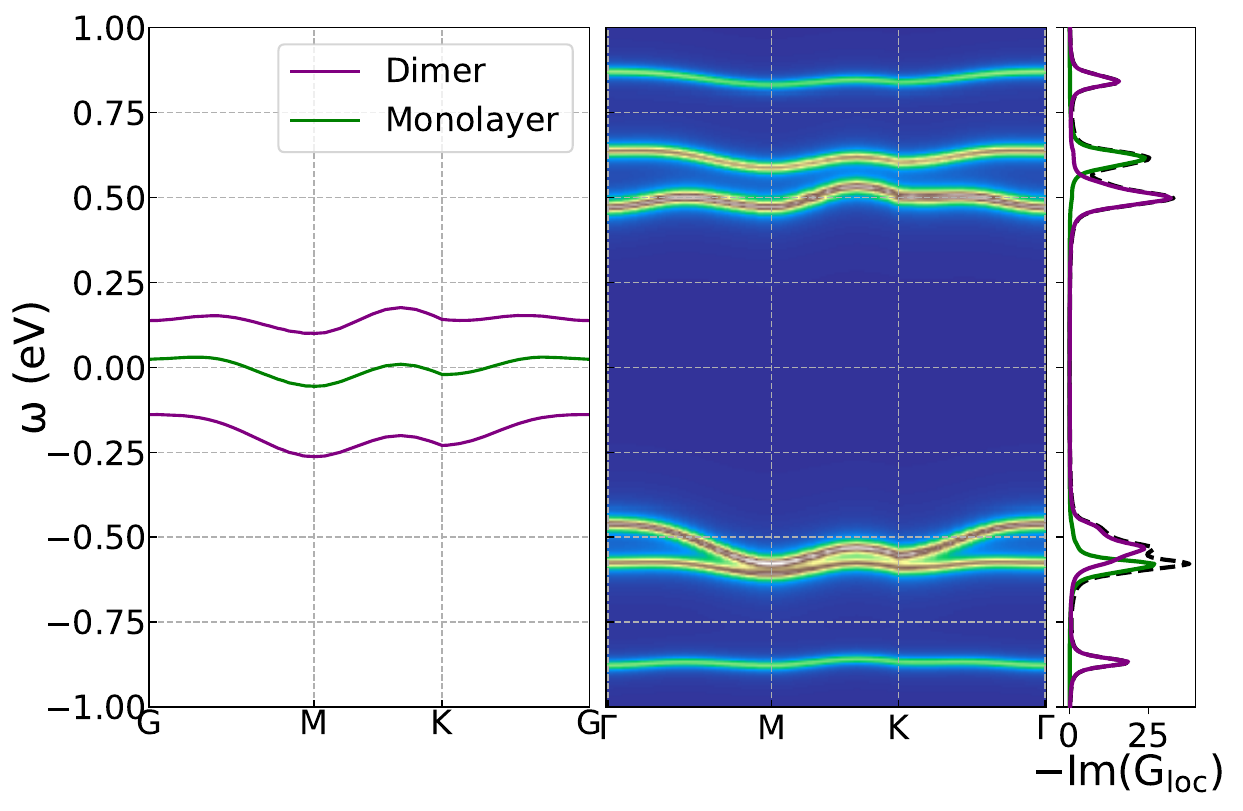}
    \caption{Trilayer DFT and Hubbard-I results. Only on-site screened coulomb interaction are taken into account. $U_0$ is taken from the bulk cRPA calculations.}
    \label{fig:trilayer}
\end{figure}
In the left panel of Fig.~\ref{fig:trilayer} we show the Wannierized DFT band structure of a trilayer \NbX{Br} using the molecular orbital basis and find three rather flat bands: a half-occupied metallic one surrounded by two fully occupied / empty ones. The metallic flat band results from the weakly hybridized ``surface'' monolayer and the other two are the bonding/anti-bonding split bands from the strongly hybridized bilayer. This is in line with the discussed band structure in Ref.~\cite{zhang_JDiode_2022}. As argued above, this mean-field picture is not properly reflecting the strong Coulomb interaction effects in the system. Therfore, we used this single-particle band structure together with the (molecular orbital) local Coulomb interaction matrix elements as obtained from our cRPA calculations (for the bulk material) to perform cluster DMFT calculations within the Hubbard-I approximation. The result is shown in the right panel of Fig.~\ref{fig:trilayer}. We see that the Coulomb interactions induce a gap with a set of three rather flat bands on either side of the Fermi energy. Projecting this interacting spectral function to the monolayer/bilayer basis shows that the gap is formed between bilayer states. In detail, the gap is formed, like in the bulk material, between the lower Hubbard band of the bonding orbital and the upper Hubbard band of the anti-bonding orbital. We also find a lower and upper Hubbard band of the initially metallic ``surface'' monolayer band.

The two systems (``surface monolayer'' and bilayer) are almost completely separated as a result of the small hybridization between them, consistent with the interpretation of a vdW material. We expect this electronic disentanglement to be a valid pictures for all finite-stacks with  unpaired surface layers. We further note that in finite-stack systems the screening to the local Coulomb interactions will be a function of the $z$-position in the stack. The local Coulomb interaction and with it the resulting correlated gaps could thus be different in thin and thick stacks and could vary with the layer distance to the top and bottom surface.

\newpage
\subsection{Magnetic Order in \NbX{X} Monolayers and Bulk\label{sec:ML_mag}}
Following the effective spin-lattice Hamiltonian given by Eq.~\ref{eq:Heisenberg} in the main text, our spin dynamics simulations  indicate that the ground state magnetic order for all bulk and monolayer of \NbX{X} exhibits conical spin spirals with in-pane propagation vector ${\bf q}$
\begin{equation}
   {\bf S}_i^A ({\bf q}, \phi^A, \theta)  =    
    \begin{pmatrix}
    \sin(\theta) \cos( {\bf q} \cdot {\bf R}_i^A +\phi^A) \\
    \sin(\theta) \sin( {\bf q} \cdot {\bf R}_i^A +\phi^A) \\
    \cos(\theta)
 \end{pmatrix}\, ,
\label{eq:spiral}
\end{equation}
where ${\bf S}_i^A$ is a classical spin of sub-lattice $A$ at site ${\bf R}_i^A$, $\theta$ is the cone angle, and $\phi^A$ the phase. In the absence of a magnetic field the spin spirals transforms from conical to flat with $\theta=\pi/2$.
In such a case, inserting Eq.~\ref{eq:spiral} into the spin-lattice Hamiltonian given by Eq.~\ref{eq:Heisenberg} results in the following expression for the energy density of the exchange interaction as a function of ${\bf q}$ and $\phi$
\begin{equation}
    E_\text{ex}({\bf q}) = -\frac{1}{2N_A} \sum_{AB} \sum_{i,j}^{N_A, N_B}J_{ij}^{AB}\cos\bigg({\bf q}\cdot{\bf R}_{ij}^{AB} + \Delta \phi^{AB}\bigg)\, ,
\label{eq:en_spiral}
\end{equation}
where the sum over $i$ is restricted to the number of magnetic sites $N_A$ in sublattice $A$ of the magnetic unit cell and $j$ runs over sites in sublattice $B$. ${\bf R}_{ij}^{AB} = {\bf R}_j^B - {\bf R}_i^A$ is a distance between two sites and $\phi^{AB}$ is a phase difference between spirals in sublatticies $A$ and $B$.
For all \NbX{X} compounds, the obtained exchange interactions \(J_{ij}\) rapidly diminish beyond the third nearest neighbors in the in-plane direction and the second nearest neighbors in the out-of-plane direction. Therefore, we truncate the sum over \(j\) to include only these neighboring distances. It is important to note that  monolayers have only one sub-lattice and \(\Delta \phi^{AB} = 0\), whereas in bulk materials there are two sub-lattices and \(\Delta \phi^{AB}\) may assume a finite value, potentially leading to magnetic ordering in the \(z\)-direction.

\subsubsection{\NbX{X} Monolayers}
\begin{figure}[h!]
    \centering
    \includegraphics[width=0.9\textwidth]{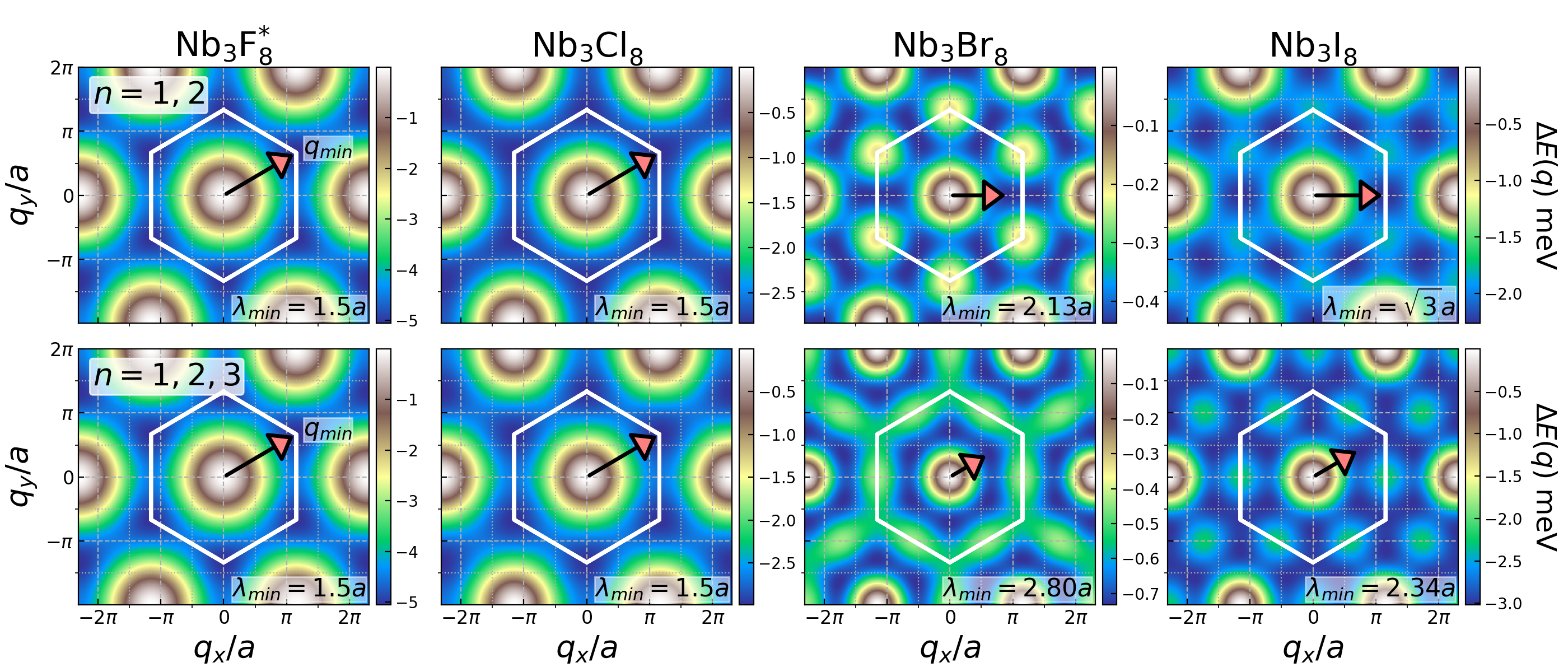}
    \caption{Energy of the exchange interaction $\Delta E({\bf q})$ as a function of the spin-spiral ${\bf q}=(q_x,q_y)$ vector for \NbX{X} monolayers. 
    The top and bottom panels illustrate the results for exchange interactions $J_{ij}^{(n)}$ truncated to two ($n=1,2$) and three ($n=1,2,3$) shells of nearest neighbors, respectively. The white hexagons represent the first Brillouin zone, while the arrows indicate the 
    ${\bf q}_{\text{min}}$ vectors corresponding to the lowest exchange interaction energy. $\lambda_{\text{min}}=2\pi/|{\bf q}|$ represents the period of the spin-spirals and $a$ is the in-plane lattice parameter.}
    \label{fig:spirals_ml}
\end{figure}

To determine the ground state parameters of the spin spirals in monolayers, we minimize the total energy described in Eq.~\ref{eq:en_spiral} with respect to the wave vector ${\bf q}$. Corresponding exchange interaction parameters $J_{ij}$ are provided in Tab.~\ref{tab:jijparameters_ML}.
Since, there is only one sublattice in monolayers, the phase \(\phi\) is irrelevant to the results.
Fig.~\ref{fig:spirals_ml} illustrates the energy of the exchange interaction as a function of the wave vector ${\bf q}$ and the truncation distance between interacting pairs, considering up to two and three shells of nearest neighbors. The \NbX{F}* and \NbX{Cl} monolayers exhibit identical propagation directions with the same value of $|{\bf q}_\text{min}|$, resulting in commensurate spin spirals with a periodicity of $1.5a$  (where $a$ is the in-plane lattice parameter), equivalent to a $120^\circ$-AFM order. While the contribution from third nearest neighbors is negligible for these two compounds, the situation changes significantly for the \NbX{Br} and \NbX{I} monolayers. Specifically, when accounting for the third nearest neighbors in \NbX{Br} and \NbX{I}, the direction of spin spiral propagation is altered, and the period of modulation increases, leading to incommensurate spin configurations.
The final information about the ground state spirals and corresponding energies of the exchange interaction are summarized in Tab.~\ref{tab:spirals_bulk_ml}.

{\renewcommand{\arraystretch}{1.2}
\begin{table}[t!]\centering
\caption{Magnetic ground state order in bulk and monolayer \NbX{X} compounds. 
Here, $\lambda = \frac{2\pi}{\bf q}$ represents the period of the spin spirals, while 
$\Delta \phi$ is the phase difference between the spirals in the two sublattices (applicable only to bulk structures). The parameter $\alpha_{\bf q}$ specifies the direction of spin spiral propagation relative to the in-plane unit cell vector ${\bf a}$, and $E_\text{ex}$ denotes the corresponding energy associated with the exchange interaction (in meV).
\label{tab:spirals_bulk_ml}}
\begin{tabular}{R{1.2cm}  R{1.2cm} R{1.2cm} R{1.2cm} R{1.2cm} | R{1.2cm} R{1.2cm} |  R{1.2cm} R{1.2cm} R{1.2cm} R{1.2cm}
}\Xhline{2\arrayrulewidth}
        & \multicolumn{4}{c|}{Bulk LT phase} & \multicolumn{2}{c|}{Bulk HT phase} & \multicolumn{4}{c}{ML}\\\cline{2-11}
        & Nb$_3$F$_8^*$
        & \NbX{Cl}
        & \NbX{Br} 
        & \NbX{I} 
        & \NbX{Cl}
        & \NbX{Br}
        & Nb$_3$F$_8^*$
        & \NbX{Cl}
        & \NbX{Br} 
        & \NbX{I}         
        \\ \hline\hline
\multicolumn{1}{l}{$\lambda$} & $1.5a$ &  $1.66a$& $3.56a$ & $3.40a$ &  $1.72a$ &  $4.00a$ & $1.5a$ & $1.5a$ & $2.8a$ & $2.34a$\\
\multicolumn{1}{l}{$\Delta \phi$} & $\pi$ &  $\pi$& $\pi$ & $\pi$ &  $\pi$ &  $\pi$ & & & &\\
\multicolumn{1}{l}{$\alpha_{\bf q}$} & 0 & 0 & 0 & 0 &  0&  $30^\circ$ & 0& 0& 0& 0\\\hline
\multicolumn{1}{l}{$E_\text{ex}$} &-3.99 & -37.61& -70.38&-185.98 &-3.62 & -3.94 & -5.06 & -3.99& -0.73& -3.02\\
\bottomrule
\end{tabular}
\end{table}}

\subsubsection{\NbX{X} Bulk}

Similar to the \NbX{X} monolayers, we investigate the magnetic properties of bulk compounds in both the low- and high-temperature phases. A notable distinction from the monolayer case is the presence of out-of-plane $J^\text{s}_\perp$ and $J^\text{s}_\perp$ exchange interactions in the bulk material, see Tab.~\ref{tab:jijparameters}.
Our spin dynamics simulations reveal that the ground state exhibits a spin spiral characterized by an in-plane propagation wave vector ${\bf q}$. Given this preference for in-plane propagation, we analyze the spirals in the two sublattices ($A = 1, 2$) within each layer, which are identified by a phase $\phi^\text{A}$. To determine the ground state parameters of these spirals, we minimize the total energy described in Eq.~\ref{eq:en_spiral} with respect to the wave vector ${\bf q}$ and the phase difference $\phi^\text{AB}$. The final results are summarized in Tab.~\ref{tab:spirals_bulk_ml}.

\begin{figure}[h!]
    \centering
    \includegraphics[width=0.75\textwidth]{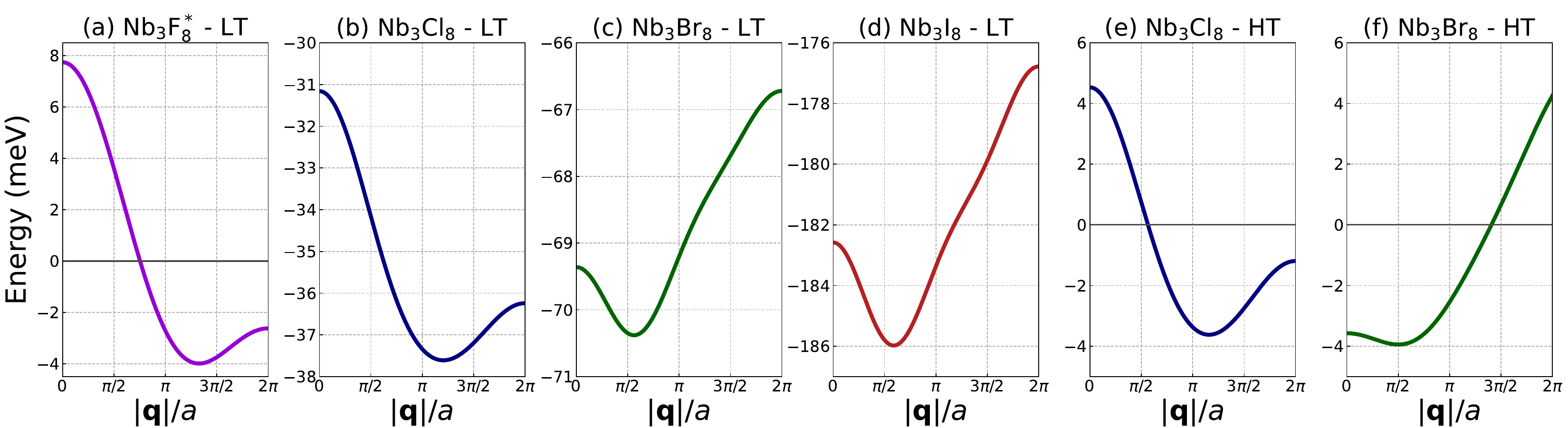}
    \caption{Energies of the exchange interactions as a function of ${\bf q}$ for (a-d)  all LT \NbX{X} compounds and (e-f) HT \NbX{Cl} and \NbX{Br}.}
    \label{fig:exchange_bulk}
\end{figure}

Fig.~\ref{fig:exchange_bulk} illustrates the energies of the exchange interactions as a function of $q$ for all LT \NbX{X} compounds, as well as HT \NbX{Cl} and \NbX{Br}. The spirals associated with these energy dispersions are characterized by minimized values of $\Delta \phi$ and $\alpha_{\bf q}$, as summarized in Tab.~\ref{tab:spirals_bulk_ml}. Since $\Delta \phi = \pi$ for all cases, the magnetic order at $q = 0$ is  AFM in the $z$-direction. From Fig.~\ref{fig:exchange_bulk} it is evident that the primary contribution to the exchange energy in LT \NbX{F} arises from in-plane spiralization. Conversely, for \NbX{Br} and \NbX{I}, the dominant contribution comes from out-of-plane AFM order, while the energy gain from in-plane spiralization is relatively minor. In the case of \NbX{Cl}, although the energy contribution from in-plane spiralization is more significant, it remains less than that from the AFM order in the $z$-direction.
Turning to the HT phases, the energy gain in \NbX{Cl}, much like that in LT \NbX{F}, is primarily due to in-plane magnetic ordering; meanwhile, for HT \NbX{Br}, it arises from the out-of-plane AFM order.

\newpage
\subsection{ARPES}\label{app:ARPES}
\begin{figure*}[h!]
    \centering
    \includegraphics[width=0.6\linewidth]{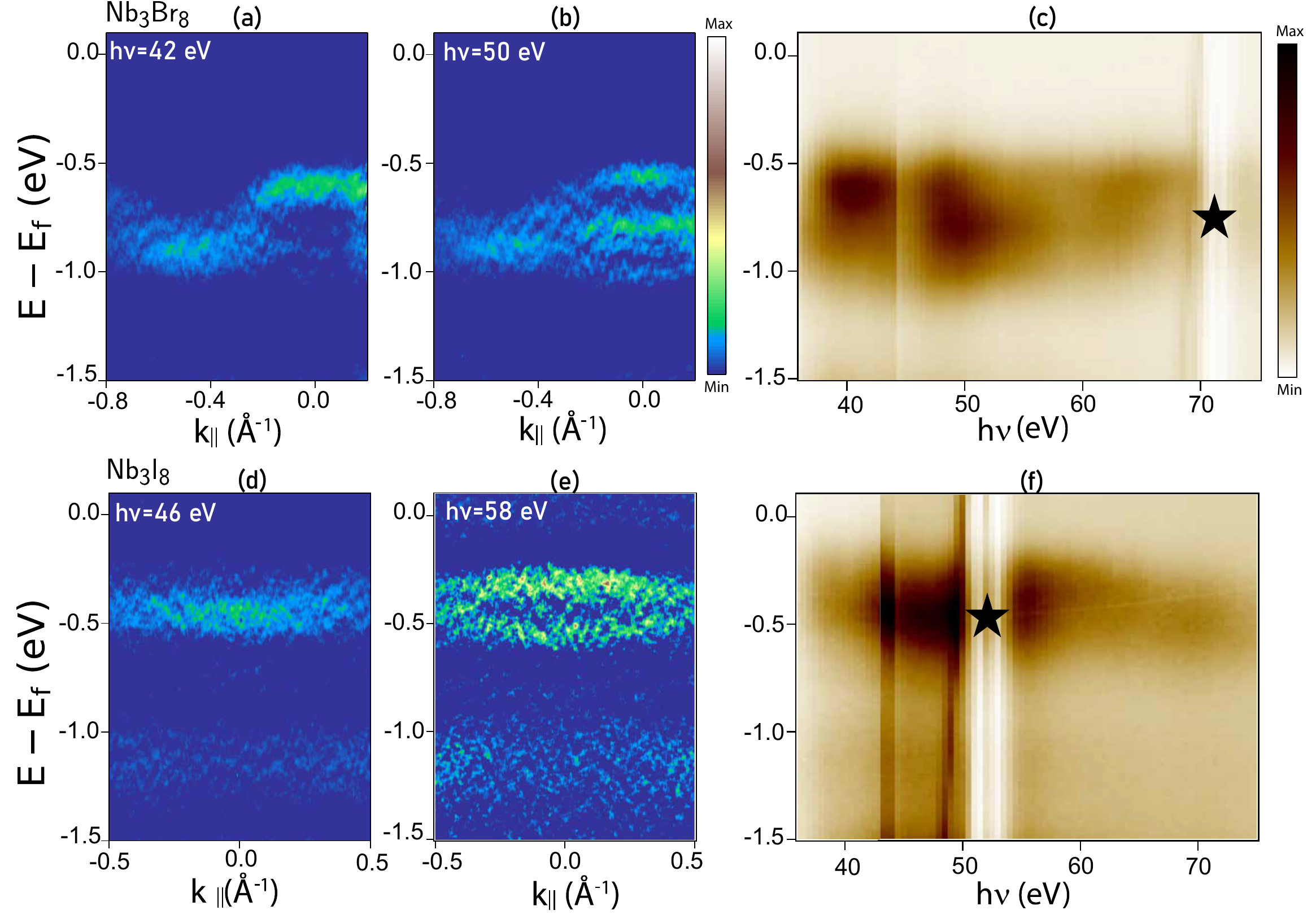}
    \caption{(a),(b),(d),(e) 2D curvature plots. (c) and (f) show the photon energy dependent ARPES data over a photon energy range of 36~eV to 75~eV at $k_\parallel = 0$.}
    \label{fig:ARPES_kz}
\end{figure*}

Figs. \ref{fig:ARPES_kz} (a,b,d,e) show the 2D curvature plots corresponding to the raw ARPES data in Figs.~\ref{fig:Hubbard-I} (e,f,g,h). Figs. \ref{fig:ARPES_kz} (c) and (f) show the raw ARPES data at $\overline{\Gamma}$ plotted against photon energy. The significant redistribution of spectral weight in energy, as visible in Figs.~\ref{fig:ARPES_kz}~(c) and (f) is interpreted as the result of probing different wavevectors along $\Gamma-A$ high-symmetry direction.

\FloatBarrier

\bibliography{bibliography}

\end{document}